\documentclass[11pt]{article}

\pdfoutput=1

\usepackage{hyperref}
\hypersetup{
  colorlinks=false  
}
\usepackage{algorithm}
\usepackage[noend]{algorithmic} 

\usepackage{caption}

\usepackage{siunitx}
\usepackage{pgf}
\usepackage{tikz}
\usetikzlibrary{arrows,automata,decorations.markings,positioning,calc}
\usepackage{graphicx}
\usepackage{amssymb}
\usepackage{amsmath}
\usepackage{url}
\usepackage[T1]{fontenc}
\usepackage{times}
\usepackage{verbatim}
\usepackage{subfig}
\usepackage{todonotes}
\usepackage{listings}
\usepackage{threeparttable}

\graphicspath{{gnuplot/}{.}}

\newcommand{\thresh}{\vartheta}

\newcommand{\PGDVD}{PGDVD}
\newcommand{\IMDBtwo}{IMDB-2gr}
\newcommand{\IMDBthree}{IMDB-3gr}
\newcommand{\PGDVDtwo}{PGDVD-2gr}
\newcommand{\PGDVDthree}{PGDVD-3gr}
\newcommand{\CensusIncome}{CensusIncome}

\newcommand{\scncnt}{\textsc{ScanCount}}
\newcommand{\looped}{\textsc{Looped}}
\newcommand{\addckt}{\textsc{TreeAdd}}
\newcommand{\kaddckt}{\textsc{SSum}}
\newcommand{\srtckt}{\textsc{SrtCkt}}
\newcommand{\schedcs}{\textsc{CSvCkt}}
\newcommand{\mgsk}{\textsc{wMgSk}}  
\newcommand{\mgopt}{\textsc{MgOpt}}
\newcommand{\dsk}{\textsc{DSk}}
\newcommand{\wheap}{\textsc{wHeap}}
\newcommand{\cdom}{\textsc{RBMrg}}  
\newcommand{\hashcnt}{\textsc{HashCnt}}
\newcommand{\wtwocta}{\textsc{w2CtA}}
\newcommand{\wtwoctn}{\textsc{w2CtN}}
\newcommand{\wtwocti}{\textsc{w2CtI}}
\newcommand{\wsort}{\textsc{wSort}}
\newcommand{\sopckt}{\textsc{SoPCkt}}
\newcommand{\wmgopt}{\textsc{wMgOpt}}

\title{Threshold and symmetric functions over bitmaps}
\author{Owen Kaser and Daniel Lemire}
\date{\today}
\begin{document}

\thispagestyle{empty}
\vspace*{0.3in}
{\centering
\begin{minipage}{4in}
\centering \LARGE Threshold and Symmetric Functions over Bitmaps\\[1ex]
\Large TR-14-001, Dept of CSAS\\[1.5ex]
\normalsize
\hfill
\parbox{1.2in}{
 Owen Kaser\\
 Dept. of CSAS\\
 UNBSJ\\
 Saint John, NB}
\hfill
\parbox{1.5in}{
 Daniel Lemire\\
LICEF, TELUQ\\
Universit\'e du Qu\'ebec\\
Montreal, QC }
\hfill\\[1ex]
\today\\[1ex]
\end{minipage}\\
}

\newpage
\setcounter{page}{1}
\maketitle

\begin{abstract}
Bitmap indexes are routinely used to speed up simple
aggregate queries in databases.  
Set operations such as intersections,
unions and complements can be represented as logical operations (\texttt{and},
\texttt{or}, \texttt{not}). However, less is known about the application of bitmap
indexes to more advanced queries. We want to extend the applicability
of bitmap indexes. As a starting point, we consider symmetric Boolean
queries (e.g., threshold functions). 
For example, we might consider stores as sets
of products, and ask for products that are on sale in 2 to 10 stores.
Such symmetric Boolean queries generalize intersection, union, and
T-occurrence queries.  

It may not be immediately obvious to an engineer how to use bitmap
indexes for symmetric Boolean queries. Yet, maybe surprisingly, we
find that the best of our bitmap-based algorithms are competitive with
the state-of-the-art algorithms for important special cases (e.g.,
MergeOpt, MergeSkip, DivideSkip, ScanCount).  Moreover, unlike the
competing algorithms, the result of our computation is again a bitmap
which can be further processed within a bitmap index.

We review algorithmic design issues such as the aggregation of many
compressed bitmaps. We conclude with a discussion on other advanced
queries that bitmap indexes might be able to support efficiently.

\let\thefootnote\relax\footnote{
This report is intended to be viewed electronically
and uses fonts and colours that make it unsuited for viewing on
paper.

}
\end{abstract}

\section{Introduction}

There are many applications to  bitmap indexes,
 from conventional databases (e.g., Oracle~\cite{874730}) all the way to
   information retrieval~\cite{Culpepper:2010:ESI:1877766.1877767} and including  on column stores~\cite{1083658}. We are primarily motivated by the application of bitmap indexes to common databases (i.e., row stores). In this case, it has long been established that bitmap indexes can speed up several queries corresponding, e.g.,  to intersections or unions (such as SELECT * WHERE A=1 AND B=2). We show that these good results extend to the more advanced queries we want to consider. 

Since bitmap indexes are widely used and basic Boolean operators over the bitmaps 
have been found useful in answering
queries, 
our broad goal is to investigate how to integrate symmetric Boolean 
functions (in particular,
threshold functions such as the Majority function) into the set of operations 
supported on bitmaps.  
This would permit queries that arbitrarily
combine such complex functions and the standard operations permitted on
bitmaps (OR, AND, XOR, perhaps NOT) and so forth.

Of course, the set of basic operations (typically depending
whether NOT is provided)  may be sufficient to synthesize any
required function, symmetric or otherwise.  However, the efficiency of
such approaches is not clear. To our knowledge, it has never been investigated in depth: the exception is Rinfret et al.~\cite{rinfret:bit-sliced-arithmetic} where two algorithms are compared on a closely related problem. 

The concrete contribution of this report is a study of an extensive set of alternative algorithms 
(see Table~\ref{tab:algos})
to compute threshold functions over bitmap indexes.
 Many of the alternatives can be generalized to 
handle arbitrary symmetric Boolean functions.
The theoretical analyses of these alternatives are summarized
in Tables~\ref{tab:complexity-uncompressed}~and~\ref{tab:complexity-rle-compressed}.
The notation used in the tables can be found in Table~\ref{tab:notation}.

\begin{table}
\begin{centering}
\caption{\label{tab:algos} Algorithms considered in this report.}
\small
\begin{tabular}{|lll|} \hline
Algorithm         & Source                                   & Section\\ \hline 
\scncnt           & \cite{Li:2008:EMF:1546682.1547171}       &  \S~\ref{sec:scancount}\\
\hashcnt          & variant of \scncnt                       &  \S~\ref{sec:scancount}\\
\mgopt            & \cite{Sarawagi:2004:ESJ:1007568.1007652} &  \S~\ref{sec:t-occurrence-algos}\\
\dsk              & \cite{Li:2008:EMF:1546682.1547171}       &  \S~\ref{sec:t-occurrence-algos}\\
\looped           &  novel                              &  \S~\ref{sec:looped-algo}\\
\addckt           &  novel                              &  \S~\ref{sec:adding-circuit-algos}\\
\kaddckt           &  related to \cite{rinfret:bit-sliced-arithmetic}                              &  \S~\ref{sec:adding-circuit-algos}\\
\schedcs          &  novel                              &  \S~\ref{sec:adding-circuit-algos}\\
\srtckt           &  novel                              &  \S~\ref{sec:sorting-circuit-algo}\\
\sopckt           &  novel                              &  \S~\ref{sec:circuit-algos}\\
\cdom             & modified from~\cite{arxiv:0901.3751}     &  \S~\ref{sec:cdom-algo}\\
\wheap            & \cite{Sarawagi:2004:ESJ:1007568.1007652} &  \S~\ref{sec:t-occurrence-algos}\\
\mgsk             & \cite{Li:2008:EMF:1546682.1547171}       &  \S~\ref{sec:t-occurrence-algos}\\
\wtwocta          & novel                               &  \S~\ref{sec:mergeable-count-algos}\\
\wtwoctn          & novel                               &  \S~\ref{sec:mergeable-count-algos}\\
\wtwocti          & novel                               &  \S~\ref{sec:mergeable-count-algos}\\
\wsort            & novel                               &  \S~\ref{sec:sorting-algo}\\

\hline
\end{tabular}\\
\end{centering}
\end{table}

We shall see that the characteristics of the bitmaps, such as the number of ones or the number of
distinct runs, have a major effect in determining the superior alternatives. 
This is determined by extensive experiments described in \S~\ref{sec:experiments}.
  To obtain results
that correspond to a practical applications of bitmap indexes, we focus on using threshold functions
over bitmap indexes to answer \emph{similarity} queries.

\paragraph{Similarity Queries:}

A similarity query presents a prototypical item.
We determine the criteria that this item meets, and then seek
all items that meet (at least) $T$ of these criteria.  For example, if a user liked a 
given movie, he might be interested in other
similar movies (e.g., same director, or same studio, or same leading star, or same date of release). As part of a recommender system,
we might be interested in identifying quickly
all movies satisfying at least $T$ of these criteria. 

Similarity queries could be used to study the vocabulary of a corpus of texts. Items are vocabulary words, and
the occurrence within a given text is a binary attribute.   Given a word such as ``pod'', we will seek words that occur
in many of the same texts that ``pod'' occurs in.   We might thus find ``whale'' and ``pea'', besides many 
words that occur in almost every text.     Once filtered of words that occur commonly across the corpus, we might produce
information that would interest a literary analyst or linguist.

Similarity queries  have previously been used with approximate 
string matching~\cite{Li:2008:EMF:1546682.1547171,Sarawagi:2004:ESJ:1007568.1007652}. 
In this case, items are small chunks
of text, and the occurrence of a particular $q$-gram is a
criterion\footnote{Their experiments used $q=3$.}.
In this previous work, each $q$-gram has a sorted list of integers that
specify the chunks of text containing it.
Ferro et al.~\cite{ferr:duplicates-qgrams} solve a very similar problem, 
but instead use a bitmap for each  $q$-gram\footnote{Their experiments
used $q=2$.}.  Montaneri and Puglisi~\cite{montanari2012near}
extend the bitmap approach to detect near-duplicate documents 
arriving over time.

A generalization of a similarity query presents \emph{several} prototypical items,
then determines the criteria met by at least one of them.   We then proceed
as before, finding all items in the database that meet at least $T$ of the
criteria.

Another generalization allows criteria to be assigned different weights, to
reflect their relative importance.  In our recommender system, we might
want to give more weight to having the right director than the right
movie studio.

Once the criteria have been defined, it is possible to use SQL to handle
the rest of the query.  For instance, one might use a 
query\footnote{MySQL syntax used.  Some databases use IIF or CASE instead of IF.}
like 
\lstset{language=SQL}
\begin{lstlisting}
SELECT * FROM movies WHERE  
  IF(Lead="Sean Penn",1,0) +
  IF(Studio="Paramount",1,0) +
  IF(Director="Ridley Scott",1,0) > 2
\end{lstlisting}

Assuming you have a bitmap index, can you do better than the row-scan that
would be done by a typical database engine?


\paragraph{Formulation}

We take $N$  \emph{sorted} sets over a universe of $r$~distinct values.
For our purposes, we represent sets as bitmaps using $r$~bits.  See
Table~\ref{tab:notation}.

  The sum of the cardinalities of the $N$ sets  is $B$. We apply a
threshold $T$ ($1\leq T \leq N$), seeking those elements that occur
in at least $T$ sets. Because the cases $T=1$ and $T=N$ correspond to intersections and unions, and are well understood, we assume that $2 \leq T \leq N-1$.  
These queries are often called $T$-overlap~\cite{li2013fast, behm2009space} or $T$-occurrence~\cite{Li:2008:EMF:1546682.1547171,jia2012eti} queries.
We show that, indeed, bitmap indexes can be used to 
speed up $T$-overlap queries (when compared with a row-store scan, see \S~\ref{sec:with-and-without-index}).

We can map a $T$-overlap query to a query over bitmaps using a Boolean
threshold function: given $N$~bits, the $T$-threshold function returns
true if at least $T$ bits are true, it returns false otherwise.
A (unary) bitmap index over a table has as many bitmaps as there are distinct
attribute values.  Each attribute value (say value $v$ of attribute $a$)
has a bitmap that encodes the set of row-ids (rids) that satisfy the criterion
$v=a$.  A $T$-overlaps query seeks rids that occur in at least $T$ of $N$ 
chosen sets.  Since each set is encoded as a bitmap, we need to compute
a bitwise threshold function over the $N$ chosen bitmaps.

Threshold functions are a subset of the symmetric Boolean functions.
They include the majority function: given $N$~bits, the majority
function returns true when $\lceil N/2\rceil$ or more bits are true, and it returns
false otherwise.

We denote the processor's native word length as
$W$ (currently\footnote{Common 64-bit PCs have SIMD instructions that work over 128-bit (SSE and AVX) and 256-bit (AVX2) vectors. These instructions might be used automatically by compilers and interpreters.}
$W=64$ is typical). An uncompressed bitmap will
have $\lceil r/W \rceil $~words. If you have $N$~bitmaps, then you
have $N \lceil r/W \rceil $~words.  To simplify, we assume $\log N < W < r$ as well 
\footnote{In this paper, $\log n$ means $\log_2 n$.} as $\log r \leq W$, which
would typically be the case in the applications we envision.  
Also, we assume that $B \geq N$, 
which would be true if there is no empty bitmap.

\begin{table}
\caption{\label{tab:notation} Notation used in analyses.}
\centering
\begin{tabular}{|cc|} \hline
Symbol & Meaning\\ \hline
$N$    & Number of bitmaps in query\\
$N_{\mathrm{max}}$    & Maximum value of $N$ allowed\\
$T$    & Minimum threshold\\
$r$    & length of bitmaps (largest index covered) \\
$B_i$  & $i^{\mathrm{th}}$ bitmap\\
$\mathcal{B}$ & $\{ B_i \}$\\
$|B_i|$& number of ones in $i^{\mathrm{th}}$ bitmap\\
$B$    & $\sum_i |B_i|$\\
$B'$   & number of ones not in $T-1$ largest bitmaps\\
$B''$  & number of ones not in $L$ largest bitmaps\\
$L$    & number of long bitmaps reserved in \dsk\\
$\textsc{RunCount}$ &number of runs of zeros and ones in collection of bitmaps\\
$\thresh(T, \mathcal{B})$ & threshold function over bitmaps\\
$\thresh(T, \{b_1, \ldots , b_N\})$ & threshold function over bits $b_i$\\
$W$    & machine word size\\\hline
\end{tabular}
\end{table}

\begin{table}
\begin{centering}
\begin{threeparttable}[b]
\caption{\label{tab:complexity-uncompressed} Time and memory complexity of some threshold algorithms over \textbf{uncompressed} bitmap indexes.
Our memory bounds exclude storage for the input and output bitmaps.
The \cdom\ is based on run-length encoding and is thus omitted. 
 Factor $rN/W$ is required to read all the input one word at a time. 
We assume that $rN/W \gg N \log N$. 
Actual bitmap
libraries may detect that a bitmap has no ones past some word and thus run faster.
}\footnotesize
\begin{tabular}{|lllll|} \hline
Algorithm          & Time complexity (big-Oh)                      & \multicolumn{2}{c}{Space complexity (big-Oh)}  & Comment \\
                   &                                               &  horizontal    & vertical        &         \\ \hline            
\scncnt            & $rN/W + B$                                    & $W$            & $r$             &         \\
\hashcnt           & expected $rN/W + B$                           & $W$            & $B$             &         \\
\wheap             & $rN/W + B\log N$                              & $N$            & $N$             &         \\
\mgsk              & $rN/W + B\log N$                              & $N$            & $N$             &  Note \tnote{1}\\
\mgopt             & $r(N-T)/W + B' (\log(N-T)+T)$                 & $N-T$          & $N-T$           &         \\
\dsk               & $r(N-L)/W + B''(\log(N-L)+L)$                 & $N-L$          & $N-L$           &         \\
\looped            & $rNT/W$                                       & $T$            & $rT/W$          &  Note \tnote{2}\\
\addckt            & $rN/W$                                        & $N$            & $rN/W$          &  Note \tnote{3}\\
\kaddckt           & $rN/W$                                        & $N$            & $rN/W$          &  Note \tnote{3}\\
\schedcs           & $rN/W$                                        & $N$            & $r\log(N)/W$    &         \\
\srtckt            & $\log^2(N) rN/W$                              & $N \log^2 N$   & $rN\log^2(N)/W$ &  Note \tnote{3}\\
\sopckt            & $N^T rN/W$                                    & $N^T$          & $r{N^T}/W$      &  Note \tnote{3}\\
\wtwoctn           & $rN/W + BN$                                   &  $WN$          & $B$             &        \\
\wtwocta           & $rN/W + BN$                                   &  $WN$          & $B$             & Note \tnote{4}\\
\wtwocti           & $rN/W + BN$                                   &  $WN$          & $B$             & Note \tnote{4}\\
\wsort             & $rN/W + B\log WN$                             &  $WN$          & $B$             & Note \tnote{5}\\

\hline
\end{tabular}
\begin{tablenotes}
\item[1] Pruning can allow skipping portions of all bitmaps, so $rN/W$ and $B$ can be
reduced, depending on data.
\item[2] While fewer than $7N$ temporary bitmaps are generated, register-allocation techniques would usually be able to
share space between temporaries.  In many cases, space for $o(N)$ bitmaps would suffice.
\item[3] Register allocation should be able to greatly reduce the actual space requirements.
\item[4] Pruning can reduce the amount of temporary data processed, but every input bitmap is completely scanned.
\wtwocti\ prunes more than \wtwocta . 
\item[5] Vertical implementation would  sort $B$ values, for a time of $O(rN/W + B\log B)$.
\end{tablenotes}
\end{threeparttable}
\end{centering}
\end{table}

\begin{table}
\begin{centering}
\begin{threeparttable}
\caption{\label{tab:complexity-rle-compressed}Time and memory complexity of some threshold algorithms over RLE-compressed bitmap indexes.
We assume that one can iterate over the ones in a RLE-compressed bitmap in $\Theta(1)$ time per one.
Horizontal or iterator-based implementations are not considered.}

\footnotesize
\begin{tabular}{|llll|} \hline
Algorithm         &  Time complexity bound (big-Oh)  & Space (big-Oh)       & Comment\\ \hline
\scncnt           &  $r+B$                           & $r$                  & Note \tnote{1}\\
\hashcnt          &  expected $B+(B/T)\log(B/T)$     & $B$                  & Note \tnote{2}\\
\wheap            & $B \log N$                       & $N$                  &        \\
\mgsk             & $B \log N$                       & $N$                  & Note \tnote{3}\\
\mgopt            & $B' (\log(N-T)+T)+B-B'$          & $N$                  & Note \tnote{3}\\
\dsk              & $B''(\log(N-L)+L)+B-B''$         & $N$                  & Note \tnote{4}\\
\looped           & $NT$ basic bitmap operations     & $T$ bitmaps          &        \\
\addckt           & $N$ basic bitmap operations      & $N$ bitmaps          & Note \tnote{5} \\
\kaddckt          & $N$ basic bitmap operations      & $N$ bitmaps          & Note \tnote{5} \\
\schedcs          & $N$ basic bitmap operations      & $\log N$ bitmaps    & \\
\srtckt           & $N \log^2 N$ basic bitmap ops    & $N \log^2 N$ bitmaps & Note \tnote{5} \\
\sopckt           & $N^T$ basic bitmap ops           & $N^T$ bitmaps        & Note \tnote{5}\\
\cdom             & $\textsc{RunCount} \log N$       & $N$                  &      \\
\wtwoctn          & $BN$                             & $B$                  &        \\
\wtwocta          & $BN$                             & $B$                  & Note \tnote{6}\\
\wtwocti          & $BN$                             & $B$                  & Note \tnote{6}\\
\wsort            & $B \log B$                       & $B$                  &        \\

\hline
\end{tabular}
\begin{tablenotes}
\item[1] Access pattern is efficient.
\item[2] Access pattern is inefficient.
\item[3] Pruning can reduce $B$.
\item[4] Pruning can reduce $B''$ and $B$.
\item[5]
With register allocation, the space complexity would typically be much lower than the value
shown, which merely bounds the number of temporary bitmaps created.
\item[6] Pruning reduces $N$.
\end{tablenotes}
\end{threeparttable}
\end{centering}
\end{table}

For the remainder of the paper we assume that a bitmap index has been
created over the table, where each attribute value is represented by
a bitmap (compressed or otherwise).  Perhaps the index is smaller 
than the original table and can fit in main memory; if not,  we assume
that at least the $N$ bitmaps involved in the threshold query can
be simultaneously stored in memory.  Thus, even if the $N$ bitmaps must be
initially loaded from disk, we only consider in-memory computation.
For even larger datasets, we could  assume that the table has been horizontally
fragmented (the fragments would be much larger than with a row store), such
that, for any fragment,  any $N$ bitmaps fit in memory.
This paper does not consider external-memory algorithms that can
compute thresholds without using much main memory.

\paragraph{Lower Bound:}

Towards a lower bound for the problem, note if the output indicates
that $X$ entries meet the threshold, we must have processed input
such that at least $TX$ predicate-satisfaction events have been observed.
If each such observation triggers $\Omega(1)$ work (as it does with
\scncnt\ (\S~\ref{sec:scancount}), when a counter is incremented), this implies an $\Omega(TX)$
lower bound.  However, this leaves open the possibility of using
bit-level parallelism (readily available in bitmap inputs) to process
several events per machine operation. (See \S~\ref{sec:circuit-algos}.)  
It also leaves open the possibility
of using Run Length Encoding (RLE), whereby many consecutive
events can be succinctly represented and processed together.
We present such an approach later in \S~\ref{sec:cdom-algo}.

While it might superficially appear that a lower-bound could be based on
a need to read the entire input, note that sometimes there is no such
need.  For instance, suppose the threshold is $T$ and we have already processed
$N-1$ of our $N$~input bitmaps. It is only necessary to inspect the 
last bitmap at positions where the current count is $T-1$.  Other positions
already
achieve the threshold, or have no possibility of achieving it.   The
strategy of skipping portions of the input has been used in prior algorithms
noted in Table~\ref{tab:algos}.  Theoretically, they typically assume
the ability to do random access on the input, so we can jump over irrelevant
portions.  The sorted-list-of-integers representation facilitates
this kind of action, as do uncompressed bitmaps.  However, word-aligned compressed
bitmap representations typically do not support the required access
in constant time~\cite{DBLP:journals/tods/WuOS06}.  Nevertheless, we have been able to adapt this idea to
compressed bitmaps: it may require more than constant time to skip over
input, but we can do the work in a small fraction of the time it would
take to process otherwise.

\section{Background}

We next consider relevant background on bitmaps, Boolean functions and two major 
implementation approaches for bitmap operations.

\subsection{Bitmaps}
\label{sec:bitmaps}

We consider compressed and uncompressed bitmaps.  The \emph{density} of a bitmap
is the fraction of its bits that are ones.  A bitmap with very low density is
\emph{sparse}, and such bitmaps arise in many applications.

\paragraph{Uncompressed Bitmaps}
An uncompressed bitmap represents a sorted set $S$ over $\{0,1,\ldots,r\}$ using 
$\lceil (r+1)/W \rceil$ consecutive words.  The $W$ bits in the first word
record which values in $[0,W-1]$ are present in $S$.  The bits in the second word
record the values in $[W,2W-1]$ that are in $S$, and so forth. 
For example, the set $\{1,2,7,9\}$ is represented as
10000110 00000010 with $W=8$; the number of ones is equal to the cardinality of the set.

Uncompressed
bitmaps have the advantages of a fixed size (updates do not change the size)
and an efficient membership test.  However, if $r$ is large, the bitmap
occupies many words---even if it is representing a set with few elements.

The Java library contains a \textsc{BitSet} class~\cite{openjdk-bitset-sourcecode} which implements an uncompressed bitmap data structure.  In most respects,
it is an uncompressed bitmap, but it does not  have a fixed size.
The number of words it occupies is $\lceil (\hat{r}+1)/W \rceil$, where $\hat{r} = \max{S}$.
Note that $\hat{r}$ might be much less than $r$; also note that if a new element
larger than $\hat{r}$ is inserted into $S$, the \textsc{BitSet} may need to grow.
Not storing the trailing empty words is an optimization that would slightly complicate
operations over a \textsc{BitSet}.  But in the case of an AND operation between
a \textsc{BitSet} representing $S_1$ (with largest value $\hat{r}_1$)
and one representing $S_2$ (with largest value $\hat{r}_2$),  bitwise AND instructions
would only need to be done over $\lceil ( \min(\hat{r}_1, \hat{r}_2) + 1)/W \rceil$~words, rather than $\lceil (r+1)/W \rceil$.  Similar optimizations are possible for
OR and XOR.  One consequence is that the speed of an operation over \textsc{BitSet}s
depends on the data involved.

\paragraph{Compressed Bitmaps}

In many situations where bitmaps are generated, the processes creating the
bitmaps generate long runs of consecutive zeros (and sometime long runs of ones).
The number of such runs is called the \textsc{RunCount} of a bitmap, or
of a collection of bitmaps~\cite{rlewithsorting}.  

Though there are alternatives~\cite {navarro2012fast}, the most
popular compression techniques are based on the (word-aligned) RLE
compression model inherited from Oracle (BBC~\cite{874730}):
WAH~\cite{wu2008breaking},
Concise~\cite{Colantonio:2010:CCN:1824821.1824857},
EWAH~\cite{arxiv:0901.3751}), COMPAX~\cite{netfli}.  These techniques
typically use special marker words to compress fill words: sequences
of $W$~bits made of all ones or all zeros. When accessing these formats, it
is necessary to read every word to determine whether it indicates a
sequence of fill words, or a \emph{dirty}~word ($W$~consecutive bits
containing a mix of ones and zeros). The EWAH format supports a limited
form of skipping because it uses marker words not only to mark the
length of the sequences of fill words, but  also 
to indicate the length of the sequences of dirty words. Because of
this feature, it is possible to skip runs of dirty words when using
EWAH.

Compressed bitmaps are often appropriate for storing sets that cannot be
efficiently handled by uncompressed bitmaps.  For instance, consider
the bitmap consisting of a million zeros followed by a million ones.
This data has two runs (\textsc{RunCount}=2) but a million ones.
It can be stored in only a few words.

RLE compressed bitmaps are not an efficient way to store
extremely sparse data that does not have dense clusters.
Sparse data with very long runs of zeros between elements will result
in a marker word and a dirty word for each one bit.  This would be even
less efficient than explicitly listing the set elements.

Software libraries for compressed bitmaps will typically include
an assortment of basic Boolean operations that operate directly
on the compressed bitmaps.  One would expect to find operations
for AND, OR, and often one finds XOR, ANDNOT, and NOT.  Some libraries
support only binary operations, whereas others support \emph{wide}
queries (for instance, a ``wide'' AND would allow us to intersect
four bitmaps in a single operation, rather than having to AND
bitmaps together pairwise).  Even if intersection is eventually
implemented by pairwise operations, the wide query allows the library
to decide how to schedule the pair-at-a-time operations, 
and this may lead to a performance gain~\cite{1316694}.

\subsection{Boolean Functions  and Circuits}

For relevant background on Boolean functions, see Knuth's book~\cite{KnuthV4A}.
Boolean functions can be computed by Boolean circuits that 
compose some basic set of operators, such as NOT
and binary AND and OR.

Some Boolean functions  are \emph{symmetric}.
These functions are unchanged under any permutation
of their inputs.  I.e., a symmetric function is completely determined
if one knows the number of ones (the Hamming weight) in its
inputs.  An example symmetric function outputs 0 iff the
Hamming weight of its inputs is a multiple of 2: this is the XOR
function.   Another
example is the function
that indicates whether the Hamming weight of its inputs
is less than 5.  The ``delta'' function indicates whether
the Hamming weight of its inputs is some specified value,
and another symmetric function  indicates whether between 5 and 8
of its inputs are ones.

Some Boolean functions are monotone;  changing an input
of such a function from 0 to 1 can never cause the output to
go from 1 to 0.  It is well known that the set of monotone Boolean
functions equals the set of Boolean functions that can be
implemented using only AND and OR operators.

\subsection{Threshold Functions}

Threshold functions are those Boolean functions that
are both symmetric and monotone.  Such a function indicates
whether the Hamming weight of its inputs exceeds some
threshold $T$.  Function $\thresh(T, \{b_1, \ldots , b_N\})$
denotes the function that outputs 1 exactly when at least $T$
of its $N$ inputs are 1.
Several well-known functions are threshold functions:
Note that $\thresh(1,\{b_1, \ldots, b_N\})$ is merely a wide OR,
$\thresh(N,\{b_1, \ldots, b_N\})$ is merely a wide AND, and
$\thresh(\lceil N/2 \rceil ,\{b_1, \ldots, b_N\})$ is 
the majority function. 
 Efficient compressed bitmap algorithms for wide
AND, OR and XOR have been studied before~\cite{1316694}.

Knuth~\cite[7.1.1\&7.1.2]{KnuthV4A} discusses threshold functions and
symmetric functions, as well as techniques by which general symmetric
functions can be synthesized from threshold functions.

There are a number of identities involving threshold functions
that might be useful to a query optimizer, if thresholds were
incorporated into database engines.
For instance, we have that 
\begin{equation*}
\overline{\thresh(T, \{b_1, \ldots, b_N\})} = \thresh(N-T, \{ \bar{b}_1, \ldots \bar{b}_N \}).
\end{equation*}

If the efficiency of a threshold algorithm depends on the density of its inputs, this 
identity can be helpful, because it can convert very dense bitmaps  into 
sparse bitmaps, or vice-versa.

In general, threshold functions may permit weights on the various inputs.  However, if
the weights are positive integers, we can simply replicate input $b_i$ so that it 
occurs $w_i$ times.  In this way we give the input an effective weight of $w_i$.  This
approach may be practical if weights are small.  Otherwise, the resulting threshold
query may be impractically wide.

Unweighted threshold functions (in the guise of $T$-overlap queries)
 have been used for approximate searching.
Specifically,  Sarawagi and Kirpal~\cite{Sarawagi:2004:ESJ:1007568.1007652}
show how to avoid unnecessary and expensive pairwise distance computations (such
as edit-distance computations) by using threshold functions to screen out
items that cannot be approximate matches.  Their observation was that 
strings $s_1$ and $s_2$ must have many ($T$)  $q$-grams in common, if they have
a chance of being approximate matches to one another.  Given $s_1$ and
seeking suitable $s_2$ values, 
we take the set of $q$-grams of $s_1$.  Each $q$-gram is associated with
a set of the words (more specifically, with their record ids) that
contain that $q$-gram at least once.  Taking these $N$ sets, we
do a threshold function to determine values $s_2$ that can be compared
more carefully against $s_1$.  Using $q$-grams, Sarawagi and Kirpal showed that
$T=|s_1| + q - 1 - k q$ will not discard any string that might be within
edit distance $k$ of $s_1$.  In applications where $k$ and $q$ are small
but the strings are long, this will create queries where $T \approx N$.

\subsection{Horizontal and Vertical Implementations}

Implementations of bitmap operations can be described as ``horizontal''
or ``vertical''.  The former implementation produces its output
incrementally, and can be stopped early.  The latter implementation
produces the entire completed bitmap.   Horizontal approaches would
be preferred if the output of the algorithm is to be consumed
immediately, as we may be able to avoid storing the entire bitmap.
As well, the producer and consumer can run in parallel.
Vertical implementations may be simpler and have less overhead.  
If the output of the algorithm needs to be stored because it is 
consumed more than once, there may be no substantial space 
advantage to a horizontal implementation. 

With uncompressed bitmaps, a particularly horizontal implementation
exists: a word of each input is read, after which one word
of the output is produced.

\section{An index is better than no index}
\label{sec:with-and-without-index}

\begin{algorithm}
\centering
\begin{algorithmic}[1]
\REQUIRE A table with $D$ attributes.  A set $\kappa$ of $N \leq D$ attributes, and for each such attribute a desired value. Some threshold $T$.
\STATE $s$ a set, initially empty
\FOR{each row in the table}
\FOR{for each attribute $k$ in $\kappa$}
\IF{attribute $k$ of the row has the desired value}
\STATE increment counter $c$
\ENDIF
\ENDFOR
\IF{$c\geq T$}
\STATE add the row (via a reference to it) to $s$
\ENDIF
\ENDFOR
\STATE return the set of matching rows $s$
\end{algorithmic}

\caption{\label{alg:rowstore} Row-scanning approach over a row store.
}
\end{algorithm}

Perhaps a simple T-occurrence query can be more effectively answered without
using a bitmap index.  In that case, our work would be of less value\footnote{
Good algorithms over bitmaps would still be essential, in the event that 
the T-occurrence operation is required over a set of computed bitmaps, for
which no corresponding table exists.}.
 Algorithm~\ref{alg:rowstore} shows such an approach, and Table~\ref{tab:with-and-without-index}
shows that it is outperformed on both compressed and uncompressed bitmaps by
one of the algorithms studied (\scncnt , which is often outperformed by
other bitmap algorithms).

If $D \gg N$ and the $N$ attributes involved in the computation are scattered
throughout the record, the row-scanning approach will not make good use of the 
memory system---few of the $N$ important attributes will be in any given
cache line.

Moreover, interesting datasets often do not fit in main memory.
Horizontal partitioning could be used, loading a group of rows,
operating on them, then loading the next group of rows.  This will
still suffer from the inefficiencies inherent in row stores.  
 
It is reassuring to see that a bitmap index can answer our queries faster than 
they would be computed from the base table.
We made 30 trials, 
using the first $10^4$ rows of a dataset (\CensusIncome, described in \S~\ref{sec:real-datasets})
with 42 attributes.
We randomly chose one value per attribute and randomly chose a threshold between 2 and 41.

In Table~\ref{tab:with-and-without-index}
we see that  using either a \textsc{Bitset} or EWAH index for this query is four times
 better than scanning the table. The advantage persists, but is smaller, when a similarity
query is done.   The results are consistent if we use the first $10^4$~rows of a 10-attribute
dataset\footnote{ The other real datasets used for
experiments in \S~\ref{sec:experiments} 
had hundreds to many thousands of binary attributes; it would have been
unreasonable to test a row store with them.  \textsc{KingJames10d}
was selected because we had used it in prior work and it was derived from
text data, as were the unsuitable datasets used in our experiments.}
(\textsc{KingJames10d}~\cite{webb2013}).


\begin{table}
\caption{ \label{tab:with-and-without-index} Time (ms) required for queries.  Top: Random attributes.  Bottom: similarity query.}
\centering
\begin{tabular}{|c|rr|} \hline
Algorithm           & \textsc{CensusIncome}     &   \textsc{KingJames10D} \\ \hline
EWAH \scncnt        &  5214                     & 1017 \\
\textsc{BitSet} \scncnt      &  5168                     & 805  \\
Row Scan (no index) & 21,275                    & 5114 \\ \hline
EWAH \scncnt        & 15,913                     & 3590 \\
\textsc{BitSet} \scncnt      & 16,999                    &  3574  \\
Row Scan (no index) & 23,129                    & 5063 \\ \hline
\end{tabular}

\end{table}

\section{Approaches for Threshold Functions}

This report focuses on threshold functions but notes approaches that generalize easily
to handle all symmetric functions.  It also notes
 approaches that can be modified to
support delta queries or interval queries.

\subsection{\cdom} 
\label{sec:cdom-algo}
 
Algorithm \cdom\ 
is a refinement of an algorithm presented in  Lemire et al.~\cite{arxiv:0901.3751},
and it works for bitmaps that have been run-length encoded.
See Algorithm~\ref{algo:genrunlengthmultiplefaster}.  The approach considers
runs similarly to integer intervals.  For each point, the set of
intervals containing it is the same as the set of intervals containing its neighbour,
unless the point or its neighbour begins a run.   Each bitmap provides a sorted sequence
of intervals. Heap $H$ enables us to quickly find, in sorted order, those points where
intervals begin (and the bitmaps involved).  At such points, we calculate the function
on its revised inputs; in the case of symmetric functions such as threshold, this can be done
very quickly.  As we sweep through the data, we update the current count.  Whenever a
new interval of ones begins, the count increases; whenever a new interval of zeros
begins, the count decreases.   For regions where no interval changes,
the function will continue to have the same value.  Assuming $\log L \leq W$,
the new value of a threshold function can be determined in $\Theta(1)$ time whenever
an interval changes.  (The approach can be used with Boolean functions in general, but 
evaluating a Boolean function over $N$ inputs may be required even when only one input
has changed, taking more than constant time.)

Every run passes through a $N$-element heap, giving a running time that is in
$O(\textsc{RunCount} \log N)$. 
One can implement the $N$ required iterators in
$O(1)$ space each, leaving a memory bound of $O(N)$.

The algorithm would be a reasonable addition to compressed bitmap index libraries that are RLE-based.   We have added it to JavaEWAH library~\cite{JavaEWAH}
as of version 0.8.

As an extreme example where this approach would excel, consider a case where each bitmap is either
entirely ones or entirely zeros.  Then \textsc{RunCount} = $N$, and in $O(N \log N$) time we can
compute the output, regardless of $r$ or $B$.

\begin{algorithm}
\centering
\begin{algorithmic}
\STATE \textbf{INPUT:} $N$ bitmaps $B_1, \ldots, B_N$ 
\STATE $I_i \leftarrow$ iterator over the runs of identical bits of  $B_i$
\STATE $\Gamma \leftarrow$ a new buffer to store the aggregate of   $B_1, \ldots, B_N$  (initially empty)
\STATE $\gamma \leftarrow$ the bit value determined by $\gamma(I_i,\ldots, I_N)$
\STATE $H \leftarrow$ a new  $N$-element min-heap storing ending values of the runs and an indicator of which bitmap
\STATE $a' \leftarrow 0$
\WHILE{some iterator has not reached the end}
\STATE let $a$ be the minimum of all ending values for the runs of $I_1, \ldots, I_N$, determined from $H$
\STATE append run $[a',a]$ to $\Gamma$ with value $\gamma$
\STATE $a' \leftarrow a+1$
\FOR {iterator $I_i$  with a run ending at $a$ (selected from $H$)}
\STATE increment $I_i$ while updating $\gamma$ 
\STATE let $b$ be the new run-ending value for $I_i$
\IF{$b$ exists}
  \STATE increase the key of the top item in $H$ from $a$ to $b$ 
\ENDIF
\ENDFOR
\ENDWHILE
\end{algorithmic}
\caption{\label{algo:genrunlengthmultiplefaster} Algorithm \cdom .
}
\end{algorithm}

The EWAH implementation of \cdom\  processes runs of clean words as described.
If the interval from $a'$ to $a$ corresponds to $N_{\textrm{clean}}$ bitmaps with clean runs, of which
$k$ are clean runs of ones, the implementation distinguishes three  cases:
\begin{enumerate}
\item $T-k \leq 0$: the output is 1, and there is no need to examine the $N  -  N_{\textrm{clean}}$~bitmaps that contain dirty words.  This pruning will help cases when $T$ is small.
\item $T-k >  N  -  N_{\textrm{clean}}$: the output is 0, and there is no need to examine the dirty
words.  This pruning will help cases when $T$ is large.
\item $1 \leq T-k \leq N  -  N_{\textrm{clean}}$: the output will depend on the dirty words.  We can
do a $(T-k)$-threshold over the $N  -  N_{\textrm{clean}}$ bitmaps containing dirty words. 

We process the  $N  -  N_{\textrm{clean}}$ dirty words as follows.
If $T-k=1$ (resp. $T-k = N  -  N_{\textrm{clean}}$), we compute the bitwise OR (resp. AND) between the dirty words. 
 Otherwise, if $T-k\geq 128$, we always use 
 \scncnt{} using 64~counters (see \S~\ref{sec:scancount}). Otherwise, we compute the number of ones (noted $\beta$) 
in the 
 dirty words. This can be done efficiently in Java 
 since the  \texttt{Long.bitCount} function on desktop processors is typically
 compiled to fast machine code. If $2 \beta \geq (N  -  N_{\textrm{clean}})(T-k)$, 
 we use the \looped{} algorithm (\S~\ref{sec:looped-algo}), otherwise we
 use \scncnt{} again. We arrived at this particular approach by trial and error: 
 we find that it gives reasonable performance.
\end{enumerate}

\subsection{Counter-Based Approaches} \label{sec:scancount} 
In information retrieval, it is common practice to solve threshold queries using sets of counters~\cite{Perry01021983}.
The simple \scncnt\ algorithm of Li et al.~\cite{Li:2008:EMF:1546682.1547171}
uses an array of counters, one counter per item.  The input is scanned, one bitmap
at a time. If an item is seen in the current bitmap, its counter is incremented. 
 A final pass
over the counters can determine which items have been seen at least $T$
times.  In our case, items correspond to positions in the bitmap.
If the maximum bit position is known in advance, if this position is not
too large,  
and if you can efficiently iterate the bit positions in a bitmap, 
then \scncnt\ is easily coded.  These conditions are frequently met when the bitmaps
represent the sets of rids in a table that is not exceptionally large.

\scncnt\ and other counter-based approaches can handle arbitrary symmetric functions, since one
can provide a user-defined function mapping $[0,N]$ to Booleans.  However,
optimizations are possible when more sophisticated counter-based 
approaches compute threshold functions.

To analyze \scncnt , note that it requires $\Theta(r)$ counters.  We assume $N < 2^W$, so
each counter occupies a single machine word.  Even if counter initialization can be avoided, 
each counter is compared against $T$.  
(An alternative is to compare only after each increment~\cite{Li:2008:EMF:1546682.1547171}, but since the counters achieve $T$ in
an unpredictable order, a sorting phase is required before the compressed bitmap index
is constructed.  In many cases, this extra overhead would be excessive.)
Also, the total number of counter increments is $B$.  Together, these imply a
time complexity of $\Theta(r+B)$ and a space complexity of $\Theta(r)$, for compressed
bitmaps.  For uncompressed bitmaps, we potentially have to wade through regions that store
only zeros and do not contribute toward $B$.  This adds an extra $Nr/W$ term to the 
time\footnote{This is easily done if one assumes an O(1) instruction to compute
the number of leading zeros in a $W$-bit word.  For instance, the x86 has a \texttt{lzcnt}
instruction and the ARM has \texttt{clz}. Otherwise, there are approaches that
use $O(\log W)$ operations~\cite[5.3]{warr:hackers-delight-2e}, and it appears the 
OpenJDK implementation of \texttt{Integer.numberOfLeadingZeros()} uses such an approach.
Without the assumption of an $O(1)$ instruction, the term $Nr/W$ would become
$\frac{Nr\log W}{W}$. 
}.
Aside from the effect of
$N$ on $B$ (on average, a linear effect), note that this algorithm does not depend on $N$.

The \scncnt\ approach fits modern hardware well:
note that the counters will be accessed in sequence, during the $N$
passes over them when they are incremented.  
Given that 16 \texttt{int} counters fit into a typical 64-byte
cache line, the performance improvement from fewer increments is likely
to be small until average selectivity of $a_j = v_j$ drops to values
near 1/16. (Far below that value and we may find two successively accessed counters are in
different cache lines.  Well above it, and two successively accessed counters are likely
to be in the same cache line, and we will not get a cache miss on both.)

Experimentally, we found that using \texttt{byte} counters when $N < 128$
usually brought a small (perhaps 15\%) speed gain compared with \texttt{int} counters.
Perhaps more importantly, 
this also quarters the memory consumption of the algorithm.
One can also experiment with other memory-reduction techniques: for instance, if $T < 128$, 
one could use a saturating 8-bit counter.
Experimentally, we found that the gains usually were less than the losses
to make the additional conditional check (needed to avoid incrementing the counter once
it had achieved either $T$ or the maximum byte values).
(In the special case where $T=1$, that is, when we want to compute
the union, the \scncnt\ array could use single-\emph{bit} counters.
These counters would effectively be an uncompressed bitmap. 
Indeed, the JavaEWAH library~\cite{JavaEWAH} uses this observation in its OR 
implementation.)

Based on our experimental results,
the \scncnt\ Java implementation used in \S~\ref{sec:experiments} switches
between \texttt{byte}, \texttt{short} and \texttt{int} counters based on
$N$, but does not use the saturating-count approach.

\scncnt\ fails when the bitmaps have extreme $r$ values.
If we restrict ourselves to bitmaps that arise within a bitmap index, this implies
that we have indexed a table with an extreme number of rows.
However, hashing can be employed to use an approach like \scncnt\  on data where $r$ is especially large.  
Algorithm \hashcnt\ uses a map from integers (items) to integers (their counts).
We only need counters for as many distinct items as actually exist in our data (which
could be at most $B$).
In Java, we can use \texttt{HashMap} from the \texttt{java.util} package.  Unfortunately, the constant-factor overheads
in hashing are significantly higher than merely indexing an array.  Also, \scncnt\ indexes
the array of counters in $N$ ascending passes, which should lead to good cache performance.
Hashing will scatter accesses to boxed Integer objects that could be anywhere in memory.
Somewhat higher-performance int-to-int hash maps are found in libraries such as GNU Trove~\cite{trove}.
However, hashing would still lead to random-looking accesses through a large array, and
poor cache performance should be anticipated.  Note also that, with compression, the
output bitmap must be constructed in sorted order.
Since there can be at most $B/T$ ones in the output bitmap, we have an $O(\frac B T  \log \frac B T)$
cost for construction.

\subsubsection{Sorting to Count}
\label{sec:sorting-algo}
Algorithm \wsort\ counts by converting each bitmap into a (presorted) list
of positions, concatenating such lists, sorting the result, and then
scanning to see which items are repeated at least $T$ times in the
large sorted list.   If we process uncompressed bitmaps horizontally,
this na\"\i ve approach may work well enough, especially given that
sorting libraries are typically very efficient.  In the worst case,
in one horizontal step our bitmap fragments have $WN$ ones, requiring
that we use
$O(WN \log (WN))$ time to sort, and requiring $O(WN)$ space.
Over the entire horizontal computation, we use $O(B \log (WN))$ time.
(We have that $B=WN$ in the our worst-case scenario.  But in others, $WN$ is an
upper bound on the number of ones in a horizontal step, and $B$ is
the total number of ones in all horizontal steps.)
However, if we are not using horizontal computation, we unpack
$B$ data items and sort them in $\Theta(B \log B)$ time.

\subsubsection{Mergeable-Count structures}
\label{sec:mergeable-count-algos}
 
A common approach to computing intersections and unions of several
sets is to do it two sets at a time. To generalize the idea to
symmetric queries, we represent each set as an array of values coupled
with an array of counters. For example, the set $\{1,14,24\}$ becomes
$\{1,14,24\}, \{1,1,1\}$ where the second array indicates the frequency
of each element (respectively). If we are given a second set
($\{14,24,25,32\}$), we supplement it with its own array of counters
$\{1,1,1,1\}$ and can then merge the two: the result is the union of
two sets along with an array of counters ($\{1,14,24,25,32\},
\{1,2,2,1,1\}$). From this final answer, we can deduce both the
intersection and the union, as well as other symmetric operations.

Algorithm \wtwoctn\ takes this approach.  Given $N$ inputs, it
orders them by increasing length and then merges each input, starting
with the shortest, into an accumulating total.  A worst-case input
would have all inputs of equal length, each containing $B/N$ items that
are disjoint from any other input.  At the $i^{\textrm{th}}$ step the
accumulating array of counters will have $Bi/N$ entries and this
will dominate the merge cost for the step.  The total time complexity
for this worst-case input is 
$\Theta(\sum_{i=1}^{N-1}Bi/N) = \Theta(BN).$  The same input ends up
growing an accumulating array of counters of size $B$.
If this algorithm is implemented ``horizontally'' on uncompressed bitmaps,
the maximum temporary space would be bounded by the maximum
number of ones that can be appear in $N$ words of size $W$.

Algorithms \wtwocta\ and \wtwocti\ are refinements.  Although they
end up reading their entire inputs, during the merging stages 
they can discard elements that cannot achieve the required threshold.
For instance, \wtwocta\ checks the accumulating counters
after each merge step.  If there are $i$ inputs left to merge,
then any element that has not achieved a count of at least
$T-i$ can be pruned.  Algorithm \wtwocti\ does such pruning
\emph{during} the merge step, rather than including elements and
then discarding them during a post-processing step after merging each pair of sets.
Because these two algorithms prune data during the processing, it might be advantageous 
to work from the small sets: this increases the probability that data 
will be pruned earlier. Hence we
order input bitmaps by increasing cardinality and then merge each input, starting
with the shortest, into an accumulating total.

In many cases, this pruning is beneficial. However, it does not
help much with the worst-case input, if $T=2$.  We cannot discard
any item until the final merge is done, because the last input set
could push the count (currently 1) of any accumulated item to
2, meeting the threshold.

\subsection{T-occurrence algorithms for integer sets}
\label{sec:t-occurrence-algos}

Others~\cite{Li:2008:EMF:1546682.1547171,Sarawagi:2004:ESJ:1007568.1007652}
have studied the case when the  data is presented as sorted lists of integers
rather than bitmaps.  The initial application was with 
approximate search, to help discover all terms with a small edit distance
to a user-supplied term.  It is not clear that these techniques could be
applied in a more general setting, to compute symmetric functions other
than threshold functions.

Specifically, we consider
\begin{enumerate}
\item \wheap~\cite{Sarawagi:2004:ESJ:1007568.1007652},
\item \mgopt~\cite{Sarawagi:2004:ESJ:1007568.1007652},
\item \mgsk~\cite{Li:2008:EMF:1546682.1547171},
\item \dsk~\cite{Li:2008:EMF:1546682.1547171}.
\end{enumerate}

For details of these algorithms, see the papers that introduced them.
The \wheap\ approach essentially uses an $N$-element min-heap that contains
one element per input, unless the input has been exhausted. 
Using the heap, it merges the
sorted input sequences.  As items are removed from the heap, we count duplicates
and thereby know which elements had at least $T$ duplicates.
This approach can be used for arbitrary symmetric functions.
It requires that we process the ones in each bitmap, inserting
(and then removing) the position of each into an $N$ element min-heap.
This determines the space bound.  If the bitmap is
uncompressed, we have an additional time cost of $Nr/W$ to find the $B$ ones.
The total time cost is thus $O(B \log N)$ for compressed bitmaps and
$O(Nr/W+ B \log N)$ for uncompressed, done vertically.  Horizontal
processing does not change the time or space use.

The remaining algorithms are also based around heaps, but are designed
to exploit characteristics of real data, such as skew.
Algorithm \mgopt\ sets aside the largest $T-1$ bitmaps.  Any item contained
only in these bitmaps cannot meet the threshold.  Then it uses an
approach similar to \wheap\ with threshold 1.  For each item found,
say with count $t$, it checks whether at least $T-t$ instance of
the item are found in the largest $T-1$ bitmaps.  The items checked for
are an ascending sequence.  If one of the largest bitmaps is asked
to check whether item $x$ occurs, and then is later asked whether 
item $y$ occurs, we know that $y>x$. Items between $x$ and $y$ in
the big list will never be needed, and can be skipped over without
inspection.  When the data is provided as a sorted list of integers,
a doubling/bootstrapping binary-search procedure can find the smallest
value at least as big as $y$ without needing to scan all values between
$x$ and $y$. We can consider the portions skipped to have been ``pruned''.

As noted in \S~\ref{sec:bitmaps},
providing random access is not a standard part of a RLE-based compressed bitmap library, 
although it is essentially free for uncompressed bitmaps.  
However, with certain compressed bitmap indexes it is possible
to ``fast forward'', skipping portions of the index in a limited way:
JavaEWAH uses the fact that we can skip runs of dirty words (e.g., when computing intersections). It should be possible, through auxiliary data structures to skip even more effectively at the expense of memory usage~\cite{moffat1996self, sanders2007intersection}. However, such auxiliary data structures do not always improve performance~\cite{Culpepper:2010:ESI:1877766.1877767}.
 The alternative is to 
convert from bitmap representation to sorted array, apply the classic
algorithm, then convert the result back into a bitmap.  The overheads of
such an approach would be expected to be unattractive.

To bound the running time, we can distinguish the $B-B'$ ones in the
$T-1$~largest bitmaps from the $B'$ ones in the remaining $N-T+1$ bitmaps.
A heap of size $O(N-T)$ is made of the $N-T+1$ remaining bitmaps, and 
$O(B')$ items will pass through the heap, at a cost of 
$O(1+\log(N-T))$ each.  As each item is removed from the heap,
it will be sought in $O(T)$ bitmaps.  With uncompressed bitmaps,
this costs $O(T)$.  

With compressed bitmaps, the situation is more complicated.
Because the items sought
are in ascending order, the $T-1$ bitmaps will each be processed
in a single ascending scan that handles all the searches.
Each of the $B-B'$ ones in the remaining bitmaps should cost us
$O(1)$ effort.  
Thus we obtain a bound of $O(B' (\log(N-T)+T)+B-B')$ for
the time complexity of \mgopt\ on an compressed bitmap.
With uncompressed bitmaps, we add a $r(N-T)/W$ term that
accounts for looking for the ones in the $N-T+1$ smallest
bitmaps, but we can deduct the
$B-B'$ term, as we do not need to wade through all positions
in the $T$ lists.  We obtain a bound of $O(r(N-T)/W + B' (\log(N-T)+T)$.
 
This analysis does not take fully into account the effect of pruning,
because even the compressed bitmaps might be able to skip many
of the $B-B'$ ones as we ``fast forward'' through the largest $T-1$
bitmaps.
Since these are the \emph{largest} bitmaps, if $T$ is close
to $N$ or if the sizes (number of ones) in the bitmaps vary
widely,  pruning could make a large difference.  This depends
on the data.


Algorithm \mgsk\ is a different modification of the heap-based algorithm.
When removing copies of an item from the heap, if there are not 
enough copies to meet the threshold, we remove some 
some extra items.  This is done in such a way that the extra items removed
(and not subsequently re-inserted) could not possibly meet the threshold. 
They are pruned.
Moreover, inspection of the heap yields a lower-bound on the next possible
item that can meet the threshold.  Each input can then be advanced
to (or beyond) that item.  Items skipped are pruned, and will avoid
the overheads of being inspected and (especially) inserted into the heap.
As with \mgopt\, the effectiveness of such pruning depends heavily 
on the characteristics of the data.  Our bound for \mgsk\  does not
reflect the improvements that pruning might bring in some cases; we
use the same bounds as \wheap.

Algorithm \dsk\ is essentially a hybrid of \mgsk\ and \mgopt .
It uses both methods of pruning.  However, rather than 
following \mgopt\ and choosing
the $T-1$ largest sets,  it chooses the $L$
largest sets.  Parameter $L$ is a tuning parameter, and choosing
an appropriate value of $L$ for a given threshold query is problematic.
The authors suggest a heuristic approach whereby another tuning parameter
$\mu$ is determined experimentally, for a given type of queries against
a given dataset.  From $\mu$ and the length of the longest input,
an appropriate value of $L$ can often be determined by a heuristic
formula.  We found this approach unsatisfactory in our experiments, as
the best $\mu$ for one query against a dataset often differed
substantially from the best $\mu$ for a similar query.
(See \S~\ref{sec:algopackage-RBMrg} for some detail on appropriate values for $\mu$ on
our datasets.)
The effectiveness of \dsk\ is definitely affected by $\mu$, though perhaps by less
than a factor of five over a reasonable range of values.
Choosing appropriate $\mu$ and hence $L$ values will be a
difficulty for any engineer planning to use \dsk , especially if the
input datasets are not available for prior analysis.
Nonetheless, \dsk\ might be considered to reflect the state-of-the-art
for the $T$-occurrence problem, at least as it arises in approximate
string searches.

The running-time bound for \dsk\ is obtained using reasoning
similar to that for \mgopt .  If there are $B-B''$ ones in the
largest $L$ bitmaps, then each of the remaining $B''$ ones
will pass through the heap and trigger searches in the $L$
largest bitmaps.  For uncompressed bitmaps, this will result
in $O(B''L)$ work for all the searches.  For
compressed bitmaps,  it will result in $O(B''L + (B-B''))$
work for all the searches.   For the right data, less than $B-B''$
items may be scanned during the searches on compressed bitmaps, as with \mgopt.
Algorithm \dsk\ also has the pruning opportunities of \mgsk ,
whereby some of the $B''$ ones are skipped.
Thus, our running-time bound for \dsk\ on uncompressed
bitmaps is $O(r(N-L)/W + B''(\log(N-L)+L))$.  
For compressed bitmaps, we do not have the $r(N-L)/W$ cost to wade through the $N-L$
smallest bitmaps, looking for ones.  However, we will potentially face a 
cost of $O(B-B'')$ while fast-forwarding through the largest $L$ lists.
As with \mgopt ,
data-dependent 
pruning could reduce the $B-B''$ term.  As with \mgsk, the multiplicative
$B''$ factor can be reduced by data-dependent pruning.
For compressed bitmaps, we end up with a bound of $O(B''(\log(N-L)+L)+(B-B''))$.

For an analysis of space, note that \mgopt\ and \dsk\ partition the inputs into
two groups.  Regardless of group, each compressed bitmap input will have an iterator constructed
for it.  For uncompressed bitmaps, only the
first group (small bitmaps) needs the iterator and  auxiliary storage.
For either kind of bitmap, the first group go into a heap that accepts one element per input.
Thus we end up with a memory bound of $O(N)$, which can be refined to $O(N-T)$ (for
\mgopt\ on uncompressed bitmaps) or $O(N-L)$ (for \dsk\ on uncompressed bitmaps).

\subsection{Boolean synthesis} 
\label{sec:circuit-algos}

 A typical bitmap implementation provides a set of
basic operations, typically AND, OR, NOT and XOR, and frequently ANDNOT. Since it is
possible to synthesize any Boolean function using AND, OR and NOT gates in
combination,  any desired bitmap function can be ``compiled'' into a sequence
of primitive operations. One major advantage is that this
approach allows us to use a bitmap library as a black box, although it is
crucial that the primitive operations have efficient algorithms and
implementations.

Unfortunately, it is computationally infeasible
to determine the fewest primitive operations except in the simplest
cases.  However, there are several
decades of digital-logic-design techniques that seek to implement circuits
using few components.   
Also, Knuth has used exhaustive techniques to determine the minimal-gate
circuits for all symmetric functions of 4 and 5 variables
(\cite[Figs.\ 9\&10,\ 7.1.2]{KnuthV4A}).  
(Following Knuth, we assume an ANDNOT operation.) 
But Intel SSE  or AVX supports AND NOT, as do several bitmap libraries (EWAH included).
Specifically, Intel has the \texttt{pandn} and \texttt{vpandn} instructions; 
however it does not look like
the standard x86 instruction set has AND NOT.
Results are given in
Table~\ref{table:knuth-exhaustive-small-thresholds} and are often
significantly smaller than
those we will obtain from our more general constructions.

\begin{table}
\centering
\caption{\label{table:knuth-exhaustive-small-thresholds} 
Number of gates in Knuth's minimum-gate constructions for $N=4$ and $N=5$}
\begin{tabular}{|c|r|r|}\hline
$T$ & $N=4$ & $N=5$ \\ \hline
1   & 3     & 4 \\
2   & 7     & {\footnotesize not shown}\\
3   & 7     & 9\\ 
4   & 3     & 10\\
5   & --    & 4\\\hline
\end{tabular}
 
\end{table}

We need to be concerned with both running time and memory consumption. 
Based on Knuth's exhaustive analysis, all 4-input symmetric ($N=4$)
functions can be evaluated in
the minimum amount of memory possible for each, without increasing the number
of operations\cite[7.1.2]{KnuthV4A}.  
Knuth does not give an algorithm for minimum-memory evaluation, 
which is probably due to its close relationship to the  known NP-hard
Register Sufficiency problem~\cite[A11.1.P01]{gare:gandj}~\cite{seth:complete-register-allocation-problems}.
(We have a restricted version of Register Sufficiency---we only need to
compute one output, rather than an output for each root of the expression
dag.  However, one could take a multi-rooted
expression dag, introduce a new node $x$ and 
make each (former) root become a child of $x$.
The resulting  dag has a single root, and the registers required for evaluation
would not have changed from the original dag.)

%
\begin{table}
\caption{\label{table:circuits-vs-optimal} 
Number of 2-input operations for $N=4$ and $N=5$, Knuth's optimal
solution, our sideways-sum adder,  our sorter, and the \looped\ approach}
\centering
\begin{tabular}{|c|rrrr|rrrr|}\hline
$T$ & \multicolumn{4}{c|}{$N=4$} & \multicolumn{4}{c|}{$N=5$}\\
    & Opt & Add & Sort & Loop & Opt       & Add   & Sort & Loop\\ \hline
1   & 3   & 11  &   3  &  3   &  4        & 14    & 12   &   4 \\
2   & 7   &  9  &   7  &  9   &  {\footnotesize not shown}& 12    & 12   &  12 \\
3   & 7   & 11  &   7  &  13  & 9         & 14    & 12   &  18 \\ 
4   & --  & --  &   -- &  --  & 10        & 11    & 12   &  22   \\ \hline
\end{tabular}
\end{table}

\begin{table}
\caption{\label{table:larger-circuits} 
Gates (i.e., operations) for adders and sorter for some larger $N$ and $T$.
Note how the sorter circuit behaves with $T$.}
\centering
\begin{tabular}{|rr|rr|r|}\hline
$N$ & $T$ & \multicolumn{2}{c|}{Adders} & Sorter\\
    &     & Tree  & S. Sum  &              \\ \hline
43  & 30  & 272   &  192   &  480  \\
85   & 12 & 562   &  398   & 1216  \\
120  & 105& 806   &  580   & 1907  \\
323  & 14 & 2226  & 1586   & 7518  \\
329  &138 & 2272  & 1620   & 9052  \\
330  &324 & 2275  & 1623   & 7549   \\
786  & 481& 5467  & 3905   & 28945  \\
786  & 776& 5461  & 3899   & 24233  \\ \hline
\end{tabular}
\end{table}

See Table~\ref{table:circuits-vs-optimal}~and~\ref{table:larger-circuits}
for information on a number of threshold circuits discussed later in this
section.
The sideways-sum adder needed 5 gates for ($N=3$, $T=2$) and the tree adder
needed only 4.  However,  for all larger values of $N$ and $T$, the sideways-sum adder
used fewer gates.   For $N>25$, it used about 40\% fewer gates. 
(Perhaps surprisingly, on \textsc{BitSet}s the dataset determined which of the
two adders was fastest, as discussed in \S~\ref{sec:bitset-results}.)
The asymptotically larger sorting-based circuits indeed are larger 
even for modest $N$, although they are optimal for the $N=4$ cases.

Consider uncompressed bitmaps using a word size $W$, 
and assume that all logic operations take unit time. Then a logic circuit with
$C$ gates can obtain a batch of $W$ threshold results with $C$ operations.
For instance, on a 64-bit architecture, our sorting circuit for 511 inputs
uses about 15\,k gates.   Thus in 15,000 bitwise operations we can compute
64 threshold results, or 234 operations per result.  Wording individually,
it would take more binary operations than this to reduce 511 inputs to a
single output.   With a larger $W$, such as provided by SSE and AVX
extensions, the advantage of this approach increases.  However, this
analysis ignores the fact that, given a limited number of registers, the CPU
must spend some time loading bit vectors from main memory and storing
intermediate results there.   Note, however, that this approach is essentially
horizontal, in that no more than $W$ bits from each input, or each intermediate
result, needs to stored at a time.


With RLE compressed bitmaps, the primitive operations do not all have the same
computational cost. For instance, consider a bitmap $B_1$ that consists of a single long run
of zeros and a bitmap $B_2$ that contains a single long run of ones, and a bitmap
$B_3$ that contains alternating zeros and ones.  The operations AND($B_1$,$B_3$) and
OR($B_2$,$B_3$) are fast operations that do not require any copying or bitwise machine
operations.   However,  OR($B_1$,$B_3$) and AND($B_2$,$B_3$) will be expensive 
operations---if $B_3$ is mutable, then a copy must be made of it.
We see that it is simplistic to assume a cost model based on the uncompressed lengths of
the bitmaps, or one that merely counts the number of bitmap operations.

\paragraph{Sum-of-Products circuits}

We implemented classic un-minimized Sum-of-Prod\-ucts circuits to allow
computation of arbitrary $N$-input Boolean functions over bitmaps. 
This became our \sopckt\ algorithm.  If the
function is known to be monotone,  negative literals can be omitted from
the Sum-of-Products circuit.  Note that threshold functions are monotone; this
realization of them consists of $N \choose T$~different $T$-input AND
gates, whose results are the input to a single $N \choose T$-input OR
gate.    These approaches can result in huge circuits, and this is
unavoidable: since none of 
the AND terms is contained within another of the AND terms,
a corollary noted by Knuth~\cite[Corollary Q, \S~7.1.1]{KnuthV4A}
of a result of Quine~\cite{quine53} implies that no other Sum-of-Products
circuit for the function can be smaller.  

If we replace each high-input gate by two input gates, we require
${N \choose T} (T-1)$~AND gates and ${N \choose T} - 1$~OR gates,
for a total of ${N \choose T} T - 1$ gates.  Comparing with Knuth's
results in Table~\ref{table:knuth-exhaustive-small-thresholds}, the result is
optimal for $N=4$ or $N=5$ if $T=1$, and reasonable when $N=4$, $T=2$
(11 gates versus 7).   If $T=N-1$ it requires $\Theta(T^2)$ gates.
Experimentally, it often performs reasonably well in these conditions, but
otherwise the $N \choose T$ factor makes the result impractical.

We also implemented techniques specialized for symmetric functions,
obtaining a sorting-based algorithm, \srtckt.  We also obtain two
adder-based algorithms, \addckt\ and \kaddckt . 

\subsubsection{Sorting}  
\label{sec:sorting-circuit-algo}
A sorting circuit~\cite[5.3.4]{KnuthV3E3} for
Booleans maps a $N$-bit Boolean input to a $N$-bit Boolean output, where the output
is $1^k0^{N-k}$ if the input had Hamming weight $k$. 
Boolean sorting circuits are usually depicted as comparator-based sorting
networks, but each comparator is simply a 2-input AND operation and a
2-input OR operation~\cite[7.1.1]{KnuthV4A}. 
It is then simple to compute whether threshold $T$ has been achieved: simply
use the $T^{th}$ output bit from the sorting circuit.  In general, for
symmetric functions one could determine whether the input had 
\textbf{exactly} Hamming weight $T$ by ANDing the $T^{th}$ output
with the negation of the $T+1^{st}$ output.  (This is
the delta function.)  One could then OR together
the appropriate collection of these ANDNOT gates to synthesize any
desired symmetric function.  Interval functions (to recognize whether
the Hamming weight of the inputs is within an interval) are also
achieved by the ANDNOT approach.

There is no known general procedure to obtain a minimum-comparator sorting
circuit~\cite[5.3.4]{KnuthV3E3}.   Asymptotically, the 
AKS construction~\cite{ajta:sorting-circuit} yields 
circuits with O($N \log N$) comparators, but the hidden constants make
this approach impractical.   For modest size inputs, the construction 
used in Batcher's Even-Odd sort is well known to use few comparators
(see~\cite[5.3.4.(11)]{KnuthV3E3})
even though asymptotically it will generate circuits of 
size $\Theta(N \log^2 N)$.   We used Batcher's sort for our \srtckt\
algorithm.  It requires only that the bitmap library
provide 2-input AND and OR primitives.  Intermediate results are
frequently reused many times.

\subsubsection{Adding}
\label{sec:adding-circuit-algos}
\begin{figure}
\centering
\begin{tikzpicture}[
	decoration={
		markings,
		mark=at position 1cm with {\arrow[black]{stealth}},
	},
	node distance=2.6cm,
	path/.style={
		->,
		>=stealth,
		postaction=decorate
	},
	every node/.style={font=\sffamily}
]
  \node[draw] (A)                    {Adder};
  \node (A1)[above of=A]           {$\ldots$};
\draw [->] (A) -- (A1)  node[pos=.5,sloped,above] {$\log n +1$};
  \node[draw] (B2)[below left of=A]                    {Adder};
  \node[draw] (B1)[below right of=A]                    {Adder};
  \node (C12)[below right of=B1]                    {$\ldots$};
  \node (C11)[below left of=B1]                    {$\ldots$};
  \node (C22)[below right of=B2]                    {$\ldots$};  
  \node (C21)[below left of=B2]                    {$\ldots$};
  \node[draw] (D121)[below right of=C12]                    {Adder};
  \node[draw] (D122)[below left of=C12]                    {Adder};
  \node[draw] (D211)[below right of=C21]                    {Adder};
  \node[draw] (D212)[below left of=C21]                    {Adder};
\draw [->] (D121) -- (C12)  node[pos=.5,sloped,above] {1};
\draw [->] (D122) -- (C12)  node[pos=.5,sloped,above] {1};
\draw [->] (D211) -- (C21)  node[pos=.5,sloped,above] {1};
\draw [->] (D212) -- (C21)  node[pos=.5,sloped,above] {1};
\draw [->] (B1) -- (A)  node[pos=.5,sloped,above] {$\log n$};
\draw [->] (B2) -- (A)  node[pos=.5,sloped,above] {$\log n$};
\draw [->] (C11) -- (B1)  node[pos=.5,sloped,above] {$\log n - 1$};
\draw [->] (C12) -- (B1)  node[pos=.5,sloped,above] {$\log n - 1$};
\draw [->] (C21) -- (B2)  node[pos=.5,sloped,above] {$\log n - 1$};
\draw [->] (C22) -- (B2)  node[pos=.5,sloped,above] {$\log n - 1$};
\end{tikzpicture}
\caption{\label{fig:tree-of-adders} A tree of adders to compute the
Hamming weight of an input, where the input length is a power of two.}\end{figure}
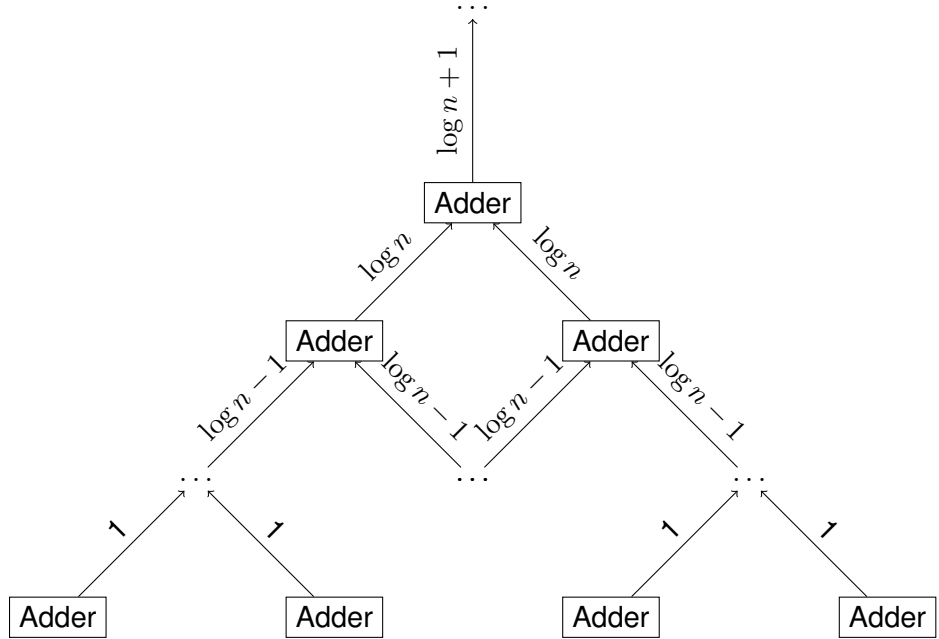

Another approach is to compute the Hamming weight of the input by
summing all the input bits.  We present two such approaches, and the
first uses a tree of adders.  
(See Fig.~\ref{fig:tree-of-adders}.) The leaves of
the tree consist of 1-bit adders (half adders, requiring an
AND and an XOR operation)
 producing a 2-bit sum.  Then these
2-bit sums are added to obtain 3-bit sums.
Multi-bit adders use the ripple-carry technique, where 
summing two $x$-bit numbers uses $2x-1$ AND operations, 
$2x-1$ XOR operations and $x-1$ OR operations. 
The final outcome
is a $\log N + 1$-bit
 binary number representing the Hamming weight of the $N$ inputs.
(When $N$ is not a power of 2, we  pad  with zero inputs and then
using a constant-propagation pass to simplify the resulting circuit.)

This $\log N + 1$-bit quantity is then compared to $T$ using
a $\geq$ circuit.  Since it operates over $O(\log N)$ inputs and
$N$ inputs involved in the Hamming weight computation, it is
not crucial to have the most efficient method of computing $\geq$.
However,
if $T$ is a power of 2, it is easy to compute the threshold
function from the Hamming weight using an OR function: 
for instance, to see if the 4-bit
number $b_3 b_2 b_1 b_0$  is at least 2,  compute $b_1 | b_2 | b_3$.

This approach can be generalized if $T$ is not a power of two;
a \emph{magnitude comparator} circuit~\cite{ashe:digital-design-book} can
determine whether the bit-string for the Hamming weight lexicographically
exceeds the bit-string for $T-1$. As various designs for such circuits
exist, we proceed to derive and analyze our circuit to compare a quantity
to a constant.

Consider two bit strings $b_{n-1} b_{n-2} \cdots b_0$ and
$a_{n-1} a_{n-2} \cdots a_0$, where we have $n= \lfloor \log 2N \rfloor$.
The first is greater if there is some $0 \leq j < n$ where
$b_j > a_j$ and the two bit strings have $b_k=a_k$ for all
$k > j$.  If we define $\mathrm{prefix\_match}(j) =
\bigwedge_{k=j+1}^{n-1} (b_k \equiv a_k)$ then
$b_{n-1} b_{n-2} \cdots b_0 > a_{n-1} a_{n-2} \cdots a_0$ can be
computed as $\bigvee_{j=0}^{n-1} \mathrm{prefix\_match}(j) \land b_j \land \neg a_j$. 
 The prefix values can be computed with 
$n$~XOR and $n$~ANDNOT operations with 
\begin{align*}
\mathrm{prefix\_match}(n)&=1, \\ 
\mathrm{prefix\_match}(k)&=\mathrm{prefix\_match}(k+1) \land \neg (b_k \oplus a_k).
\end{align*}
Altogether  $5n-1$~bitwise
 operations (AND, ANDNOT, XOR, OR) are used to determine the truth value of the inequality $b_{n-1} b_{n-2} \cdots b_0 > a_{n-1} a_{n-2} \cdots a_0$.

We can do better because $T-1$ (whose binary representation is 
the second bit string, $a_{n-1}\cdots a_0$)
is a constant.  
\begin{itemize}
\item First, OR terms drop out for positions $j$ where $a_j=1$,
leaving us with $\bigvee_{\mbox{$j|a_j = 0$}} \mathrm{prefix\_match}(j) \land b_j. $
\item   Second, the previous expression does not need prefix\_match for the
trailing ones in $T-1$; they no longer appear in our expression and there is
no 0 to their right. (The prefix\_match value required for a 0 is indirectly
calculated from prefix\_match values for all positions to the left of the 0.)
\item Third, 
we can redefine
$\mathrm{prefix\_match}(j)$ as  
$\bigwedge_{\stackrel{j<k<n}{ \land a_k=1}} b_k.$ Thus, if $T-1 = 101100_2$,
$\mathrm{prefix\_match}(2)$ no longer checks that the other bit string
starts with 101.  Instead, it matches $101$ or $111$.  In the latter case,
the bit string is already known to exceed $T-1$.
\end{itemize}

For an example, consider $T-1 = 0010100111_2$ with $n=10$.
We compute 
$
b_9 \lor b_8 \lor 
(\mathrm{prefix\_match}(6) \land b_6) \lor
(\mathrm{prefix\_match}(4) \land b_4) \lor
(\mathrm{prefix\_match}(3) \land b_3), $ where
$\mathrm{prefix\_match}(6) = b_7$, 
$\mathrm{prefix\_match}(4) = b_5 \land b_7 = b_5 \land \mathrm{prefix\_match}(6)$
and  $\mathrm{prefix\_match}(3) = b_5 \land b_7 = \mathrm{prefix\_match}(4) $.
Thus, we get the formula 
$
b_9 \lor b_8 \lor 
(b_6 \land b_7 ) \lor
( b_4 \land b_5 \land b_7 ) \lor
(b_3 \land b_5 \land b_7 ). $ Since $b_5 \land b_7$ is a shared
sub-expression, we need only 8 operations in this case, which  less than $n=10$.

With these optimizations, computing all required
$\mathrm{prefix\_match}$ values needs one AND for every 1~bit
except for the trailing ones and the first one: for each of these 1~bits
no gate is required. 
We also need a single wide OR that takes an input for each of the zeros (there
must be a 0 and a 1 in the binary representation of  $T-1$) in
$T-1$.  That OR input is computed as a 2-input AND; as a minor optimization,
the leading zeros in $T-1$ can omit the AND, because they
would use a $\mathrm{prefix\_match}$ value that would be 1.
To count operations, let $\nu(T-1)$ denote the Hamming weight
when $T-1$ is written in binary, let nto($T-1$) denote the number
of trailing ones, and nlz($T-1$) denote the
number of leading zeros.  Therefore, the number of zeros in $T-1$ is $n-\nu(T-1)$.
Thus, computing the $\mathrm{prefix\_match}$ values needs 
$\max(0,\nu(T-1)-\mathrm{nto}(T-1)-1)$ ANDs.  
We need to ``OR'' $n-\nu(T-1)$ items, requiring
$n-\nu(T-1)-1$ two-input OR gates.  The items are computed by 
$n-\nu(T-1)-\mathrm{nlz}(T-1)$ AND gates. In total we use
at most  $2n - \nu(T-1) - \mathrm{nlz}(T-1) - \mathrm{nto}(T-1)-1$
 operations; i.e.,
between $0$ (when $T-1=011\cdots1$) and 
$2n-3$ (when $T-1=100\cdots0$).
The similar comparator circuit presented by
Ashenden~\cite{ashe:digital-design-book} can also
be simplified when $a$ values are constant.  The worst case value of
$T$ gives a circuit whose size is similar to ours, but the
best case is not as good.

There is an alternative way to compare
two $\log N$-bit quantities, $T$ and the Hamming weight, with
approximately $7 \log N$ gates: use an adder to sum $-T$ and the Hamming
weight; then inspect the sign of the result.
Considering the algorithms introduced later in this  section,
we used this approach in the implementation of \schedcs , but otherwise
our \addckt\ and \kaddckt\ algorithms used the previous approach.

For general symmetric functions, one could use a sum-of-products
circuit over the $1+\lfloor \log N \rfloor$-bit number that gives the Hamming weight of
the inputs.  Knuth~\cite[7.1.2]{KnuthV4A} observes that since he has
calculated the minimum number of gates (7) to realize any 4-input
Boolean function, we immediately can realize any symmetric Boolean
function of $N\leq 15$ inputs using $7+c(N)$ gates, where $c(N)$ is the number
of gates used by our adder tree. 

To determine $c(N)$ when $N=2^k$ is a power of two, note that we use
$2^{k-1}$~half-adders in the first level of the adder tree.  The
$2^{k-2}$~resultant 2-bit numbers are added in the second levels, using
a half-adder and a full adder for each 2-bit sum.  In the third level,
we require $2^{k-3}$ 3-bit adders, each containing two full adders and a
half adder.  In general, the $i^{\mathrm{th}}$ level contains
$2^{k-i}$ adders that each require $i-1$~full adders and a half adder.
The final level is level $k = \log N$.  
Since a half-adder requires two gates and a full adder requires five,
we obtain a total gate count of 
\[
c(2^k) = \sum_{i=1}^{k} 2^{k-i} (5(i-1)+2) = 7 N - 5 \log N - 7.
\]

For the first few powers of $2$ this yields 

\hfill\begin{tabular}{clllll}
$N=$     & 2 & 4 & 8 & 16 & 32\\
$c(N)=$  & 2 & 11& 34& 85 & 192\\
\end{tabular}.\hfill{}

\subsubsection{Sideways Sum}
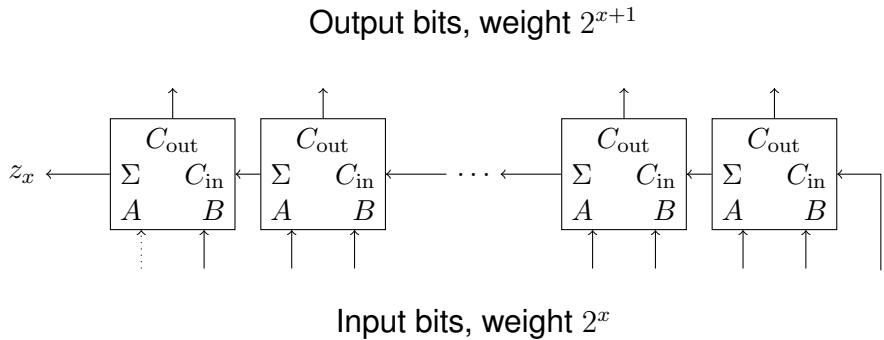
\begin{figure}
\centering
       
\begin{tikzpicture}[
	decoration={
		markings,
		mark=at position 1cm with {\arrow[black]{stealth}},
	},
	node distance=2cm,
	path/.style={
		->,
		>=stealth,
		postaction=decorate
	},
	every node/.style={font=\sffamily}
]

\node (A) {$z_x$};
\node[draw] (B) [right of=A] {\begin{minipage}{1.4cm}\centerline{$C_{\mathrm{out}}$}$\Sigma$\hfill{}$C_{\mathrm{in}}$\\$A$\hfill{}$B$\end{minipage}};
\node[draw] (C) [right of=B] {\begin{minipage}{1.4cm}\centerline{$C_{\mathrm{out}}$}$\Sigma$\hfill{}$C_{\mathrm{in}}$\\$A$\hfill{}$B$\end{minipage}};
\node (D) [right of=C] {$\ldots$};
\node[draw] (E) [right of=D] {\begin{minipage}{1.4cm}\centerline{$C_{\mathrm{out}}$}$\Sigma$\hfill{}$C_{\mathrm{in}}$\\$A$\hfill{}$B$\end{minipage}};
\node[draw] (F) [right of=E] {\begin{minipage}{1.4cm}\centerline{$C_{\mathrm{out}}$}$\Sigma$\hfill{}$C_{\mathrm{in}}$\\$A$\hfill{}$B$\end{minipage}};
\node (FF) [below right of=F] {};

\draw [->] (B) -- (A)   {};
\draw [->] (C) -- (B)   {};
\draw [->] (D) -- (C)   {};
\draw [->] (E) -- (D)   {};
\draw [->] (F) -- (E)   {};
\draw [->] (FF.north) -- +(0,0) |- (F) ; 

\node(DD) [above of=D,minimum width=10cm] {\large Output bits, weight $2^{x+1}$};

\draw [->] (B.north) to ($(B.north)+(-0cm,0.4cm)$);
\draw [->] (C.north) to ($(C.north)+(-0cm,0.4cm)$);
\draw [->] (E.north) to ($(E.north)+(-0cm,0.4cm)$);
\draw [->] (F.north) to ($(F.north)+(-0cm,0.4cm)$);

\node(CC) [below of=D,minimum width=10cm] {\large Input bits, weight $2^x$};

\draw [->,dotted] ($(B.south west)!0.50!(B.south)+(-0cm,-0.51cm)$) to  ($(B.south west)!0.50!(B.south)$ ) ;

\draw [->] ($(C.south west)!0.50!(C.south)+(-0cm,-0.51cm)$) to  ($(C.south west)!0.50!(C.south)$ ) ;

\draw [->] ($(E.south west)!0.50!(E.south)+(-0cm,-0.51cm)$) to  ($(E.south west)!0.50!(E.south)$ ) ;

\draw [->] ($(F.south west)!0.50!(F.south)+(-0cm,-0.51cm)$) to  ($(F.south west)!0.50!(F.south)$ ) ;

\draw [->] ($(B.south east)!0.50!(B.south)+(-0cm,-0.51cm)$) to  ($(B.south east)!0.50!(B.south)$ ) ;

\draw [->] ($(C.south east)!0.50!(C.south)+(-0cm,-0.51cm)$) to  ($(C.south east)!0.50!(C.south)$ ) ;

\draw [->] ($(E.south east)!0.50!(E.south)+(-0cm,-0.51cm)$) to  ($(E.south east)!0.50!(E.south)$ ) ;

\draw [->] ($(F.south east)!0.50!(F.south)+(-0cm,-0.51cm)$) to  ($(F.south east)!0.50!(F.south)$ ) ;
\end{tikzpicture}
\caption{\label{fig:sideways-sum} 
One layer in the sideways sum circuit, similar to that discussed by Knuth.
Note that this is not a ``ripple carry'' arrangement: the \textbf{sum} bit
from a full adder becomes the carry-in input of the next full (or half) adder.
Whether the leftmost adder is a full or a half adder depends on whether the
number of input bits is even.}
\end{figure}

Knuth describes an alternative ``sideways sum'' 
circuit to compute the Hamming weight of
a vector of bits~\cite[7.1.2]{KnuthV4A}. 
This circuit  consists of $O(\log N)$ levels.
The first level takes a collection $C_1$ of bits of weight 1,
and as output produces a single weight-1 bit, $z_0$,  and
a collection $C_2$ of bits of weight 2.   The Hamming weights are
preserved: i.e., $z_0+2\  \mathrm{hamming}(C_2) = \mathrm{hamming}(C_1)$.
The second level takes $C_2$ and produces a single weight-2 bit, $z_1$,
and a collection $C_3$ of bits of weight 3. 
 Again, Hamming weights
are preserved: $2 z_1 + 4 \ \mathrm{hamming}(C_3) = 2 \  \mathrm{hamming}(C_2)$.
Note that bits $z_1$ and $z_0$ are the least significant bits
of $\mathrm{hamming}(C_1)$.
Subsequent levels are similar, and the Hamming weight of the $N$
input bits is specified by the $z$ bits.   Fig.~\ref{fig:sideways-sum}
shows how input bits of weight $2^x$ are used to compute the single
weight-$2^x$ bit $z_x$ and the output bits of weight $2^{x+1}$.

Compared to the tree of adders, the sideways-sum arrangement of
full- and half-adders
results in 
somewhat fewer gates (bounded by $5N$ rather than $7N$).  For $N$
a power of 2, it has $s(N)$ gates, where

\hfill\begin{tabular}{clllll}
$N=$     & 2 & 4 & 8 & 16 & 32\\
$s(N)=$  & 2 & 9 & 26& 63 & 140\\
\end{tabular}.\hfill{}

If all bitwise operations take the same amount of time, this circuit should
always outperform the tree of adders.  However, when computing over
compressed bitmaps, we did find 
cases where certain inputs favoured the larger circuit
(\S~\ref{sec:experiments}).

\subsubsection{Compiling or interpreting circuits}
Implementation using the circuit-based approach can be done in
several ways.  Regardless, at some point an appropriate circuit
needs to be constructed.  We used a directed, acyclic graph whose
nodes are labelled by the associated Boolean operator.  The $N$ inputs
to this circuit are associated with the $N$ bitmaps.

A straightforward vertical implementation then simulates the circuit,
processing the circuit's nodes in topological order, propagating
values from inputs towards the output.  When a node is processed, its
inputs will be associated with bitmaps.  We apply the node's operation
to these bitmaps, and then associate the resultant bitmap to the
node's  
output.  Since the output of a node can become the input of
several nodes, some book-keeping would be necessary to know when the
storage occupied by an intermediate bitmap can be reclaimed.

An alternative approach generates straight-line code
from the circuit (again, processing the nodes in topological order).
See Appendix~\ref{app:gencode} for sample C++ code.  With this approach,
we can do a compiler-type analysis to determine the point when
an intermediate value is last used.  The straight-line code could
then include operations to reclaim storage occupied by no-longer-useful
intermediate results.  Having obtained such code, it can be compiled
and executed.  An initial implementation that machine-generated Java
code ran into a limitation of the Java Virtual Machine: there is a 
fixed, and relatively small, maximum size for the JVM byte code of
any method.

Our third approach generates our own byte-code (operations for
AND, OR, XOR, reclaim, and so forth).  A trivial byte-code interpreter
then processes this straight-line byte code.  This is the implementation
adopted for the experiments.

One might also consider a horizontal implementation.   It would be 
straightforward to produce groups of $W$ bits of an uncompressed ``array of bits''
bitmap by any of the approaches given above. With compressed bitmaps,
it is considerably more difficult.  One reason is that two RLE-compressed
bitmaps may never have runs that begin at the same location.
Another is that operations over RLE-compressed bitmaps gain efficiency by
skipping portions of their input.  For instance, if $B_1$ begins with a
run of a thousand zeros, an AND operation using $B_1$ may skip the
corresponding portion of its other input.

The JavaEWAH library allows iterator-based basic operations (AND, OR, etc.), 
which take and produce iterators.  Each iterator has has the ability to
skip portions, as noted above.
If the circuit were a tree (no output connects to more than one input)
we can compose iterators according to the circuit.  For instance, 
the circuit corresponding to 
\begin{verbatim}
AND(OR(b1,b2),b3)
\end{verbatim}
gives rise to something like
\begin{verbatim}
EWAHIterator it_b1 = b1.iterator()
EWAHIterator it_b2 = b2.iterator()
EWAHIterator temp1 = iteratorOR(it_b1, it_b2)
EWAHIterator it_b3 = b3.iterator()
EWAHIterator result = iteratorAND(temp1,it_b3)
\end{verbatim}

The \texttt{result} iterator can then be made to yield its values.
Unfortunately, this approach does not handle shared sub-expressions
well. (Consider what would happen if the last statement were
\texttt{iteratorAND(temp1, it\_b1)}.)   A solution would need to buffer the
results produced by an iterator, in case that iterator's results are
needed by several consumers.  Moreover, different portions of the 
iterator's output may be skipped by the different consumers.  While it is
not impossible to envision solutions to this problem, they seem infeasibly
expensive.  We have implemented an alternative approach that unfolds
a circuit with sharing into a tree---one that can be exponentially
larger.  This approach is infeasible for the adder and sorting circuits.
However, the (mostly impractical) Sum-of-Products circuit is a tree,
except near the leaves.  The resulting iterator-based threshold function
could be used when $N \choose T$ is small.  Since this is rarely the case,
we do not report performance results of this approach.

\subsubsection{Tabulating circuits}
\label{sec:tabulating}
There may be considerable overhead in constructing the required circuit,
whether it be an adder, a sorter or the sum-of-products construction.
In our implementations, we first construct the circuit. In the case of
the sorting circuit, we are interested only in the $T^{\mathrm{th}}$ 
output.  Yet we have generated all the other outputs, as well as the
subcircuits that affect only the unused outputs.  In the case of the
tree-adder circuit  we introduce constant
zero inputs, which end up being passed into our gates.  Some of the
circuit constructions can generate unary AND, OR or XOR gates.
Therefore, after constructing a circuit with these deficiencies, we
then apply a variety of post-processing optimizations.  First, a
reachability analysis removes all gates that are not in the ``fan''
of the one desired output: i.e., a gate whose output has no path
to the overall circuit output is useless and it is removed.
We then apply a simple constant-propagation pass that simplifies the
circuits using such rules as $\mathrm{AND}(x_1,\ldots, x_k,0) \rightarrow 0$ and
$\mathrm{OR}(x_1,\ldots, x_k, 0) \rightarrow \mathrm{OR}(x_1, \ldots, x_k)$.
These simplifications can produce unary gates, which can be removed:
$\mathrm{XOR}(x_1) \rightarrow x_1$ and so forth.

We do not propose that circuits be computed dynamically, at run time.
Instead, circuits can be pre-computed and stored.  Indeed, we do not
need to generate a circuit for all $\Theta(N^2_{\mathrm{max}})$ 
possible combination of $N$ and $T$.  
Observe that if we have a circuit for $N=16, T=8$, we can
use it to solve a variety of other threshold problems.
For instance, if we wish to compute $N=10, T=7$, we can add five additional
padding bitmaps that contain only zeros.  We can also add one additional
padding bitmap that contains only ones.  We now have 16 bitmaps, and
the threshold of 8 can be achieved with the padding bitmaps iff 
the threshold of 7 could be achieved on the original 10 bitmaps.

The general rule is that a circuit for $N$ and $T$ can be used, with
appropriate padding, to compute any threshold problem with $N'$
inputs and a threshold of $T'$, iff
\begin{enumerate}
\item $N' \leq N$ and
\item $T' \in [T-(N-N'), T]$ 
\end{enumerate}

If we want to compute
all values of $T$ from 2 to $N_{\mathrm{max}}-1$, note that none of
these $(N,T)$ pairs covers another.  However, together they do cover
all other pairs.  Thus, we can tabulate $N_{\mathrm{max}}-2$ circuits.

Nevertheless, if $N' \ll N_{\mathrm{max}}$, we will do a great deal of extra
work to answer the query.
We can tabulate $O(N_{\mathrm{max}} \log N_{\mathrm{max}})$ circuits if we
choose all $T$ values 
for $N = 2, 4, 8, \ldots, 2^{\lfloor \log N \rfloor}, N$.
This guarantees that we can answer any query for $N'$ with a value of
$N$ where $N' \leq N < 2N$.

We can further reduce the size of the tabulated circuits by taking advantage
of the similarity between the Sorter circuits  for $(N,T)$ and
$(N,T')$.  Both are based on minor optimizations applied to the same
$N$-input Even-Odd mergesort circuit:  if we know a particular value 
of $T$,  we currently
prune out the portions of the circuit that compute the outputs used for
other values of $T$.    Instead, we can forgo this optimization
With one circuit for
$N$ inputs, we compute all possible $T$ values.
Choosing 
$N = 2, 4, 8, \ldots, 2^{\lfloor \log N_{\mathrm{max}} \rfloor}, N_{\mathrm{max}}$, we see
that $\Theta(\log N_{\mathrm{max}})$ circuits can be tabulated and handle all
queries.   The \emph{total} size of these circuits will
$\Theta(N_{\mathrm{max}} \log^2 N_{\mathrm{max}} )$.

The Adder circuits also offers a similar opportunity.
Ignoring any pruning of portions unused for a particular value of $T$,
the basic adder tree (or sideways-sum circuit) 
 is the same for all $N$-input Adder circuits.
If we then compare for $\geq T$ using the $O(\log N)$-gate
subtract-and-check-the-sign circuit, the exact same circuit can be
used for all $T$, since the $\Theta(\log N)$ bits for $-T$ can be
inputs.  Thus, the total tabulated size to handle one choice of
$N$ and any value of $T$ is $\Theta(N)$.  Choosing
$N = 2, 4, 8, \ldots, 2^{\lfloor \log N_{\mathrm{max}} \rfloor}, N_{\mathrm{max}}$,
we have a total size bound of 
$\Theta(N_{\mathrm{max}})$ for all our tabulated 
circuits.

\subsection{\looped\ Algorithm}
\label{sec:looped-algo}

We can develop a threshold algorithm for individual binary variables
$b_1, b_2, \ldots, b_N$ by observing that\\ 
$\thresh(1, \{b_1\}) = b_1   $ and\\
$\thresh(T, \{b_1\}) = \mathrm{false}$ if $T>1$ and\\
$\thresh(T, \{b_1, b_2, \ldots, b_N\}) = \thresh(T, \{b_1, \ldots , b_{N-1}\}) + 
   \thresh(T-1),\{b_1, \ldots , b_{N-1}\}) \cdot b_N$

In words:  we can achieve a given threshold $T$ over $N$ bits, either by achieving it over $N-1$ bits, or by
having a 1-bit for $b_N$ and achieving threshold $T-1$ over the remaining $N-1$ bits. 

We can then use bit-level parallelism to express this  as a computation over bit vectors
and a loop can compute the result specified by the recurrence.  Note that $\Theta(NT)$
bit-vector operations are used, but we need only $O(T)$ working bitmaps during the
computation, in addition to our $N$ inputs.

Essentially, Algorithm~\ref{alg:looped} tabulates $\thresh(T,\{b_1,\ldots,b_N\})$ (using a table with rows for $N$ and columns for $T$)
by filling in the table row-wise.  At the end of the $k^{\mathrm th}$ iteration, 
$C_j$ stores $\thresh(j,\{ b_1, \ldots, b_k\})$.

Although one can understand this as a computation over a recursively defined
circuit, the tabulation-based implementation avoids circuit-based overheads
present in the adder or sorting circuits.

The number of binary bitmap operations is $2NT-N-T^2+T-1$ 
and
depends linearly on $T$, which is unusual compared with
our other algorithms. 
Referring to Table~\ref{table:circuits-vs-optimal}, we see that even for
$(N=4,T=3)$ this  circuit uses twice as many  bitmap operations
as \kaddckt.
Nevertheless, in  experiments we have seen
that this algorithm can sometime outperform \kaddckt.
Indeed, the number of bitmap operations is not necessarily a good predictor
 of performance  when using compressed bitmaps. It depends on the dataset.  

\begin{algorithm}
\centering
\begin{algorithmic}[1]
\REQUIRE $N$ bitmaps $B_1, B_2, \ldots, B_N$, a threshold parameter $T$
\STATE create $T$ bitmaps $C_1, C_2, \ldots, C_T$ initialized with false bits
\STATE $C_1 \leftarrow B_1$
\FOR{$i \leftarrow 2$ \textbf{to} $N$}
\FOR{$j \leftarrow \min(T,i)$ \textbf{down to} 2}
\STATE    $C_j \leftarrow C_j \lor (C_{j-1} \land B_i) $
\ENDFOR
\STATE $C_1 \leftarrow C_1 \lor B_i$
\ENDFOR
\STATE return $C_T$
\end{algorithmic}
\caption{\label{alg:looped} \looped\ algorithm.}
\end{algorithm}

\subsubsection{\schedcs\ algorithm}

The \schedcs\ algorithm also does not use an explicit
circuit, but is inspired by a carry-save increment approach described
by Ellingsen~\cite{elli:scheduled-vertical-counter}. As with \kaddckt\
and \addckt, we construct a ``vertical counter'' and then look for
items whose counts exceed $T$.  The approach is to maintain $\lfloor
\log 2N \rfloor$  counter digits, using a redundant encoding scheme
where each counter digit can have value 0, 1 or 2, encoded respectively
as 00, as 01 or 10, and as 11.  Hence we need two bitmaps for each counter
digit.   The redundantly encoded vertical counter is used as an accumulator,
to which the input bitmaps are successively added.  With a conventional
binary encoding rather than the redundant encoding, we would have to
do about $5 \log N$ bitmap operations every time a new bitmap is added
to the counter---we need to update all bitmaps in the counter and each such
update needs the five operations that make up a full adder.
The advantage of the
redundant encoding is that it permits one to  (usually) avoid propagating
changes to all counter digits. Frequently, changes need to be made
to the least significant digit.   Half as frequently, changes need to be made
to the next-least significant digit, and so forth.  
Even though the individual positions
may be maintaining different count values, it is sufficient to propagate changes
on a fixed schedule that sees an amortized 2 digits updated for every bitmap
added to the accumulator. (This fixed schedule corresponds to the 
bit changes from a counter that increments by one in every
cycle, and the amortized number of such digit updates is asymptotically
2 per cycle~\cite{CLRSbook}.) 
Thus we need approximately $2 \times 5 \times N$ bitmap
operations to build the final vertical counter using this redundant
encoding.

 It is easy to recover a vertical counter that uses a conventional
binary encoding from this redundant encoding.  For each digit, we
merely sum the encoding's two bits, plus any carry from converting the
previous digit.  The result is the binary digit, plus possibly a carry
for the next stage.  This requires $O(\log N)$ bitmap operations.

Then any method for comparing the counter against $T$ can be used.
As an alternative to the approach used for $\kaddckt$, we show
the approach where we subtract $T$ from the counter and check that the
result is non-negative.  A minor optimization is possible:
of the difference bits $V_i$, only the sign bit ($V_{1+\lfloor \log 2N \rfloor}$)
needs to be computed.  Thus, the body of the final \textbf{for} loop either 
consists of an AND operation or consists of an OR and an ANDNOT operation.

Pseudocode for this approach is given in Algorithm~\ref{alg:carry-save-vert}.  
We used an
additional small optimization in our implementation that is 
not reflected in the pseudocode: when the first two bitmaps are 
added into the total, they directly become the first encoded digit.  Then
the \texttt{time} variable starts out at 1 rather than 0.

Given the ease of implementing the approach, it appears that it may be
a reasonable alternative to \kaddckt\ (which uses $\approx 5N$ bitmap operations)
or \addckt\ (which uses $\approx 7N$ bitmap operations), depending on the cost of
the operations.

\begin{algorithm}
\begin{algorithmic}  
\REQUIRE input bitmaps $B_1, B_2, \ldots , B_N$, integer threshold $T$ with
$2 \leq T < N$
\STATE create $1+\lfloor \log 2N \rfloor$ bitmap pairs
$\langle C^1_1, C^2_1 \rangle,
\langle C^1_2, C^2_2 \rangle, \ldots, 
\langle C^1_{\lfloor \log 2N \rfloor}, C^2_{\lfloor \log 2N \rfloor} \rangle,$ 
\hspace*{1cm}$\langle C^1_{\lfloor \log 2N \rfloor+1}, C^2_{\lfloor \log 2N \rfloor+1} \rangle$
of empty bitmaps (initialized with false bits)
\STATE  $t \leftarrow 0$ \COMMENT{$t$ is ``time''}
\COMMENT {Add each bitmap to accumulating carry-save counter}
\FOR{$i \leftarrow 1 $ \TO $N$} 
\STATE{$c \leftarrow B_i$}
\STATE $t \leftarrow t + 1; \ \ x \leftarrow $ number of trailing zeros in  $t$
\FOR{$p \leftarrow 1 $ \TO $x$}
\STATE $a \leftarrow C^1_p ;\ \ b \leftarrow C^2_p;\ \  C^1_p \leftarrow$ an empty bitmap
\STATE $s \leftarrow a \oplus b;\ \ C^2_p \leftarrow s \oplus c;\ \ c \leftarrow  (a\land b)\lor(c\land s)$
\ENDFOR
\STATE $C^1_{x+1} \leftarrow c$ 
\ENDFOR
\COMMENT{convert redundant representation to two's complement}
\STATE create  $1+\lfloor \log 2N \rfloor$ empty bitmaps $V_1, V_2, \ldots, V_{1+\lfloor \log 2N \rfloor}$ 
\STATE $c_{\textrm{in}} \leftarrow$ an empty bitmap
\FOR{$i \leftarrow 1$ \TO $\lfloor \log 2N \rfloor$}
\STATE $a \leftarrow C^1_i ;\ \ b \leftarrow C^2_i$
\STATE $s \leftarrow a \oplus b;\ \ V_i \leftarrow s \oplus c_{\textrm{in}};\ \ c_{\textrm{in}} \leftarrow  (a\land b)\lor(c_{\textrm{in}}\land s)$
\ENDFOR
\COMMENT{Compare the vertical counter to $T$ (subtract $T$ and check sign)}
\STATE{Let $T_{\lfloor \log 2N \rfloor+1}, T_{\lfloor \log 2N \rfloor},\ldots, T_2, T_1$ be the bits in the two's complement binary representation of $-T$} 
\STATE $c_{\textrm{in}} \leftarrow$ an empty bitmap
\FOR{$i \leftarrow 1$ \TO $1+\lfloor \log 2N \rfloor$}
\STATE $a \leftarrow V_i$
\IF{$T_i = 1$}
\STATE $s \leftarrow \neg a;\ \ V_i \leftarrow s \oplus c_{\textrm{in}};\ \ c_{\textrm{in}} \leftarrow  a \lor(c_{\textrm{in}}\land s)$
\ELSE 
\STATE $s \leftarrow a;\ \ V_i \leftarrow s \oplus c_{\textrm{in}};\ \ c_{\textrm{in}} \leftarrow  c_{\textrm{in}}\land s$
\ENDIF
\ENDFOR
\COMMENT{use the sign bit of the difference we just computed}
\STATE return $\neg V_{1+\lfloor \log 2N \rfloor}$
\end{algorithmic}
\caption{\label{alg:carry-save-vert} \schedcs\ algorithm; see~\cite{elli:scheduled-vertical-counter} for the first parts.}
\end{algorithm}

\section{Detailed Experiments}
\label{sec:experiments}

We conducted extensive experiments on the various threshold algorithms,
using both uncompressed bitmaps and EWAH compressed bitmaps.
Our experiments involved both synthetic and real datasets.   
On the real datasets,
the various bitmaps in the index vary drastically in characteristics such
as density.  
We discuss this in more detail
before giving the experimental results. 

\subsection{Platform:} 
Experimental results were gathered on a two identical
desktops with Intel Core~i7 2600 (3.40\,GHz, 8\,MB of L3~CPU cache)
3.4\,GHz processors with
16\,GB of memory (DDR3-1333 RAM with dual channel). Because all algorithms
are benchmarked after the data has been loaded in memory, disk performance is
irrelevant.

One system was running Ubuntu 12.04LTS with Linux kernel 3.2, and
the other ran Ubuntu  12.10 with Linux kernel 3.5. 
 During experiments,
we disabled dynamic overclocking (Turbo~Boost) and dynamic frequency scaling (SpeedStep).
Software was written in Java (version 1.7), compiled and run using 
OpenJDK (IcedTea 2.3.10) and the OpenJDK 64-bit server JVM.
We believe that results obtained on either system are comparable;
even if there is some speed difference, the result should not
favour one algorithm over another.
We used the JavaEWAH software library~\cite{JavaEWAH}, version 0.8.1,
for our EWAH compressed bitmaps. It includes support for the  \cdom{} 
algorithm.
Our measured times were in wall-clock milliseconds.
All our software is single-threaded. 

Our measured times were in wall-clock milliseconds.
Since our reported times vary by six orders of magnitude (from about
$10^{-1}$ to $10^5$ ms) we used an adaptive approach, rather than repeat
the computation a fixed number of times.  To do this, the computation
was attempted.  If it completed in less than $10^3$ ms, we ran the computation
twice (restarting the timing).  
If that completed too fast, we then repeated the computation four times,
and so forth, until  the computation had been repeated enough times to
take at least $10^3$ ms.

\subsection{Real data}
\label{sec:real-datasets}

Real data tests were done using  \IMDBthree, \IMDBtwo,
\PGDVDthree, \PGDVDtwo, \PGDVD\ and \CensusIncome.\footnote
{See \url{http://lemire.me/data/symmetric2014.html}.}  
Some statistics for each dataset are given in Table~\ref{tab:real-data-char}.
 
The first two are
based on descriptions of a dataset used in the work of Li et al.~\cite{Li:2008:EMF:1546682.1547171}, 
in an application looking for actor names that are at a small edit distance from a (possibly
misspelt) name.  The list of actor names used is obtained from the Internet Movie Database project.
Although the data is available for noncommercial use, the terms of its distribution
may not permit us to make the data publicly available.  Thus, Appendix~\ref{sec:imdb-appendix}
gives a detailed explanation of how this dataset was obtained. (Others can follow
the steps and obtain a similar dataset.)   Each actor name became a record and we extracted its
$q$-grams  (for $q \in \{2, 3\}$). 
Essentially, for each $q$-gram that occurs in at least one name, we 
recorded the list of record-ids where the actor's name 
contained at least one occurrence of the $q$-gram.  Whereas the work of 
Li et al.~\cite{Li:2008:EMF:1546682.1547171} would have computed results over
the sorted lists of integers, we followed the approach of Ferro et al.~\cite{ferr:duplicates-qgrams}
and represented each sorted list of integers by a bitmap: in our case either an uncompressed bitmap (\textsc{BitSet}) or
an EWAH compressed bitmap.  Whereas Li et al. chose $q=3$, Ferro et al. chose $q=2$.
We used both values of $q$.    There is at least one bigram that is \emph{extremely} common
in this data: most actor names are given as a surname, a comma, a space, and a forename.
Such a format will contain the comma-space bigram.   Therefore, there is at least one
bitmap that is very dense, and this may help explain why algorithms did not always
behave similarly on the two datasets.

The \PGDVDthree\ and \PGDVDtwo\ datasets are similar to the IMDB datasets, 
except instead of actor
names, we formed $q$-grams from chunks of text from the Project Gutenberg
DVD~\cite{GutenbergDVD}.  Each chunk was obtained by concatenating
paragraphs until we accumulated at least 1000 characters.  We rejected
any paragraph with over 20\,000 characters---this protected us from 
some content on the Project Gutenberg DVD.  (Unfortunately, the DVD
contains not only the
desired natural-language texts of out-of-copyright books, but it also
contains the human genetic code and also the digits of $\pi$ and other
irrational constants.)

The \PGDVD\ dataset is based on the vocabulary present in the 
Project Gutenberg DVD.  The text in the 11,118 suitable text files
files on the DVD was tokenized into maximal sequences
of alphabetic characters.  Tokens longer than 20 characters were discarded; the
human genome files created such long tokens.   The remaining distinct tokens (there were 
more than 2 million of them) were considered
to be words, although not all of them are in English.   We formed a dataset with
a row for every word, and a column (corresponding to a bitmap)  for each of the
11,118~files.  The bitmap for a file contained a 1 whenever that file contained
the associated word at least once. Such bitmaps should be highly compressible, 
because words that first occur within a
given text will occupy contiguous portions of the bitmaps.

Whereas an inverted index over the DVD would
have a set for each word (indicating the files that contain the word), we have
a set for each file.  Thus our \PGDVD\ dataset can be considered an ``uninverted''
index, or the transpose of an inverted index.   Similarity queries on such a dataset
will yield words that often occur in the same files/books as a query word.  We
suspect that a useful tool for a literary analyst could use this approach, and 
the development of this dataset was motivated by some of our 
prior work~\cite{KaserKeithLemire2006}.

\begin{table}
\caption{\label{tab:real-data-char}Characteristics of real datasets.
Overall bitmap density is the number of ones, divided by the product of
the number of rows and the number of bitmaps.
}
\begin{tabular}{|crrrrr|} \hline
Dataset         & Rows     & Attributes & Bitmaps & \multicolumn{2}{c|}{Average Density}\\ 
                &          &            &         &   Overall & Similarity \\
\hline
\IMDBtwo        & 1,783,816  &  4276      &  4276   &  $4.5\times 10^{-3}$   & $1.3 \times 10^{-1}$\\
\IMDBthree      & 1,783,816  &  50,663     &  50,663  &  $4.1 \times 10^{-4}$  & $3.0 \times 10^{-2}$  \\
\PGDVD          & 2,439,448  & 11,118      & 11,118   &  $2.9 \times 10^{-4}$  & $6.1 \times 10^{-3}$  \\
\PGDVDtwo       & 3,513,575  & 755        &  755    & $2.8 \times 10^{-1}$   & $7.0 \times 10^{-1}$  \\ 
\PGDVDthree     & 3,513,575  & 20,247      &  20,247  & $2.8 \times 10^{-2}$   & $3.6 \times 10^{-1}$   \\
\CensusIncome   & 199,523   & 42         & 103,419  &  $4.1 \times 10^{-4}$  & $5.6 \times 10^{-1}$ \\ \hline 
\end{tabular}
%
%



\end{table}

We also chose a more conventional dataset with many attributes, \CensusIncome~\cite{arxiv:0901.3751,MLRepository}.
A conventional bitmap index was built, having a bitmap for every
attribute value.  One attribute is responsible for 99,800 of the bitmaps; the remaining
3619 bitmaps are much denser than these 99,800.

\subsection{Synthetic data}

We also experimented with smaller synthetic datasets.  One difficulty
is that threshold queries for moderately large $T$ almost always
return an empty set (except for the row selected for the
similarity query), which may not be very realistic.

Each random bitmap represented a set with 10,000 entries, created by a uniform
random process or a clustered~\cite{Anh:2010:ICU:1712666.1712668} process.
With both the uniform and the clustered process, three datasets were generated: 
a dense one, where the set elements ranged between 0 and 29,999; a
moderately sparse one, where the set elements ranged between 0 and 999,999; and
a sparse one, where the set elements ranged up to 9,999,999.  
The names of the synthetic datasets encode the parameters:
[Clustered;1111;10000;$r$] and [Uniform;1111;10000;$r$] ($|B_i| = 10000$
and value 1111 was used to seed the random-number generator.)

The moderate
or sparse uniform datasets would not be good candidates for EWAH compression:
when the elements present in a set are sorted,  an element and its successor
will usually not be within 64 values of each other, and therefore most set elements
require a whole machine word for storage (plus the overhead to specify the
run of zeros).  Thus the moderate and sparse uniform data would be better
stored as sorted lists of integers.

Our dense uniform datasets are
well suited for bitmap representation.  Since runs of zero (or one)  will rarely be
long, bitmap compression is not useful.

\subsection{Queries Used}
\label{sec:queries}

For our experiments, we will have an intended value of $N$.
A similarity query takes some row-id and determines the bitmaps whose
associated sets contain that rid.  If there were exactly $N$ bitmaps, we
would then use them in a threshold computation.  Even if $T=N$, we
will obtain a nonempty answer to a threshold computation, as the original rid 
meets the threshold.

Unfortunately, this may be fewer or more than
$N$ bitmaps.   If there are more than $N$
bitmaps, we simply take the first $N$.   
If we have $N'$ bitmaps, where $N' < N$, we will take $\lfloor N/N' \rfloor$ 
or $\lceil N/N' \rceil$ copies of each bitmap, in order to obtain a collection
of $N$ bitmaps.   The trick of taking repeated copies is a known approach
to obtaining a \emph{weighted} threshold computation by transforming it to
our unweighted thresholds~\cite{KnuthV4A}.  In our case, we are simulating
having weights $\lfloor N/N' \rfloor$ and $\lceil N/N' \rceil$. 

We have a second approach to boost the number of bitmaps in a similarity query.
Instead of taking a single row-id, we take 10 or 100.  For each row id,
we proceed as before, determining the set of bitmaps that have 1 for that
row-id.  The union of all these sets of bitmaps is then considered, and we
either select the first $N$ of them (if the union has more than $N$), or we
take multiple copies as needed to obtain a collection of $N$.
We use the notation ``similarity(10)'' and ``similarity(100)'' to
indicate the use of this second approach.   Note that threshold computations
with $T \approx N$ will now have the (very likely) possibility of returning
empty results.

Note that the frequencies of various $q$-grams differ widely, and the frequencies
of words in the \PGDVD\ dataset will also follow Zipf's law.  An implication
is that if we randomly select a word from the dataset and take its $q$-grams,
we will, on average, be looking at frequent $q$-grams.  Thus the last two columns
in Table~\ref{tab:real-data-char} differ: the first shows the average density
of bitmaps from the dataset, whereas the second shows the average density of the
bitmaps actually selected by similarity queries.

There is one subtle but important difference, involving binary attributes,  between similarity queries
against \CensusIncome\ 
and the other datasets.   The other datasets are based only
on binary (Boolean) attributes, and the similarity queries are based only on \textit{true} values.
For example, a similarity query on string `Sam' will involve the bitmap for bigram `am', but 
for the edit-distance application of Sarawagi and Kirpal, 
it will not involve a bitmap for ``does not contain bigram `pq'''.
However, if we had encoded the relationship of strings and bigrams in a table, there would
have been an attribute for each possible bigram.  Two values (\textit{true} and \textit{false})
would appear in the column.  Thus, there would be a (presumably sparse) bitmap for the true
value, and a (presumably dense) bitmap for the false value.   A similarity query for 'Sam'
over such a table \emph{would} involve the bitmap for ``does not contain bigram `pq''', because
`Sam' does meet the condition of this bitmap.   Since the
Sarawagi and Kirpal approach is based only on the bigrams actually present, our solution has been
to treat binary attributes specially.  Yet there are other cases (perhaps a gender attribute with
values `M' and `F') where binary attributes should be treated the same as other attributes.   
This may indicate
that similarity queries should allow users to specify how to handle  'False' values of Boolean 
attributes.

Note that including the dense bitmaps would have drastically favoured algorithms 
that do not iterate over the 1 bits (such as the circuit-based approaches or \cdom).  

\paragraph{Competitions:} We assess the effectiveness of the various
algorithms by measuring their wall-clock times on a variety of test runs,
each called a competition.  A competition consists of a dataset,
specific values of $N$ and $T$, and a randomly selected rid (or set of rids)
that will determine which $N$ bitmaps are selected in a similarity query
\footnote{In our experiments
the choice of $N$, random seed and dataset uniquely determines 
the set of bitmaps chosen.  This means that as $T$ is varied, we can
cleanly see the effect of $T$ \emph{on that particular set of  $N$ bitmaps},
as long as we use the same seed for the random number generator.
However, if we hold $T$ constant and look at the effect of varying $N$,
we may get a collection of three sparse bitmaps for $T=3$, but four dense
bitmaps for $T=4$, and so forth.  To reduce this effect, we use 
10 different runs and report their averages when showing
how time varies with $N$.
}.

Each algorithm is assigned a rank that depends on its speed, with rank 0 being the
fastest algorithm.   Each algorithm is also compared, for each
competition, to see whether
it is no more than 50\% or 100\% slower than the rank-0 algorithm.

Results are then tabulated for a related collection of competitions.
Currently, we have one collection (\textsc{Small-Competitions})
 of competitions to reflect cases
where $N$ is small.  The queries correspond to a single row. 
Pairs $N$ and $T$ are as follows: (4,3), (8,3), (8,4), (8,6), (8,7),
(16,3), (16,4), (16,5), (16,6), (16,9), (16,12), (16,13), (16,14),
(16,15), (32,3), (32,4), (32,6), (32,9), (32,13), (32,15), (32,19),
(32,21), (32,28), (32,30), (32,31).
These pairs are generated by doubling $N$ and using $T$ values that are
either $T'$ or $N+2-T'$ where $T'=3, 4, 6,\ldots$ is generated by
$T'_1=3$ and $T'_k = \lfloor \frac 3 2 T'_{k-1} \rfloor$.  This choice
of $T$ values favours especially large and especially small thresholds.
Especially large thresholds, for moderate sizes of $N$,
 should be typical of certain applications,
including that of Li et al.~\cite{Li:2008:EMF:1546682.1547171}.
On these small competitions, we used similarity queries for
\IMDBtwo, \IMDBthree, \PGDVD\ and \CensusIncome .

We also run \textsc{Medium-Competitions}, a collection of competitions 
using a wider range of $N$ and $T$ values.  
The values of $N$ are still obtained by repeated doubling, and the
values of $T$ are still obtained as before.  However, 
 $N$ goes up to 128 (and $T$ can go up to 127).  For \textsc{Medium-Competitions},
we used similarity(10) queries for \IMDBtwo, \IMDBthree, \PGDVD\ and \CensusIncome .

Finally, we have \textsc{Large-Competitions}, where $N$ can go up to
512 and $T$ can go to 511.  
We used similarity(100) queries for \IMDBtwo, \IMDBthree, \CensusIncome\ and \PGDVD,
but similarity queries were used for \PGDVDtwo\ and \PGDVDthree .
Due to the nature of our synthetic datasets, it does not really matter
which type of query we used, except we  note that we did not duplicate 
any bitmaps.

\subsection{Scalability of Algorithms}

As previewed in Tables~\ref{tab:complexity-uncompressed}~and~\ref{tab:complexity-rle-compressed},
the various algorithms' running times are all affected\footnote{
For table entries (such as that for \scncnt ) where $B$ is given but $N$ is not explicit, note that
$B$ grows as $N$ grows: given a set of $N$ bitmaps with $B$ ones, if a new nonzero bitmap
is added, the total number of ones increases.} by $N$.
Some are affected by $T$ and others are highly sensitive to the
characteristics of the datasets being processed.

Although we later present more details on the effect of these factors on each
algorithm, a few examples on compressed bitmaps first illustrate these effects.

\begin{figure}
\begin{centering}
\subfloat[linear axes]{
\includegraphics[width=0.8\textwidth]{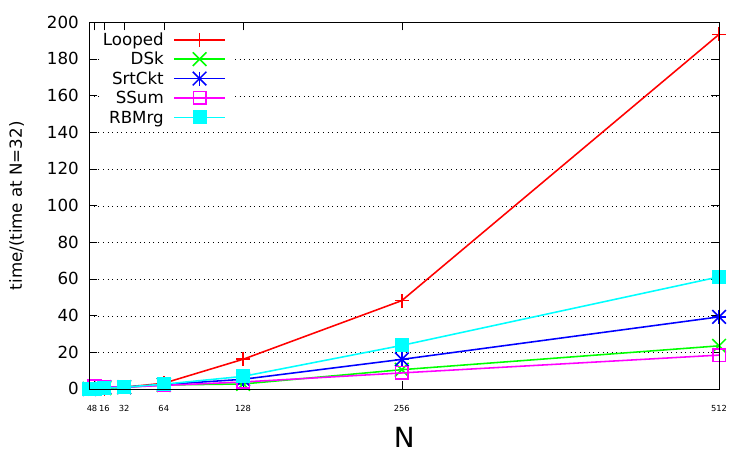}
}\\
\subfloat[logarithmic axes]{
\includegraphics[width=0.8\textwidth]{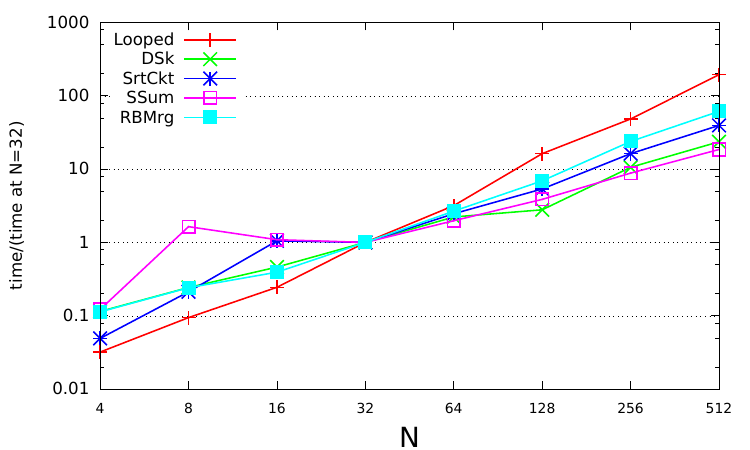}
}
\caption{\label{fig:timesVsN}Effect of N on running times of several algorithms
on EWAH compressed bitmaps.  $T=N/2$, and the chosen dataset
is \PGDVDthree.   To focus on scalability, each algorithm's
running time is normalized so that it takes unit time when $N=32$.
}
\end{centering}
\end{figure}

Fig.~\ref{fig:timesVsN} shows how several algorithms behaved as $N$ was
changed.  For each value of $N$, we took 10 similarity queries and averaged
their running times, holding $T$ constant at $N/2$.
Note that we did \emph{not} count the time to compute the appropriate threshold
circuit, for the circuit-based approaches.  This is justified by the discussion 
in \S~\ref{sec:tabulating}.  To focus on the growth rate, the time for an algorithm
has been normalized by dividing it by that algorithm's speed when $N=32$.

\begin{figure}
\begin{centering}
\includegraphics[width=\textwidth]{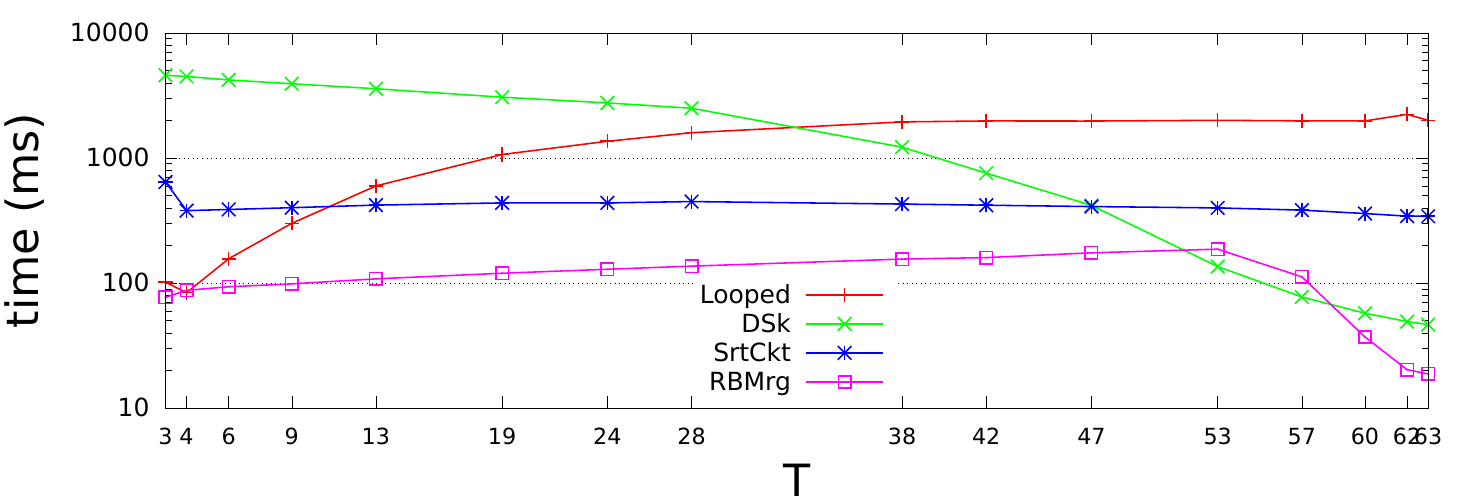}
\caption{\label{fig:timesVsT}Effect of T on running times of several algorithms on EWAH compressed bitmaps.  N=64, and the dataset
is \PGDVDthree.  \kaddckt\ took about 200\,ms except for $T=3$, where it took
about 400\,ms.}
\end{centering}
\end{figure}

Fig.~\ref{fig:timesVsT} shows the effect of varying $T$, on one particular set of 64
bitmaps. Absolute times are shown, but on a logarithmic scale.

\subsection{`w' and Non-`w' Algorithms}

We have two main implementation styles.  Algorithms whose names begin with
`\textbf{w}' were implemented by converting the compressed bitmaps into sorted lists of
integers, and then processing these lists.  Some of the algorithms in question
require the manipulation of such lists and seem to demand all of their 
data in memory in such a format.  Examples would be \wtwocti\  or
\wsort .  Thus, there is no native bitmap implementation of such an algorithm.
Other algorithms were implemented using both implementation styles, and
thus have a `w' version and a native bitmap version.
We did not implement a native bitmap version of \mgsk, having observed that
\begin{itemize}
\item 
the other `w' implementations almost always outperformed it, and
\item
the overheads of converting a bitmap into a sorted list of integers always
outweighed whatever gains might be obtained by processing integer arrays.
\end{itemize}

With these `w' algorithms, there are limitations to which datasets
and queries can be processed, due to the requirement of storing the input as
(frequently huge) integer lists.  
Algorithm \wsort\ particularly had difficulties, and many data points
for $N \geq 128$ are missing. \mgsk\ had trouble for $N=512$ on \PGDVDthree.
Even when larger $N$ values were  handled by such algorithms,
the running times were often much higher than one might anticipate, and
also the CPU utilization for the single-threaded tests were frequently noted to 
exceed 100\%.  This presumably indicates concurrent work by the garbage collector,
and such results may not represent the behaviour we would find on a
machine with more memory.

\subsection{Uncompressed bitmaps}
\label{sec:bitset-results}
While our primary interest is in word-aligned RLE bitmaps, it is
useful to look at the simpler case of uncompressed bitmaps.  
Therefore, we experimented with algorithms operating over \textsc{java.util.BitSet}.
We implement \looped, \dsk\ with various choices of parameter $\mu$, 
\scncnt\  (see Appendix~\ref{app:scancnt-impl}),
\addckt, \kaddckt, \sopckt\ and \srtckt .

The technique used by \scncnt\  to iterate through only the set bits
means that it does less work when there are few set bits. 
  Thus we experiment with
queries where the \textsc{BitSet}s are taken from the index, as well as 
cases where the same \textsc{BitSet}s are negated prior to threshold
computations.  

For display purposes we are comparing only a few algorithms, and thus 
we can associate each
one with a colour.  Then for each dataset we
generate two different sequences of queries.
The first randomly chooses $\log  N$ uniformly  from 2.0 to 10.0.
From that, we compute the nearest integer $N$.  We then randomly choose an 
integer threshold $T$ from 2 to $N-1$ inclusive.  Finally, we choose $N$ bitmaps
corresponding to a similarity query.  Note that if a particular value of $N$
is chosen several times, we do not (usually) generate the same $N$ bitmaps.
The query is executed using each of the algorithms, and we record which one
was fastest.  We also keep track of the total time taken by each
algorithm.  Since the choice of $N$ and $T$ often determines which algorithm
is fastest, we place the winner's coloured marker on an $N$,$T$ grid.

The second sequence of queries is similar, except that every bitmap has
been negated.

\begin{table} 
\caption{\label{tbl:total-bitset-workload-rowsample} Total workload times (seconds), uncompressed bitmaps (\textsc{BitSet}) }
\small
\begin{tabular}{|crrr|rrr|} \hline
Algorithm & \multicolumn{3}{c|}{As Given}   & \multicolumn{3}{c|}{Negated}\\
          &IMDB-2         & IMDB-3         & \PGDVD         & IMDB-2          & IMDB-3        &\PGDVD \\ \hline
\scncnt   & 23.1          &  8.6           & 3.3            & 112.7           & 124.2         & 105.3   \\
\looped   & 208.2         & 203.0          & 113.9          & 219.1           & 218.1         & 170.4   \\
\dsk      & 61.5          &  18.4          & 8.0            & 619.0           & 586.3         & 493.0   \\
\addckt   & \textbf{ 6.4} &  \textbf{6.1}  & \textbf{3.0}   & \textbf{6.4 }   & \textbf{ 6.5} & \textbf{3.9} \\
\kaddckt  & 6.5           &  6.4           & 3.8            & 7.2             & 7.9           & 4.8          \\
\srtckt   & 26.6          &  25.8          & 16.3           & 26.0            & 25.6          & 21.1  \\
\wmgopt   & 106.1         &  28.6          & 12.0           & --              & --            & --           \\
\wtwocti  & 55.5          &  16.3          & 5.0            & --              & --            & --           \\\hline
\end{tabular}\\
\end{table}

\begin{figure}\centering
\subfloat[\IMDBtwo \label{fig:winBitset-imdbtwo-no-negate-no-simple}]
  {\includegraphics[width=.47\textwidth]{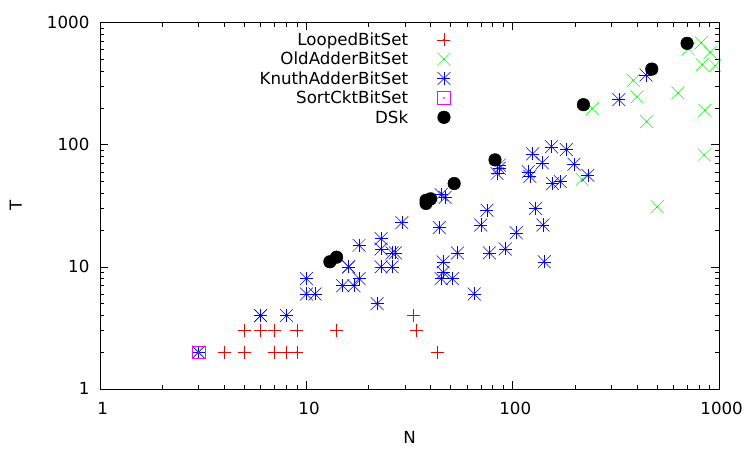}}
\subfloat[\IMDBthree \label{fig:winBitset-imdbthree-no-negate-no-simple}]
  {\includegraphics[width=.47\textwidth]{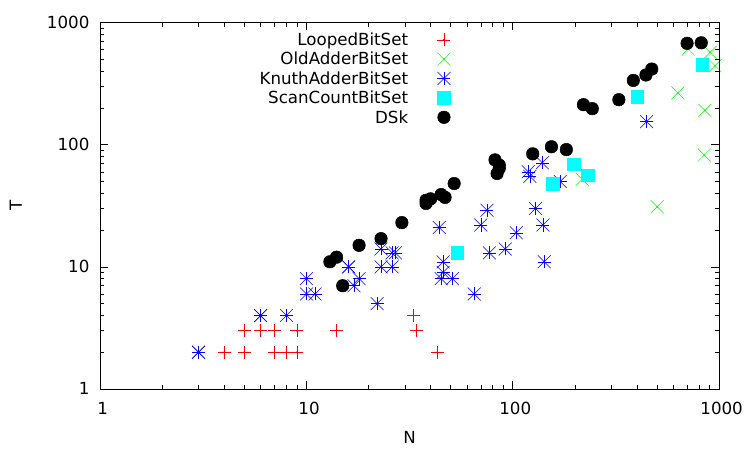}}\\
\subfloat[\PGDVD \label{fig:winBitset-pgdvd-no-negate-no-simple}]
  {\includegraphics[width=.47\textwidth]{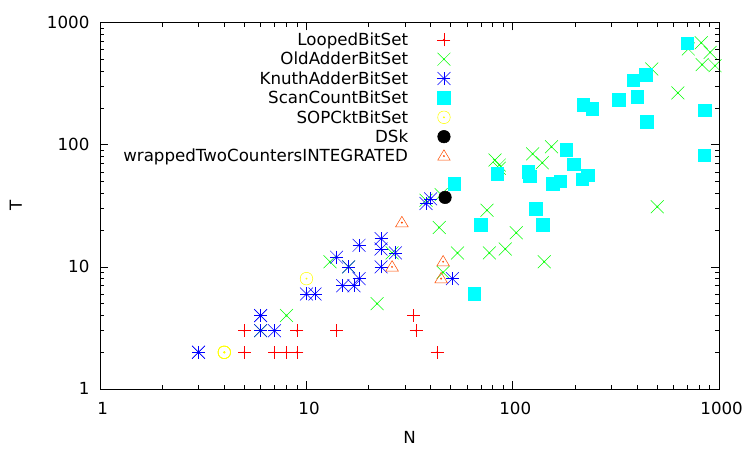}}
\subfloat[\IMDBtwo\ (negated) \label{fig:winBitset-imdbtwo-negate-no-simple}]
  {\includegraphics[width=.47\textwidth]{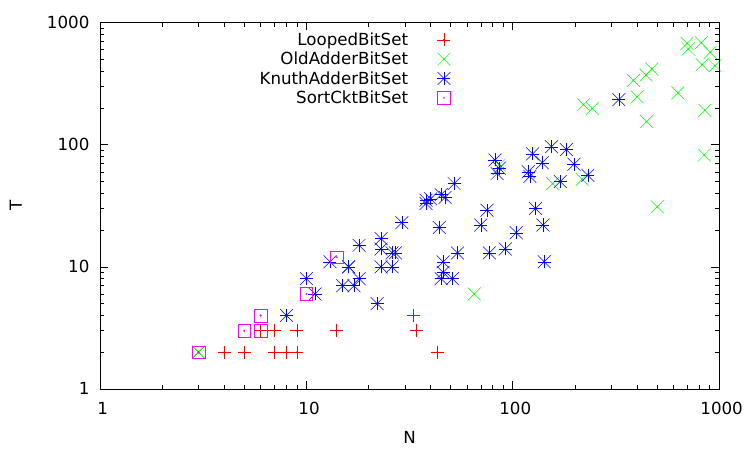}}
\caption{\label{fig:bitset-no-simple} The fastest algorithm for each of the 100 queries in the workload
are shown.  Similarity queries were used over \textsc{BitSet}s.}
\end{figure}

Table~\ref{tbl:total-bitset-workload-rowsample} gives results 
for similarity queries.
We would usually 
select the bitmap for the comma-space bigram in each test for \IMDBtwo , and this bitmap's high density hurts
\scncnt\ enough that \addckt\ can overcome it.  
These effects are shown visually in 
Fig.~\ref{fig:bitset-no-simple}.
Each marker shows which algorithm was fastest for a query of a given $N$ and $T$ in the workload.
As noted, we presume  the comma-space bigram's bitmap was involved in most queries for \IMDBtwo.  Thus
we see the adder circuits performed well for moderate and large values of  $T$, whereas \scncnt\ 
never was best. 
For small values of $T$,  \looped\ is usually the fastest technique when $T \leq 3$.
With \IMDBthree\ and \PGDVD\ we see the adders competing with  \scncnt\  once $N$ is larger than about 60.
Depending on the dataset, the pruning-based approaches \dsk\ and \wtwocti\ are sometimes
effective (\dsk\ requires $T \approx N$ to be successful, whereas \wtwocti\ can
handle cases when $T$ and $N$ differ more).

Using \textsc{BitSet}, we found that the sideways-sum adder was frequently 
slower than the tree adder for certain datasets.  For instance, with
\PGDVD\ and $N=119$, $T=60$ it took 55.1\,ms for the sideways-sum adder,
but only 46.6\,ms for the other.  Other combinations of
$N$ and $T$ were similar.    We found this surprising,
because the tree adder requires about 40\% more operations
than the sideways-sum adder.  Investigating the OpenJDK source code for
\textsc{BitSet}~\cite{openjdk-bitset-sourcecode}, we found that operations
such as AND and OR are optimized to avoid work after the word containing
the last set bit in a bitmap.  By setting a sentinel bit after the last
valid row-id in the index, we can force all \textsc{BitSet}s to have their maximum
length.  In that case, both kinds of adder became significantly slower:
for $N=119$ and $T=60$ the sideways-sum  and tree-adder speeds were
71.8\,ms and 113\,ms, respectively.  Thus, the speed ratio between adders 
was much closer to what we expected.   Since \textsc{BitSet} is not a fixed-length
vector of bits, we see how its implementation makes the determination
of the faster algorithm become data dependent.  While we know the tree-adder 
circuit manipulates more \textsc{BitSet}s, presumably they were much shorter ones
when processing our datasets for small values of $N$:   
the pattern in Fig.~\ref{fig:bitset-no-simple} is that \kaddckt\ 
was the faster Adder circuit for
smaller values of $N$, whereas \addckt\ was faster for large $N$. 

When the \textsc{BitSet}s were negated, we obtained extremely dense bitmaps.
It is not surprising that \scncnt\ was negatively affected
by the density of the bitmaps, as it iterates over the ones.  
It may be surprising that \looped, \addckt, \kaddckt\ and 
\srtckt\  were
affected, because should have been (more or less) oblivious to their data.  
However, the \textsc{BitSet} representation, as implemented in OpenJDK,
 only stores bitmap words until the word with
until the largest one bit.  In negating the bitmaps, we used \texttt{BitSet.flip(0,numRows)}.   Suppose that
\texttt{numRows} was 1000, but we had a bitmap whose last set bit was 20.  After the flip, it would have required 50 times more
data to store.  Presumably it would also have taken
 longer to process.  

The poor performance for \looped\ arose because the workload  included queries with large values of $T$,
and the algorithm's running time depends strongly on $T$.  However, it frequently outperformed the other
algorithms when $T$ was extremely small.  Unfortunately, the total workload times were dominated by the times
of the queries where $N$ and $T$ were large\footnote{
The query-generation process tends to produce large $T$ values when $N$ is large.
}.  For similar reasons, it appears that \addckt\ was always to be 
preferred to \srtckt .   However, for smaller $N$ and $T$, \srtckt\ frequently outperformed  \addckt .

Under \textsc{BitSet} negation,
we see a pattern for \IMDBtwo\  (Fig.~\ref{fig:winBitset-imdbtwo-negate-no-simple})
much like Fig.~\ref{fig:winBitset-imdbtwo-no-negate-no-simple}, except that \dsk\ is \emph{never}
best.  (Due to memory limitations, none of the `w' algorithms could be run.)
 Results for \textsc{BitSet} negation on  \IMDBtwo\ and \PGDVD\ were almost indistinguishable from the pattern
seen in Fig.~\ref{fig:winBitset-imdbtwo-negate-no-simple}, except in which adder circuit 
was better (they preferred the sideways-sum adder).

%

\paragraph{Discussion:} If one does not know much about the data and does not know
 the values of $N$ and $T$
to expect, the adder circuits are safe bets (see 
Table~\ref{tbl:total-bitset-workload-rowsample}.
If $T$ is known to be very small, \looped\ is a reasonable choice, and \srtckt\ may also
fare well.  Algorithms \scncnt, \dsk\ and the `w' algorithms can be good choices if the data is
known to be very sparse.  For \wtwocti\ and \dsk , we need to have effective opportunities for
pruning, and thus $T$ needs to be very large.   However, \scncnt\ is too slow when the data
is dense,  \dsk\ can perform very poorly unless $T \approx N$, and the `w' algorithms will fail if
the integer-list representation requires too much memory 
(each bit can turn into a 32-bit quantity, for a 32-fold space expansion).

\subsection{Comparing Algorithms on EWAH}

Our experiments first focus on the individual algorithms.
For each algorithm, 
we show how it scales with $N$ and $T$ on similarity queries involving
both synthetic and real datasets.
We omit \sopckt\ because it cannot be run with most $N$ and $T$ combinations.
We omit \wtwoctn\ and \wtwocta\ because the better pruning done by 
\wtwocti\ means they are usually worse than \wtwocti.

The experiments varying $T$ were done with one particular set of bitmaps.
It corresponds to $N=32$ and \textsc{Small-Competitions}.  
The logarithmic $y$ scale can make differences look small, but this is needed
to show the drastic differences that $T$ can make for some algorithms.

To study scaling with $N$, we ran a series of competitions using $N=4,8,16,\ldots$ 
up to $512$
for $T=3$, $T=N/2$ and $T=N-1$.   Ten similarity queries (each with one prototypical item)
were generated for each value of $N$.  The 10 times were then averaged, and the averaged
time was then normalized by dividing it by the average time for $N=32$.

To allow a comparison between algorithms, we show heat maps indicating
the conditions ($N$ and $T$) when that algorithm performed well on our
real datasets.  (We also show when its performance is \emph{terrible},
which we define as 10 times longer than the fastest algorithm.) 
This data included \textsc{Small-Competitions}, \textsc{Medium-Competitions},
\textsc{Large-Competitions} as well as the competitions used when
varying $N$.

Unfortunately, the heat map
aggregates results from our various real datasets,
and the algorithm performance can
be greatly influenced by the dataset used.  The heat maps do not provide
a way to see the effect of the dataset.  Thus, these heat maps
might be appropriate for a workload in which all datasets were used equally,
and with $N$ and $T$ distributions that were independent of the dataset.

After these per-algorithm results are given and discussed, \S~\ref{sec:workload} 
ranks the algorithms by supposing a workload that gives equal importance
to queries over each dataset, but in which a fixed $N$ and $T$ are used.
A stacked bar chart is then used.  An algorithm's overall slowness 
on the workload is shown by the overall height of the bar.  However, it
is illustrative to see the individual contributions from the various datasets.
From this we can assess the  sensitivity of the various algorithms to
the characteristics of their datasets.
We show such stacked bar charts for a few representative $(N,T)$ combinations.

Appendix~\ref{sec:badness-plots} provides yet another view into the outcome
of the competitions.  Here, results for the various $(N,T)$ combinations
are aggregated, but we focus on the individual suboptimalities
(  (measured time - best time)/best time ) observed.   They are shown
as points on a per-algorithm badness column.
The problem with
this view is that one cannot tell which $N$ and $K$ values correspond to
a given point.

\newcommand{\figwith}[1]{\
  \begin{figure}\
  \includegraphics[width=\textwidth]{#1}\
  \caption{\label{fig:#1}#1.}\
  \end{figure}}

\newcommand{\figwithzoom}[1]{\
  \begin{figure}\
  \includegraphics[width=0.5\textwidth]{#1-zoom} \
  \includegraphics[width=0.5\textwidth]{#1} \
  \caption{\label{fig:#1-zoom}#1.}\
  \end{figure}}

\newcommand{\smallfigwith}[1]{
  \begin{figure}
  \includegraphics[width=.5\textwidth]{#1}
  \caption{\label{fig:#1}#1.}
  \end{figure}}

\clearpage

\newcommand{\algopackage}[2]{
  \subsubsection{Algorithm #2} 
  \label{sec:algopackage-#1}
  The effect of varying $T$ on #2 are shown in Figs.~\ref{fig:synth-varyT-One-32-#1} and
\ref{fig:varyT-One-32-#1}.
\begin{enumerate}
\item
The effects of varying $N$ are shown in Figs.~\ref{fig:synth-varyN-T=3-#1-rel}--\ref{fig:synth-varyN-T=N-1-#1-rel}
and \ref{fig:varyN-T=3-#1-rel}--\ref{fig:varyN-T=N-1-#1-zoom-rel}.
\item

The percentage of 
\begin{enumerate}
\item cases where the algorithm achieved the fastest results is shown 
in Fig.~\ref{fig:pctWins-#1};  
\item cases where it was within 50\% of the fastest result 
(including the fastest cases) is in Fig.~\ref{fig:pctGoods-#1}; 
\item cases within 100\%  is
shown in Fig.~\ref{fig:pctOkays-#1}.  
\item disastrous cases is shown in Fig.~\ref{fig:pctSkunks-#1}.
In such cases, the algorithm took at least 10 times as long as the fastest algorithm. 
\end{enumerate}
\end{enumerate}
\begin{figure}
\centering
\subfloat[Vary $T$, $N=32$]{\label{fig:synth-varyT-One-32-#1}
\includegraphics[width=.47\textwidth]{synth-varyT-One-32-#1}}
\subfloat[Vary $N$, $T=3$]{\label{fig:synth-varyN-T=3-#1-rel}
\includegraphics[width=.47\textwidth]{synth-varyN-T=3-#1-rel}}\\
\subfloat[Vary $N$, $T\approx N/2$]{\label{fig:synth-varyN-T=middleT-#1-rel}
\includegraphics[width=.47\textwidth]{synth-varyN-T=middleT-#1-rel}}
\subfloat[Vary $N$, $T=N-1$]{\label{fig:synth-varyN-T=N-1-#1-rel}
\includegraphics[width=.47\textwidth]{synth-varyN-T=N-1-#1-rel}}
\caption{\label{synth-fig:varyN-#1-rel} Varying $T$ and $N$ on #2, synthetic data (effect of varying $N$ uses time relative to $N=32$)}
\end{figure}
\begin{figure}
\centering
\subfloat[Vary $T$, $N=32$]{\label{fig:varyT-One-32-#1}
\includegraphics[width=.47\textwidth]{varyT-One-32-#1}}
\subfloat[Vary $N$, $T=3$]{\label{fig:varyN-T=3-#1-rel}
\includegraphics[width=.47\textwidth]{varyN-T=3-#1-rel}}\\
\subfloat[Vary $N$, $T\approx N/2$]{\label{fig:varyN-T=middleT-#1-rel}
\includegraphics[width=.47\textwidth]{varyN-T=middleT-#1-rel}}
\subfloat[Vary $N$, $T=N-1$]{\label{fig:varyN-T=N-1-#1-rel}
\includegraphics[width=.47\textwidth]{varyN-T=N-1-#1-rel}}
\caption{\label{fig:varyN-#1-rel-real} Varying $T$ and $N$ on #2, real data (effect of varying $N$ uses time relative to $N=32$)}
\end{figure}
\ 
\begin{figure}
\centering
\subfloat[$T=3$]{\label{fig:varyN-T=3-#1-zoom-rel}
\includegraphics[width=.32\textwidth]{varyN-T=3-#1-zoom-rel}}
\subfloat[$T\approx N/2$]{\label{fig:varyN-T=middleT-#1-zoom-rel}
\includegraphics[width=.32\textwidth]{varyN-T=middleT-#1-zoom-rel}}
\subfloat[$T=N-1$]{\label{fig:varyN-T=N-1-#1-zoom-rel}
\includegraphics[width=.32\textwidth]{varyN-T=N-1-#1-zoom-rel}}
\caption{\label{fig:varyN-#1-zoom-rel} Effect of $N$ (times relative to $N=32$) for smaller values of $N$, on #1}
\end{figure}
\
\begin{figure}
\centering
 \subfloat[Wins]{\label{fig:pctWins-#1}
\includegraphics[width=.47\textwidth]{pctWins-#1}
}
 \subfloat[Good]{\label{fig:pctGoods-#1}
\includegraphics[width=.47\textwidth]{pctGood-#1}
}\\
 \subfloat[Fair]{\label{fig:pctOkays-#1}
\includegraphics[width=.47\textwidth]{pctOk-#1}
}
 \subfloat[Terrible]{\label{fig:pctSkunks-#1}
\includegraphics[width=.47\textwidth]{pctSkunks-#1} 
}
\caption{Heat map showing when to use and when to avoid #1 \label{fig:heatmaps-#1}}
\end{figure}
\clearpage
}

\algopackage{ScnCnt}{\scncnt}

Several algorithms showed little effect from $T$.  This include \scncnt , 
\wsort, \wheap, 
 \hashcnt\  and (to some extent) \addckt.

\scncnt\ had  strong overall performance, and was strongest
when $N\geq 32$ and $T$ was neither particularly big or particularly small.
The number of terrible cases was few, and they tended to arise when $N$ was
large and $T < N/2$.   Since \scncnt\  was not much affected by $T$,  these
results tend to highlight values of $N$ and $T$ where competitors were
particularly strong. 

Synthetic data clearly shows the dependence on $r$, and also dense cases had an
unusual situation at $N=4$, $N=8$ (and peaking at $N=16$) where the
algorithm was substantially slower than for $N=32$.  This effect remains
unexplained.  Nothing similar was observed on real data.

\algopackage{RBMrg}{\cdom}

The most complex behaviour with respect to $T$ was with \cdom\ (Fig.~\ref{fig:varyT-One-32-RBMrg}.)
  On one real dataset, its performance
was not affected by $T$.  On two others, time dropped substantially when $T$
was large enough.  The growth rates of \CensusIncome , \PGDVDtwo\ and \PGDVDthree\ resembled
the growth observed for the dense synthetic data: an increase in running time, followed by a 
reduction at the largest $T$.   \IMDBtwo\ and \IMDBthree\ more resembled the moderately sparse
synthetic data.

Synthetic data shows that \cdom's run-based performance benefited from the
clustered dense data, which presumably had few runs. 
The improvement with
$T$ may be surprising, as the input runs processed are clearly independent of
$T$.  However, we compute an output, which is more quickly done if there
are fewer runs.  When $T$ is large, we are typically 
just coalescing runs of zeros, a fast operation for a compressed bitmap.
Similarly, when $T$ is small, we may be quickly coalescing runs of ones.
For in-between values of $T$ we have to solve a threshold problem involving
the dirty words.

The complexity of the output bitmap, in terms of the number of its runs, could
be as high as the total \textsc{RunCount} of its inputs.  Thus, while the
processing of the input's \textsc{RunCount} runs may be unaffected by $T$,
the number of output runs (after coalescing) may vary dramatically.  If the
cost of constructing a compressed bitmap is substantially higher than the
cost of reading one, we could see results where the threshold makes these kinds of
differences (factors near 10). 

Its strengths seem to be with $N=16$ and $N=32$ for larger values of $T$.
It was only infrequently terrible, even more so than \scncnt.

\algopackage{DSk}{\dsk}    
The algorithms that use pruning (\dsk, \mgsk, \mgopt ) benefited
greatly from large thresholds.  Typical differences were between 1 and 2
orders of magnitude.  However, the shape of the drop-off curve varied
by dataset.  

Synthetic data shows that the density of a dataset was important in determining
how steep the drop-off curve from pruning is. (We see this both in
Figs.~\ref{fig:synth-varyT-One-32-DSk} and~\ref{fig:varyN-T=N-1-DSk-rel}.)
It seems that denser data led to more benefit from pruning.

When $T=N-1$, we see a very complex behaviour on several real datasets that
peaked early with $N$. \PGDVDtwo\ peaks at 16 and then became faster with
larger $N$, until $N=128$.  Thereafter, its time very gradually increased with $N$.

\dsk\ depends significantly on the choice of an appropriate parameter
$\mu$ that controls $L$.  The suggested approach from~\cite{Li:2008:EMF:1546682.1547171} was to determine
this experimentally, for each dataset, using a collection of representative
queries. 
We chose a single sample of size $N$ as discussed
above, then ran queries with a collection of different threshold values.
The selected values of $\mu$ are shown in Table~\ref{tab:mu-values}.
Note that the best value for
$\mu$ depends significantly on the threshold, and thus the values in the
tables are thus compromises that may work poorly for many thresholds.
For instance, when $N=512$, $T=369$, \IMDBthree\ preferred $\mu=0.00490$.
Yet the approach in~\cite{Li:2008:EMF:1546682.1547171} would have us use $\mu=0.139$ whenever 
$N=512$, because many smaller thresholds preferred it.

\begin{table}
\caption{\label{tab:mu-values}Values of $\mu$ recommended.}
\begin{tabular}{c|rrr}
\hline\hline
  Dataset        & 1 row, $N=16$ & 10 rows, $N= 128$  & 100 rows, $N=512$\\ \hline
  \PGDVD         &  0.0343       &  0.0226            & 0.5820\\          
  \CensusIncome  &  0.0432       &  0.0162            & 0.0144\\ 
  \IMDBtwo       &  0.0508       &  0.0495            & 0.0536\\
  \IMDBthree     &  0.1170       &  0.0731            & 0.1390\\ \hline
\end{tabular}
\end{table}

We see a wide range of $\mu$ values.  For experiments, we chose
four representative
values:  $\mu \in \{ 0.005, 0.02, 0.05, 0.1 \}$, labelling
\dsk\ with these choices as
DSk.005, DSk.02, DSk.05 and DSk.1.
We also report (as DSk) on the
\emph{best} result achieved by any of the 4 $\mu$ values, which is
what is shown here.  This gives the algorithm an unfair advantage,
as we do not know how to obtain these results without multiple attempts
using different values of $\mu$.  In our experience, the difference between
a good and a poor choice of $\mu$ (within the range discussed above) is usually
less than an order of magnitude.  

The heat maps show that \dsk\ was a good choice when $T$ was close to $N$.
However, when this was not the case, DivideSkip was frequently terrible.

\algopackage{MgOpt}{\mgopt}   

\mgopt has performance similar to (but weaker than) \dsk.  It
seems to have more trouble on non-dense synthetic data than \dsk\ when
$T\approx N/2$.  
On real data, with $T=N-1$ it has many
of the same behaviours as \dsk .

\algopackage{w2CtI}{\wtwocti}

Algorithm \wtwocti\ 
appears
to have had superlinear (in $N$) running time on synthetic data.
Recall that its worst-case running time is $\Theta(BN)$ and that
$B$, on average, increases linearly with $N$. Thus it might not be
surprising to see super-linear increases.  We also see that this algorithm
was fastest for dense data.  With such data, there is a good chance for items
to have occurred in both lists being merged, resulting in a smaller output and
less data being processed in future merges.
Similar effects were observed (but are not shown)
 for \wtwocta\ or \wtwoctn\, which do less (or no)
pruning.

All `w' implementations suffered from memory problems that affected their
performance when $N$ was large, for some of the larger datasets.
In particular, on \PGDVDtwo, the computation for \wtwocti\  took excessive time
for $N=256$ and failed for $N=512$.

On real data, \wtwocti\ 
heat maps show that
this was a very poor algorithm for large $N$ and small $T$.
It did have some wins and some good behaviour when $T \approx N$.

\algopackage{Looped}{\looped}
As expected, \looped\ usually became worse with $T$, although the effect was
much more pronounced on real data.  The large dip from $T=3$ to $T=4$ 
on the dense synthetic data is unexplained. The algorithm for \looped\ does
the operations for $T=3$ in the course of computing $T=4$.  We see the same
surprising effect on real data, for \CensusIncome.
When $N$ is varied with $T=3$ we again see unusual behaviour for the 
dense synthetic data.   This may be due to the efficiency with which 
JavaEWAH can OR bitmaps that have long runs of ones.
Otherwise, for synthetic data 
we see curves that make sense, given our $O(NT)$ analysis.

Heatmaps confirm that the approach, which does not win particularly often, does
so for smaller values of $T$, even with larger $N$.  
It is almost always terrible when both $N$ and $T$
are large.

\algopackage{wMgSk}{wMergeSkip}  

The `w' version of MergeSkip is outperformed by the `w'
version of DivideSkip (which is in turn outperformed
by the bitmap version of DivideSkip).
On synthetic and real data, we see that larger $T$ values improve time: pruning
is working.

On real data, we see the memory problems faced by many `w' implementations: for
several datasets, $N=512$ results could not be obtained.

Examining heatmaps, we see the algorithm was never best.
Again, its (limited) strength is along the $T \approx N$
diagonal and it is frequently terrible.

\algopackage{SrtCkt}{\srtckt}

On synthetic data, \srtckt\ is not much affected by $T$, but it prefers
dense data.   Such bitmaps compress better and there is more opportunity
for bit-level parallelism than with extremely sparse data. 
For dense synthetic data, there is an initial peak when $N$ varies, similar
to \scncnt .

On some real datasets, varying $T$ makes a significant difference without an obvious
trend.
 Since the circuits
differ according to $T$, this may not be too surprising.  However, the same
circuit behaves differently for different datasets.  

On real datasets, increasing $N$ appears to lead to superlinear time increases.  We may be
seeing the extra $\log^2 N$ factor, or the slowdown may be due to memory use (\srtckt 
seems to require more space for its temporary results than the adder circuits).

\srtckt\ is strongest when both $N$ and $T$ are small.  It becomes increasingly terrible with $N$ and
is frequently terrible when $N=512$, but is
rarely terrible when $N \leq 32$.

\algopackage{TreeAdd}{\addckt} 

On synthetic data, \addckt\ preferred dense data, but its scaling was
very little affected by density.  It had erratic behaviour for the
three smallest values of $N$ (4, 8 and 16) on dense synthetic data,
and real data shows similarly erratic behaviour for small $N$.

This circuit is a good choice when $N$ and $T$ are both small. It is infrequently
terrible; when it is, the situation is that the pruning-based algorithms have become
especially strong.   \addckt\ has not really degraded.

\algopackage{SSum}{\kaddckt} 
The sideways-sum circuit  behaved much differently for dense
synthetic data than did \addckt .  The odd behaviour for clustered
dense data, as $T$ varies, is unexplained.  There seems to be no similar
behaviour on our real data, however.
The heatmaps show that it won many competitions across a wide range
of $T$ and $N$, and was at least frequently fair (except when $N$ was large and
$T$ was small, when \looped\ often made it look worse than fair). 
Terrible cases were limited (but more frequent when $N=512$) and appeared
mostly on the diagonal, where pruning algorithms could sometimes make it look
bad.  

\algopackage{CSvCkt}{\schedcs}

The carry-save adder approach was not much affected by $T$ or by whether the
synthetic data was uniform or clustered.  It was somewhat slower than the
other adder approaches and therefore had fewer wins. It was more frequently
terrible (but in similar situations) when compared to the other adders.

\algopackage{wSort}{\wsort}

Algorithm \wsort\ is not much affected by density, for synthetic data.  This
makes sense.  The behaviour is very consistent as $N$ is increased.  Since 
we use $T$ only after most of the work (sorting) has been done, we should not
be surprised to see that $T$ has fairly little effect.

Despite being a `w' algorithm, it won a few cases for smaller $N$.  However, this algorithm
lead to more memory-exhaustion failures than any other, as \PGDVDthree\ and \PGDVDtwo\ are not
successful for $N\geq 256$.

\algopackage{HashCnt}{\hashcnt}

On synthetic data, \hashcnt\ preferred dense data. Fewer distinct counters
needed to be maintained, and thus the required hash table was smaller, leading
to cache benefits.

The \hashcnt\ approach was usually terrible.   This is somewhat surprising, considering how
strong \scncnt\ was.  If the range of items was too big for \scncnt , a typical
programmer would consider this route instead.

To see why this approach might be so ineffective, we have to compare the costs of hashing the counter
number to the cost of indexing into an array.  Also, the \texttt{HashMap} implementation
will require boxing and unboxing integer values, although that can be avoided
with more specialized int-to-int hash libraries.   Worse, the hashing
process means that the efficient ascending scan through counter numbers will 
turn into irregular accesses through the heap, even with the more 
specialized libraries.
Finally, constructing the compressed bitmap index requires iterating over the 
keys in sorted order.   Sorting them is expensive.

\algopackage{wHeap}{\wheap} 

On synthetic data, the characteristics of the data were insignificant. 

The time complexity has a $\log N$ factor,  which presumably accounts for the
shape of the curves on synthetic data.   As one of the `w' algorithms, we 
could not run all datasets with queries for $N=512$.
The \wheap\ algorithm was usually terrible.

\subsection{Workload}
\label{sec:workload}

Although we have seen the algorithms in isolation,
we still want some way to say that ``Algorithm X
is better than Algorithm Y''.  To do this, we can examine
total time taken on a particular set of queries (workload).
However, we wish to control for the different sizes of the
datasets---otherwise, performance on small datasets will be
disregarded.  Therefore, we use a workload that assumes
fast-running (for \emph{some} algorithm) queries are issued more 
often than queries that
are slow (for \emph{all} algorithms).

\begin{figure}
\subfloat [Full y range]{
\includegraphics[width=\textwidth]{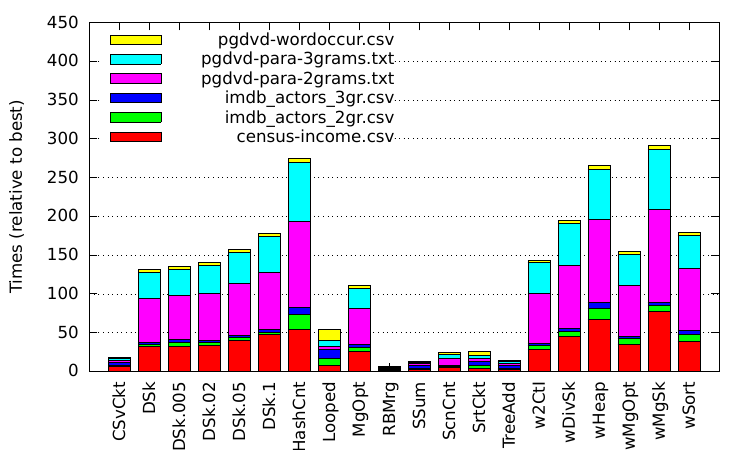}}\\
\subfloat [Reduced y]{
\includegraphics[width=\textwidth]{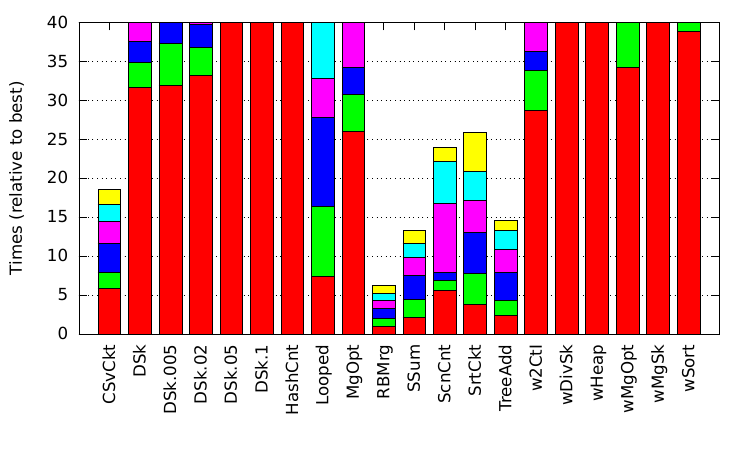}}
\caption{\label{fig:all-speeds-One-at-3219}Relative speeds of algorithms over a collection of queries with $T=16$ and $N=32$. }
\end{figure}

Fig.~\ref{fig:all-speeds-One-at-3219} shows how the various algorithms fare
for a threshold of 19 over 32 bitmaps.  
Since the datasets are of
varying size and we do not want the largest datasets to dominate
the effects, we normalize the running times on a dataset D by dividing the
running time by the fastest running time of any algorithm on D. 
I.e., if $t_{\mathcal{A}}(D)$ denotes the actual running time when
algorithm $\mathcal{A}$ processes a  given
collection of queries  on dataset $D$, then its normalized time
is $$t^{\mathrm{norm}}_{\mathcal{A}}(D) = t_{\mathcal{A}}(D) / \min_{\mathcal{A}'}t_{\mathcal{A'}}(D).$$

This 
is similar to saying that we have a workload where queries against the
smaller datasets are more frequent.  It makes equal each dataset, in that the
 total time spent on all queries (executed by the  fastest algorithm)
for two different datasets would be equal. 
If one algorithm were fastest in all 6 cases, its total bar height would be
6.  We see no bar is quite that low, but \scncnt ,
\cdom, \addckt and \srtckt\ look
promising.  (Many of the other algorithms look bad because they performed
very poorly on one dataset, \PGDVDtwo.)

Looking at bars whose colours are unequal,
we identify several algorithms that are sensitive to the characteristics
of the datasets (for instance, the various wrapped algorithms seem to
do more poorly on \PGDVDtwo\ and \PGDVDthree\ than other datasets,
whereas \looped\ finds \IMDBthree\  difficult. \cdom, the adder and sorting
circuits, and \scncnt\ seem less sensitive to the datasets than
other algorithms.

\begin{figure}
\subfloat [Full y range]{
\includegraphics[width=\textwidth]{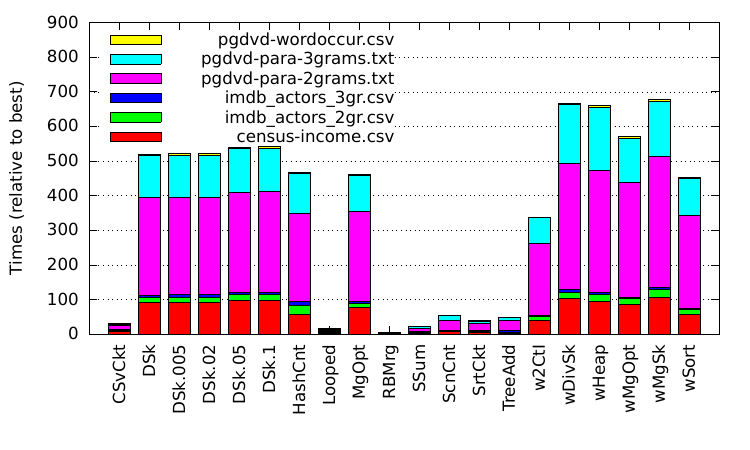}}\\[1em]
\subfloat [Reduced y]{
\includegraphics[width=\textwidth]{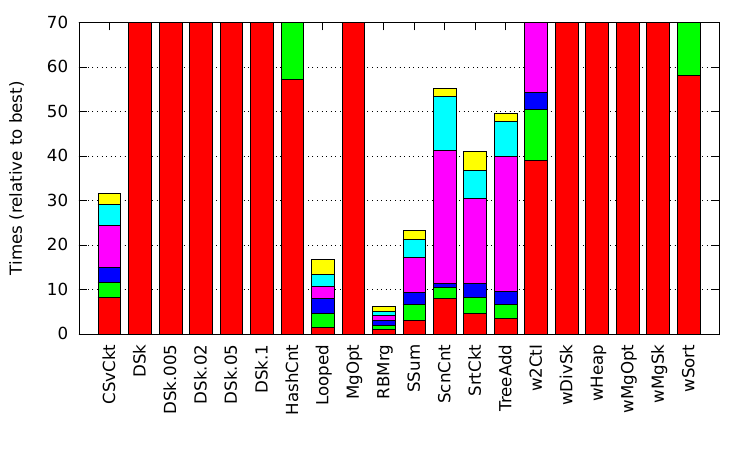}}
\caption{\label{fig:all-speeds-One-at-3203}Relative speeds of algorithms over a collection of queries with $T=3$ and $N=32$. }
\end{figure}

\begin{figure}
\subfloat [Full y range]{
\includegraphics[width=\textwidth]{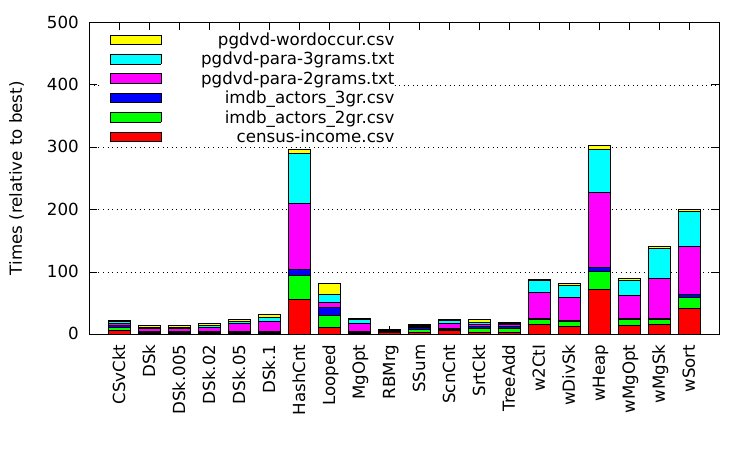}}\\
\subfloat [Reduced y]{
\includegraphics[width=\textwidth]{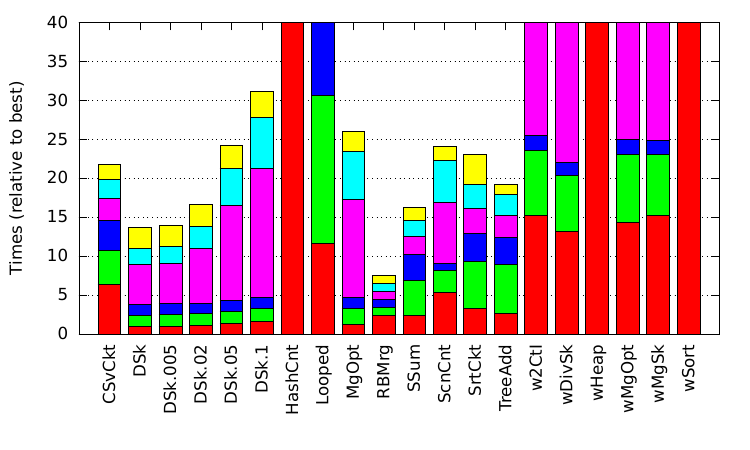}}
\caption{\label{fig:all-speeds-One-at-3231}Relative speeds of algorithms over a collection of queries with $T=31$ and $N=32$. }
\end{figure}

However, if we repeat the process for thresholds 3  or 31 on 32 bitmaps,
results differ (see Figs.~\ref{fig:all-speeds-One-at-3203}~and~\ref{fig:all-speeds-One-at-3231}.  
For $T=31$, \dsk\  works particularly  well and 
\looped\  is about 10 times slower than optimal, whereas
for $T=3$ \looped\  performs best and \dsk\  is roughly 40
(250/6) times slower than optimal.
(Note that the $y$ range changes in these figures).
Algorithms \cdom\ and \looped\ are
relatively insensitive to the dataset.   For $T=31$ we see that the value $\mu$ for \dsk\
affects whether the \PGDVDtwo\ dataset is difficult.

\subsection{Discussion}

If we are to recommend only one threshold algorithm for
similarity queries on
run-length compressed bitmaps, then \cdom\ should be used.
It was frequently the fastest algorithm, and it rarely
behaved disastrously.   However, if $T$ is very small,
\looped\ may perform better.   As well, the sideways-sum
adder circuit (\kaddckt) performed well overall and was infrequently
disastrous.  Both \looped\ and \kaddckt\ have the advantage
that they can use an unmodified (perhaps closed-source) bitmap
library, whereas \cdom\ would require adding code to the
bitmap library.  The \looped\ algorithm has the advantage
of simplicity, since a substantial implementation effort
is required with any of the circuit-based approaches.

The simple \scncnt\ depends heavily on the dataset.  On some
datasets, its performance is frequently best, whereas in
others, it is frequently terrible.  Its use is risky.   The
hash-map variant, \hashcnt, is almost always disastrous.

For similarity queries over run-length compressed bitmaps, 
the `w' implementation approach (transform the data into
lists of sorted integers) cannot be recommended.  We observed
many cases where the list of integers was much larger, thus
creating memory problems.  Also, the time cost in doing the
conversion is too high.

Well-known pruning-based algorithms such as \mgopt\ and \dsk\ 
are not suitable for general threshold computations over bitmaps. In
cases where $T \approx N$, they often do well (but many times,
so does \cdom ).   There are some applications where the requirement
for a large $T$ can be met. For instance,
using the formula in~\cite{Li:2008:EMF:1546682.1547171}
(which they credit to Sarawagi and Kirpal~\cite{Sarawagi:2004:ESJ:1007568.1007652}) with strings of
length 64, if we are interested in finding
the strings of edit distance at most two from some target
using trigrams, the appropriate threshold is $64+3-1-2*3=60$.

\subsubsection{Performance counters results.}

Execution times include factors other than the number of 
instructions executed.  Pipeline stalls due to cache or translation-lookaside-buffer (TLB) misses
can lead to unexpectedly poor running times.  To assess our
implementations, we generated 10  ``Random attributes''  queries
(such as those used in the top of Table~\ref{tab:with-and-without-index})
 and used
the Linux \texttt{perf stat} tool  
to monitor various cache-miss rates
as well the overall instructions-per-cycle figure of merit.  
Our focus was on the most promising algorithms.
Although the
data generation phase was included, it took much less than 
\SI{1}{\second}, and the
query was repeatedly executed for \SI{100}{\second}. 
Query input volumes ranged between \SI{103}{kB} and \SI{43}{MB}, so the
computation cannot be done entirely in cache.  
The
formulae used to calculate the L2 and L3 miss rates were 
non-obvious and found
on an Intel forum~\cite{cepe:sandy-bridge-cache-formulae,inte:dev-man-perfcodes}, which indicates
there is no way to calculate the L1 miss rate.
  Results are in
Table~\ref{tbl:perfctrs}.

\begin{table}
\caption{\label{tbl:perfctrs} Hardware performance counter results.
}
\centering
\footnotesize
\begin{tabular}{lSSSSS} \hline 
Algorithm         & IPC & \multicolumn{4}{c}{Miss rates (\%)}  \\ 
                  &     & \multicolumn{1}{c}{L2}&\multicolumn{1}{c}{L3}&\multicolumn{1}{c}{TLB-load}&\multicolumn{1}{c}{TLB-store}\\ \hline 
\scncnt           & 1.9  & 5.0    & 0.31    & 4.2e-3 & 7.4e-4     \\ 
\wtwocti          & 2.0  & 11.9   &  0.90   & 3.6e-2 & 6.7e-2    \\ 
\mgopt            & 2.2  & 1.7    &  0.16   & 5.7e-2 & 8.0e-4    \\ 
\dsk              & 2.2  & 1.8    & 0.17    & 3.8e-2 & 7.4e-4     \\ 
\kaddckt          & 2.1  & 5.9    & 0.43    & 7.1e-3 & 2.1e-2    \\ 
\looped           & 2.1  & 8.4    &  0.31   & 5.0e-3 & 2.3e-2    \\  
\cdom             & 2.1  & 6.1    & 0.17    & 1.9e-1 & 4.4e-3     \\  
\hline  
\end{tabular}
\end{table}

The table shows remarkable consistency in Instructions Per Cycle (IPC).
Algorithms \dsk\ and \mgopt\ put the least load on the memory system,
whereas---unsurprisingly---\wtwocti\ put the most load on it.     The \cdom\ algorithm
had the worst TLB performance,
 but very good results for L3.  
Understanding the behaviour of an algorithm on different cache levels  is important for parallel computing.
In a multithreaded environment, we will want to do most of our computation
within L1 and L2 (on our processor,  L3 cache is shared
between all cores, unlike L1 and L2).

\section{Conclusion and Future Work}

This report has considered algorithms for threshold (and more generally,
symmetric Boolean functions) on bitmaps, both compressed (EWAH) and
uncompressed (\textsc{BitSet}). The algorithms both include known algorithms 
that have been adapted so they work with bitmaps (rather than sorted lists
of integers), and also include a number of novel algorithms.

Experiments were conducted using small synthetic datasets, as well as
several moderately large real datasets.  The use of similarity queries
and the choices of real datasets mean that our results can be related
to those in previous work, such as that by Li et al. 

We have seen that the most straightforward algorithm, \scncnt, can often
be outperformed on similarity queries over bitmap indexes.  
This observation is consistent with previous results reported
by Li et al.\ 
when the data was represented as sorted lists of
integers.   Nevertheless,
it is not disastrous if the bitmaps are sufficiently sparse.
We have also seen that the other algorithms from Li et al.\ are
highly dependent on the characteristics of the data and the selected
threshold.   When $T$ is near $N$, they are frequently good.  However, 
for other values of $T$ they can behave very poorly.  For EWAH compressed
bitmaps, the \cdom\ algorithm exploits the runs in the data and rarely
behaves badly.  For both compressed and uncompressed (\textsc{BitSet}) bitmaps, the algorithms
based on Boolean circuits perform well.  Of \addckt , \kaddckt\ and \srtckt,
we recommend \kaddckt .  It uses fewest operations for large $N$ and seems to
be less dependent on the data characteristics than \addckt\ or \srtckt.
These algorithms are not affected much by the choice of $T$, and can make
use of an unmodified bitmap library.
A disadvantage is that considerable implementation effort is required.
An alternative adder design, \schedcs, has a much simpler implementation.
However, its performance does not match \kaddckt\ or \addckt .  It should be
possible to implement \kaddckt\  following the general style of \schedcs ,
rather than using the explicit circuit approach.  
The simple \looped\ algorithm also can use an unmodified bitmap library,
and it is an excellent choice when $T$ is very small.  While it becomes worse
as $T$ grows, the growth on actual data may be significantly better than our
theoretical worst-case analysis suggests.

For future work, we should broaden the threshold computations to include
those needed to answer queries other than similarity queries.  We have seen
that similarity queries tend to use bitmaps that are significantly denser
than the average bitmap present in an index.  We have identified another
class of queries as being interesting, wherein a random attribute value is
selected for each of $N$ attributes, as we did in
Table~\ref{tab:with-and-without-index}. This will end up choosing sparser bitmaps
and is likely to favour  \scncnt .

Our work has considered $N$ values up to 512.  Yet we use
datasets whose indexes have
tens of thousands of bitmaps, and datasets with millions of bitmaps are not
out of the question.  Would there be applications where $N=1,000,000$ would
be useful?  If so, which algorithms should be used?  The circuit-based
algorithms will become infeasible for extremely large $N$, but which of the
others can be used?  Can new algorithms be developed for this case?

We have seen that several algorithms need certain characteristics of the
data in order to be successful.  To date, we have mostly looked at the
sparseness of the bitmaps or their \textsc{RunCount}.  Are there other
data characteristics that can determine whether a given algorithm should
be recommended?  If so, can we find ways to compute these characteristics or
maintain them in the DBMS?

Finally, algorithms can be parallelized, and while most of our
threshold computations take only a few milliseconds, we have some that
may take tens of seconds.  If we try extremely large $N$
values, this may increase.  At some point, it may become important to
have one threshold computation run faster than is possible using a
single core.  For multicore processing, a particular challenge is that
all cores compete for access to L3 and RAM. E.g., this means that it
is best if intermediate results fit in L2 cache.  

\bibliographystyle{plain} 
\bibliography{../../bib/longtitles,../../bib/lemur}

\begin{thebibliography}{10}

\bibitem{ajta:sorting-circuit}
M.~Ajtai, J.~Koml\'{o}s, and E.~Szemer{\'e}di.
\newblock Sorting in c log n parallel steps.
\newblock {\em Combinatorica}, 3(1):1--19, 1983.

\bibitem{Anh:2010:ICU:1712666.1712668}
Vo~Ngoc Anh and Alistair Moffat.
\newblock Index compression using 64-bit words.
\newblock {\em Software---Practice and Experience}, 40(2):131--147, 2010.

\bibitem{874730}
G.~Antoshenkov.
\newblock Byte-aligned bitmap compression.
\newblock In {\em Data Compression Conference (DCC'95)}, page 476, Washington,
  DC, USA, 1995. IEEE Computer Society.

\bibitem{ashe:digital-design-book}
Peter~J. Ashenden.
\newblock {\em Digital Design ({Verilog}): An Embedded Systems Approach Using
  {Verilog}}.
\newblock Elsevier, 2007.

\bibitem{behm2009space}
Alexander Behm, Shengyue Ji, Chen Li, and Jiaheng Lu.
\newblock Space-constrained gram-based indexing for efficient approximate
  string search.
\newblock In {\em Proceedings IEEE 25th International Conference on Data
  Engineering (ICDE'09)}, pages 604--615. IEEE, 2009.

\bibitem{cepe:sandy-bridge-cache-formulae}
Shannon Cepeda.
\newblock How to get the {L1,L2} cache miss of an intel[sic] i5 {Sandy Bridge}
  (response).
\newblock online, Intel Developer Zone,
  \url{http://software.intel.com/en-us/forums/topic/280087}, 2012.
\newblock last checked 2014-02-17.

\bibitem{Colantonio:2010:CCN:1824821.1824857}
Alessandro Colantonio and Roberto Di~Pietro.
\newblock Concise: Compressed 'n' composable integer set.
\newblock {\em Information Processing Letters}, 110(16):644--650, July 2010.

\bibitem{CLRSbook}
Thomas Cormen, Charles Leiserson, Ronald Rivest, and Clifford Stein.
\newblock {\em Introduction to Algorithms}.
\newblock MIT Press, 3rd edition, 2009.

\bibitem{Culpepper:2010:ESI:1877766.1877767}
J.~Shane Culpepper and Alistair Moffat.
\newblock Efficient set intersection for inverted indexing.
\newblock {\em ACM Transactions on Information Systems}, 29(1):1:1--1:25,
  December 2010.

\bibitem{trove}
Rob Eden.
\newblock {GNU TROVE} high performance collections for {Java}.
\newblock online: \url{https://bitbucket.org/robeden/trove/}.
\newblock last checked 2014-02-17.

\bibitem{elli:scheduled-vertical-counter}
Erling Ellingsen.
\newblock Bit tricks, part {III}: Fast vertical counter.
\newblock online:\url{http://www.steike.com/code/bits/vertical-counter/}, 2009.
\newblock last checked 2014-02-17.

\bibitem{ferr:duplicates-qgrams}
Alfredo Ferro, Rosalba Giugno, Piera~Laura Puglisi, and Alfredo Pulvirenti.
\newblock An efficient duplicate record detection using q-grams array inverted
  index.
\newblock In {\em 12th International Conference on Data Warehousing and
  Knowledge Discovery (DaWaK'10), LNCS 6263}, pages 309--323. Springer, 2010.

\bibitem{MLRepository}
A.~Frank and A.~Asuncion.
\newblock {UCI} machine learning repository.
\newblock \url{http://archive.ics.uci.edu/ml} (checked 2014-02-17), 2010.

\bibitem{netfli}
F.~Fusco, M.~P. Stoecklin, and M.~Vlachos.
\newblock {NET-FLi}: On-the-fly compression, archiving and indexing of
  streaming network traffic.
\newblock {\em Proceedings of the VLDB Endowment}, 3:1382--1393, 2010.

\bibitem{gare:gandj}
Michael~R. Garey and David~S. Johnson.
\newblock {\em Computers and Intractability: A Guide to the Theory of
  {NP}-Completeness}.
\newblock W. H. Freeman, New York, 1979.

\bibitem{openjdk-bitset-sourcecode}
GrepCode.
\newblock Grepcode: java.util.bitset(.java)-class-source code view.
\newblock online:
  \url{http://grepcode.com/file/repository.grepcode.com/java/root/jdk/openjdk/7-b147/java/util/BitSet.java#BitSet}.
\newblock last checked 2014-02-17.

\bibitem{inte:dev-man-perfcodes}
{Intel Corporation}.
\newblock {\em Intel 64 and IA-32 Software Developer's Manual: System
  Programming Guide, Part 2}, volume~3B.
\newblock Intel Corporation, September 2013.
\newblock \url{http://www.intel.com/Assets/PDF/manual/253669.pdf} last checked
  2014-02-17.

\bibitem{jia2012eti}
Lianyin Jia, Jianqing Xi, Mengjuan Li, Yong Liu, and Decheng Miao.
\newblock {ETI}: an efficient index for set similarity queries.
\newblock {\em Frontiers of Computer Science}, 6(6):700--712, 2012.

\bibitem{KaserKeithLemire2006}
Owen Kaser, Steven Keith, and Daniel Lemire.
\newblock The {LitOLAP} project: Data warehousing with literature.
\newblock In {\em Proceedings, CaSTA'06 : The 5th Annual Canadian Symposium on
  Text Analysis}, pages 93--96, 2006.

\bibitem{KnuthV3E3}
Donald~E. Knuth.
\newblock {\em Searching and Sorting}, volume~3 of {\em The Art of Computer
  Programming}.
\newblock Addison-Wesley, Reading, Massachusetts, 1997.

\bibitem{KnuthV4A}
Donald~E. Knuth.
\newblock {\em Combinatorial Algorithms, Part 1}, volume~4A of {\em The Art of
  Computer Programming}.
\newblock Addison-Wesley, Boston, Massachusetts, 2011.

\bibitem{rlewithsorting}
Daniel Lemire and Owen Kaser.
\newblock Reordering columns for smaller indexes.
\newblock {\em Information Sciences}, 181(12):2550--2570, June 2011.

\bibitem{arxiv:0901.3751}
Daniel Lemire, Owen Kaser, and Kamel Aouiche.
\newblock Sorting improves word-aligned bitmap indexes.
\newblock {\em Data \& Knowledge Engineering}, 69(1):3--28, 2010.

\bibitem{JavaEWAH}
Daniel Lemire, Cliff Moon, David McIntosh, Robert Becho, Colby Ranger, Veronika
  Zenz, and Owen Kaser.
\newblock {JavaEWAH} - {GitHub} page.
\newblock online: \url{https://github.com/lemire/javaewah}, 2014.
\newblock last checked 2014-02-13.

\bibitem{Li:2008:EMF:1546682.1547171}
Chen Li, Jiaheng Lu, and Yiming Lu.
\newblock Efficient merging and filtering algorithms for approximate string
  searches.
\newblock In {\em Proceedings of the 2008 IEEE 24th International Conference on
  Data Engineering (ICDE'08)}, pages 257--266, Washington, DC, USA, 2008. IEEE
  Computer Society.

\bibitem{li2013fast}
Mengjuan Li, Lianyin Jia, Jinguo You, Jianqing Xi, HaiFei Qin, and Rui Zeng.
\newblock Fast {T}-overlap query algorithms using graphics processor units and
  its applications in web data query.
\newblock {\em World Wide Web}, pages 1--17, 2013.

\bibitem{moffat1996self}
Alistair Moffat and Justin Zobel.
\newblock Self-indexing inverted files for fast text retrieval.
\newblock {\em ACM Transactions on Information Systems}, 14(4):349--379, 1996.

\bibitem{montanari2012near}
Daniele Montanari and Piera~Laura Puglisi.
\newblock Near duplicate document detection for large information flows.
\newblock In {\em IFIP WG 8.4, 8.9, TC 5 International Cross Domain Conference
  and Workshop on Availability, Reliability, and Security, CD-ARES 2012, LNCS
  7465}, pages 203--217. Springer, August 2012.

\bibitem{navarro2012fast}
Gonzalo Navarro and Eliana Providel.
\newblock Fast, small, simple rank/select on bitmaps.
\newblock In {\em Proceedings, 11th International Symposium on Experimental
  Algorithms (SEA 2012), LNCS 7276}, pages 295--306. Springer, 2012.

\bibitem{Perry01021983}
Shirley~A. Perry and Peter Willett.
\newblock A review of the use of inverted files for best match searching in
  information retrieval systems.
\newblock {\em Journal of Information Science}, 6(2-3):59--66, 1983.

\bibitem{GutenbergDVD}
{Project Gutenberg Literary Archive Foundation}.
\newblock July 2006 {Gutenberg} {DVD}.
\newblock \url{http://www.gutenberg.org/wiki/Gutenberg:The_CD_and_DVD_Project},
  2006.
\newblock (Last checked 2014-02-13).

\bibitem{quine53}
Willard van~Orman Quine.
\newblock Two theorems about truth functions.
\newblock {\em Bolet\'{\i}n de la Sociedad Matem\'atica Mexicana}, 10, 1953.

\bibitem{rinfret:bit-sliced-arithmetic}
Denis Rinfret, Patrick O'Neil, and Elizabeth O'Neil.
\newblock Bit-sliced index arithmetic.
\newblock In {\em Proceedings of the 2001 ACM SIGMOD International Conference
  on Management of Data}, pages 47--57. ACM, May 2001.

\bibitem{sanders2007intersection}
Peter Sanders and Frederik Transier.
\newblock Intersection in integer inverted indices.
\newblock In {\em 2007 Proceedings of the Ninth Workshop on Algorithm
  Engineering and Experiments (ALENEX07)}, volume~7, pages 71--83,
  Philadelphia, PA, USA, 2007. SIAM.

\bibitem{Sarawagi:2004:ESJ:1007568.1007652}
Sunita Sarawagi and Alok Kirpal.
\newblock Efficient set joins on similarity predicates.
\newblock In {\em Proceedings of the 2004 ACM SIGMOD International Conference
  on Management of Data}, pages 743--754, New York, NY, USA, 2004. ACM.

\bibitem{seth:complete-register-allocation-problems}
Ravi Sethi.
\newblock Complete register allocation problems.
\newblock {\em SIAM Journal on Computing}, 4:226--248, 1975.

\bibitem{1083658}
Mike Stonebraker, Daniel~J. Abadi, Adam Batkin, Xuedong Chen, Mitch Cherniack,
  Miguel Ferreira, Edmond Lau, Amerson Lin, Sam Madden, Elizabeth O'Neil, Pat
  O'Neil, Alex Rasin, Nga Tran, and Stan Zdonik.
\newblock {C-Store}: a column-oriented {DBMS}.
\newblock In {\em VLDB'05, Proceedings of the 31st International Conference on
  Very Large Data Bases}, pages 553--564, New York, NY, USA, 2005. ACM.

\bibitem{warr:hackers-delight-2e}
Henry~S. Warren, Jr.
\newblock {\em Hacker's Delight}.
\newblock Addison Wesley, 2nd ed. edition, 2013.

\bibitem{webb2013}
Hazel Webb, Daniel Lemire, and Owen Kaser.
\newblock Diamond dicing.
\newblock {\em Data \& Knowledge Engineering}, 86:1--18, 2013.

\bibitem{wu2008breaking}
K.~Wu, K.~Stockinger, and A.~Shoshani.
\newblock Breaking the curse of cardinality on bitmap indexes.
\newblock In {\em SSDBM'08: Proceedings of the 20th International Conference on
  Scientific and Statistical Database Management, LNCS 5069}, pages 348--365.
  Springer, 2008.

\bibitem{1316694}
Kesheng Wu, Ekow Otoo, and Arie Shoshani.
\newblock On the performance of bitmap indices for high cardinality attributes.
\newblock In {\em VLDB'04, Proceedings of the 30th International Conference on
  Very Large Data Bases}, pages 24--35. Morgan Kaufmann, 2004.

\bibitem{DBLP:journals/tods/WuOS06}
Kesheng Wu, Ekow~J. Otoo, and Arie Shoshani.
\newblock Optimizing bitmap indices with efficient compression.
\newblock {\em ACM Transactions on Database Systems}, 31(1):1--38, 2006.

\end{thebibliography}

\appendix

\section{IMDB dataset}
\label{sec:imdb-appendix}
The IMDB dataset was used in~\cite{Li:2008:EMF:1546682.1547171}, 
but the details necessary to prepare an identical copy of the dataset were not
given.
The IMDb organization makes its data available for noncommercial use, 
but it is not free data. 
The conditions also explicitly forbid using it to make an online database: 
this may prevent us from distributing the final dataset we obtained.
However, the following steps can be followed:
\begin{enumerate}
\item Download \verb+actors.list.gz+ from \url{ftp://ftp.sunet.se/pub/tv+movies/imdb/}.  
In autumn 2013, the gzipped file was 223\,MB; uncompressed it was 803\.,MB.
Li et al. noted the actors list
was 22MB  ~\cite{Li:2008:EMF:1546682.1547171}. 
\item Find the line \verb!Name\s+Title!, skip the following line, then process
lines until a line consisting of at least 60 -'s is reached (several hundred lines from the end
of the file.  
\item For the processed part of the file, take every line  that is  non-blank in column 1, 
and extract from column 1 to the first tab character as an actor name, using the sed pipeline\\
{\footnotesize
\verb+sed '/^	.*/d' actorslistmiddle.txt  | sed '/^\s*$/d' | sed 's/	.*//'+ \\}where the apparent
spaces are tabs. 
In autumn 2013, the result was 35\,MB with
1783816 actors.  No other attempts to clean data or regularize punctuation should be made.  The
number of unique bigrams
 was 4276  and the number of unique trigrams was 50663.  
The number of unique bigrams is larger than one
might expect, because the actor list
is international and has many non-English names and stage names.
Names can include punctuation and other symbols.
Also, our n-grams are case sensitive.

Note that the comma-space bigram occurs in almost every name due to 
the heavy use of the "surname, firstname" pattern.
There may be some other extremely common bigrams. The majority of
n-grams occur very rarely, however.
\end{enumerate}

The database has grown: \cite{Li:2008:EMF:1546682.1547171} report 1,199,299 names
with total size 22\,MB, and the number of unique grams was 34737.  (They used 3-grams.)
We determined that the bitmap indexes would have about 35 million set
bits.  Considering the 1.7 million actors, this means about 20 bigrams or trigrams per actor on average, 
very similar to the 19 reported by Li et al.~\cite{Li:2008:EMF:1546682.1547171}.

\section{Compiled code}
\label{app:gencode}

Our experiments compiled Boolean circuits into a number
of different targets: Java (using either the EWAH
or BitSet libraries), C++ (uncompressed bitmaps stored in arrays),
or a custom byte-code that a Java program interprets (using either
\texttt{BitSet} or EWAH).  We finally
used the last approach, but the other targets are human-readable 
and carry out the same logical operations.

The C++ code has a each bitmap as a fixed-size array of 64-bit words.
Using AVX, we could
instead work with 256 (or 512, with AVX2) bit words.

For $N=5$, $T=2$ and the \kaddckt\ circuit we get the following.   The
function is vertical, in that it takes complete input vectors and
produces a complete output vector.  Internally, the computation is carried
out in a memory-efficient, horizontal (word at a time) manner.

{\footnotesize
\begin{verbatim}
void compiledFunc(long long **inputs, int vectLen, 
long long *output) {
int ctr=0;
 for (ctr=0; ctr < vectLen; ++ctr) {
  long long gate2=inputs[0][ctr];
  long long gate3=inputs[1][ctr];
  long long gate4=inputs[2][ctr];
  long long gate5=inputs[3][ctr];
  long long gate6=inputs[4][ctr];
  long long gate7=gate2 ^ gate3;
  long long gate8=gate7 ^ gate4;
  long long gate9=gate7 & gate4;
  long long gate10=gate2 & gate3;
  long long gate11=gate9 | gate10;
  long long gate12=gate8 ^ gate5;
  long long gate14=gate12 & gate6;
  long long gate15=gate8 & gate5;
  long long gate16=gate14 | gate15;
  long long gate17=gate11 & gate16;
  long long gate18=gate11 ^ gate16;
  long long gate19=gate17 | gate18;
   output[ctr] = gate19;
  }
}
\end{verbatim}
}

\newpage
The same example compiled to Java, for EWAH, is given below.
Internally, a vertical implementation is used.  A last-use analysis removes
references to bitmaps that are no longer needed, to reduce memory use. 
A more sophisticated approach would remove the null
assignments for  gates 2--6 (since references to these bitmaps
would be held by the caller), and  17 and 18 (since these references
will be lost before any more memory is required).  

{ \footnotesize
\begin{verbatim}
public static EWAHCompressedBitmap compiledFunc(
                                 EWAHCompressedBitmap [] inputs) {
 EWAHCompressedBitmap g2 = inputs[0];
 EWAHCompressedBitmap g3 = inputs[1];
 EWAHCompressedBitmap g4 = inputs[2];
 EWAHCompressedBitmap g5 = inputs[3];
 EWAHCompressedBitmap g6 = inputs[4];
 EWAHCompressedBitmap g7 = EWAHCompressedBitmap.xor(g2,g3);
 EWAHCompressedBitmap g8 = EWAHCompressedBitmap.xor(g7,g4);
 EWAHCompressedBitmap g9 = EWAHCompressedBitmap.and(g7,g4);
 g4= null;
 g7= null;
 EWAHCompressedBitmap g10 = EWAHCompressedBitmap.and(g2,g3);
 g2= null;
 g3= null;
 EWAHCompressedBitmap g11 = EWAHCompressedBitmap.or(g9,g10);
 g9= null;
 g10= null;
 EWAHCompressedBitmap g12 = EWAHCompressedBitmap.xor(g8,g5);
 EWAHCompressedBitmap g14 = EWAHCompressedBitmap.and(g12,g6);
 g6= null;
 g12= null;
 EWAHCompressedBitmap g15 = EWAHCompressedBitmap.and(g8,g5);
 g5= null;
 g8= null;
 EWAHCompressedBitmap g16 = EWAHCompressedBitmap.or(g14,g15);
 g14= null;
 g15= null;
 EWAHCompressedBitmap g17 = EWAHCompressedBitmap.and(g11,g16);
 EWAHCompressedBitmap g18 = EWAHCompressedBitmap.xor(g11,g16);
 g11= null;
 g16= null;
 EWAHCompressedBitmap g19 = EWAHCompressedBitmap.or(g17,g18);
 g17= null;
 g18= null;
 return  g19;
}
\end{verbatim}
}

\section{Suboptimality plots}
\label{sec:badness-plots}

We compare, for each competition,   each algorithm's running time 
to the best  running time.  The suboptimality is the additional running time
the algorithm requires, compared to the best running time of any of the
algorithms, divided by that best running time.
For each algorithm, we show the distribution of its suboptimality scores
on the competitions.  Since the $y$ axis is logarithmic,
we do not show the cases where the algorithm's running time was within
10\% of the best running time.
The order of algorithms on the $x$ axis is by increasing mean suboptimality.

\subsection{Wide range of $N$}

\begin{figure}
\includegraphics[width=\textwidth]{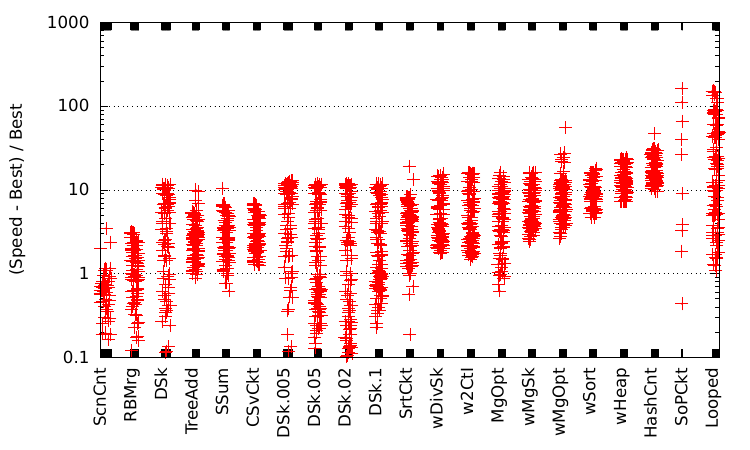}
\caption{\label{wideFigStart}Suboptimality (for cases when the algorithm was at least 10\%
slower than the fastest algorithm). \IMDBtwo , $N \leq 512$,
Similarity(100) queries.}
\end{figure}

\begin{figure}
\includegraphics[width=\textwidth]{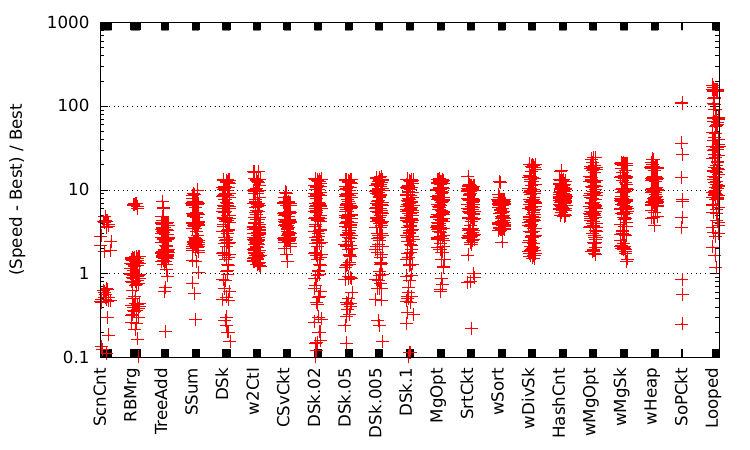}
\caption{\label{wideFigEnd}Suboptimality. \PGDVD, $N \leq 512$, Similarity(100) queries.}
\end{figure}

Some cases were omitted, when too much memory was required.
See Figs.~\ref{wideFigStart}--\ref{wideFigEnd}.

\begin{figure}
\includegraphics[width=\textwidth]{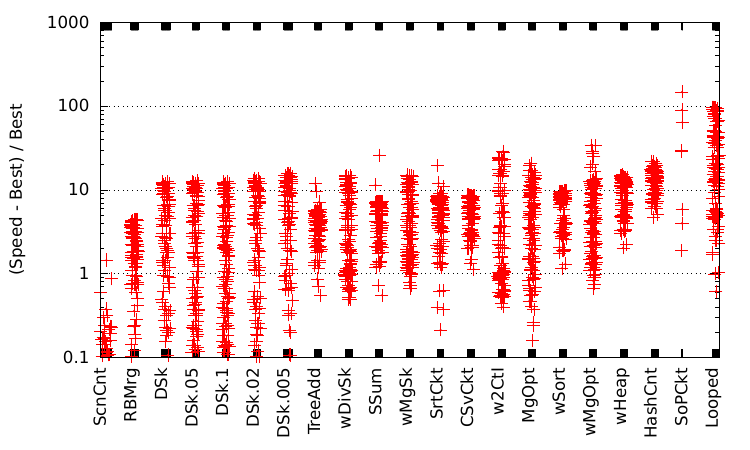}
\caption{Suboptimality. \IMDBthree, $N \leq 512$, Similarity(100) queries.}
\end{figure}

\begin{figure}
\includegraphics[width=\textwidth]{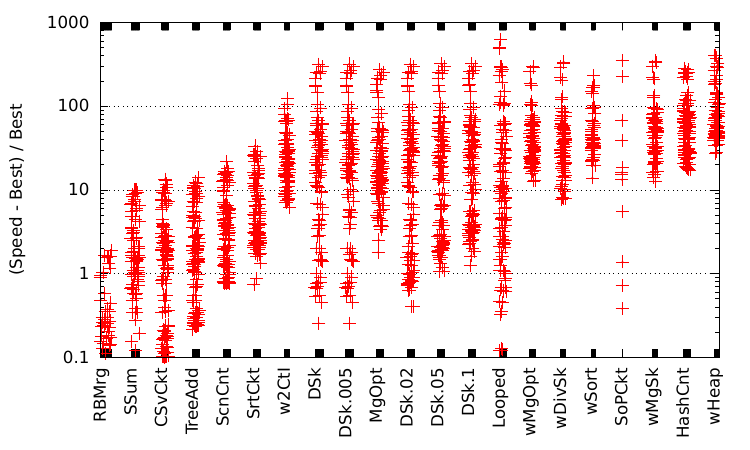}
\caption{Suboptimality. \PGDVDthree , $N \leq 512$, Similarity queries.}
\end{figure}

\begin{figure}
\includegraphics[width=\textwidth]{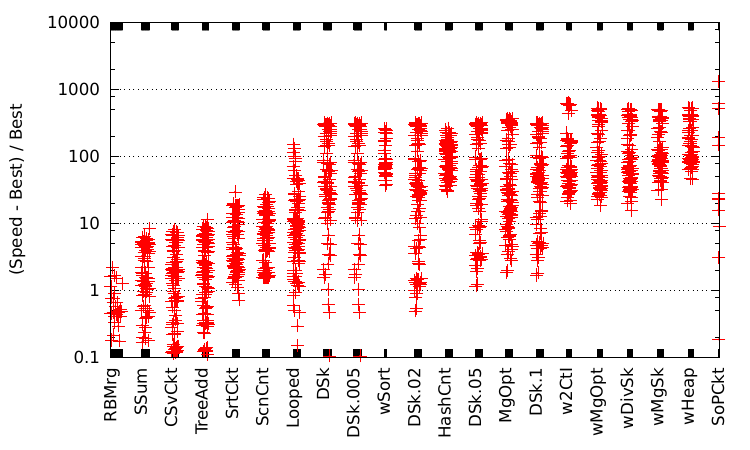}
\caption{Suboptimality. \PGDVDtwo , $N \leq 512$, Similarity queries.}
\end{figure}

\clearpage

\subsection{Narrower range of $N$}

\begin{figure}
\includegraphics[width=\textwidth]{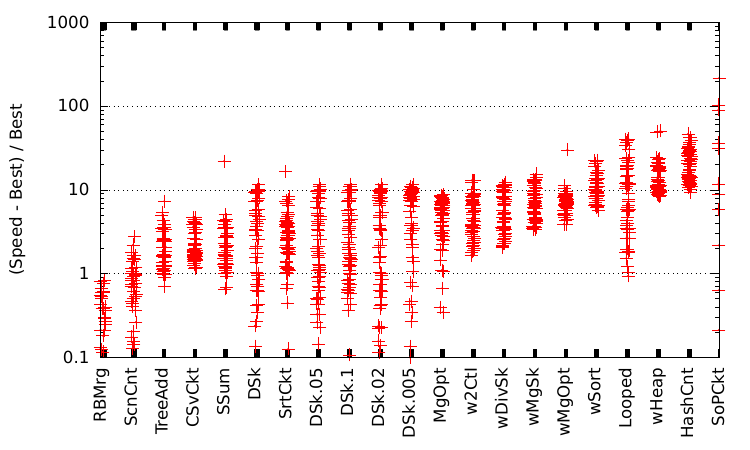}
\caption{\label{narrowerFigStart}Suboptimality. \IMDBtwo, $N \leq 128$, Similarity(10) queries.}
\end{figure}

\begin{figure}
\includegraphics[width=\textwidth]{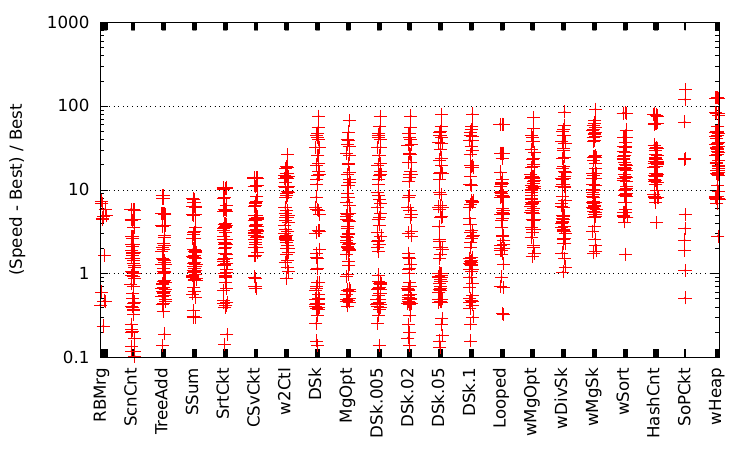}
\caption{\label{narrowerFigEnd}Suboptimality.  \CensusIncome, $N \leq 128$, Similarity(10) queries.}
\end{figure}

See Figs.~\ref{narrowerFigStart}--\ref{narrowerFigEnd}.

\begin{figure}
\includegraphics[width=\textwidth]{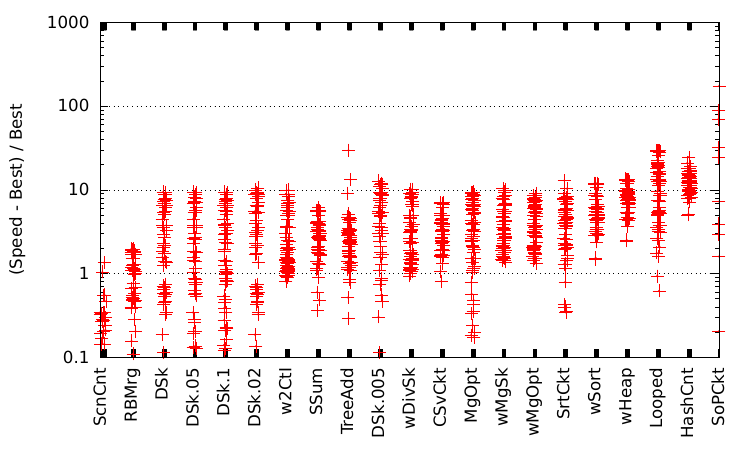}
\caption{Suboptimality. \IMDBthree, $N \leq 128$, Similarity(10) queries.}
\end{figure}

\begin{figure}
\includegraphics[width=\textwidth]{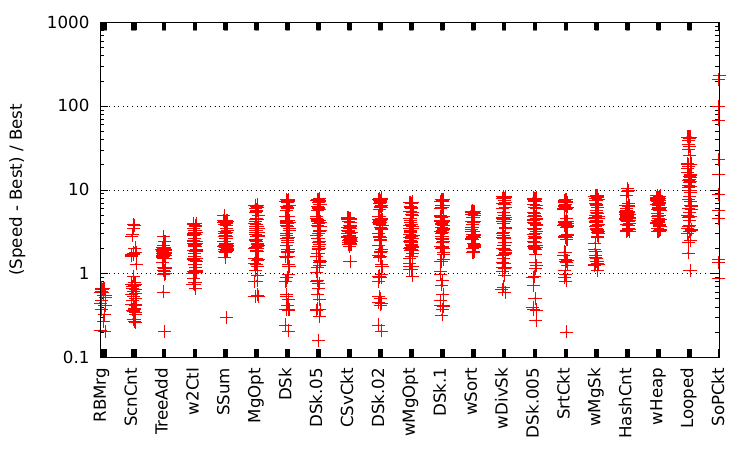}
\caption{Suboptimality. \PGDVD, $N \leq 128$, Similarity(10) queries.}
\end{figure}

\clearpage
\subsection{Narrowest range of $N$}

\begin{figure}
\includegraphics[width=\textwidth]{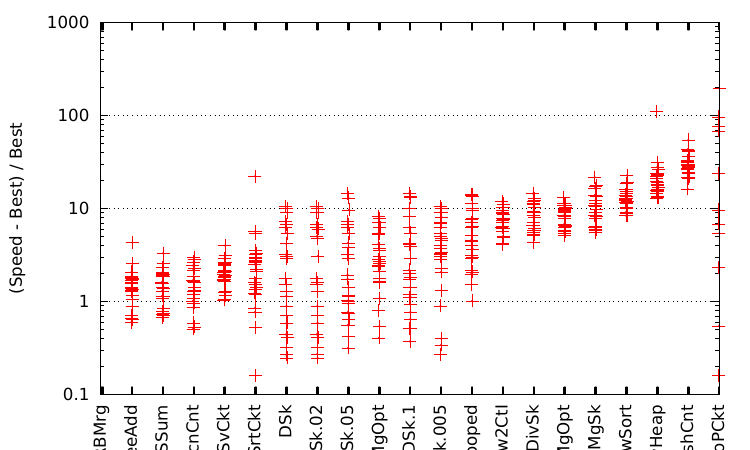}
\caption{\label{narrowestFigStart}Suboptimality. \IMDBtwo, $N \leq 32$, Similarity queries.}
\end{figure}

\begin{figure}
\includegraphics[width=\textwidth]{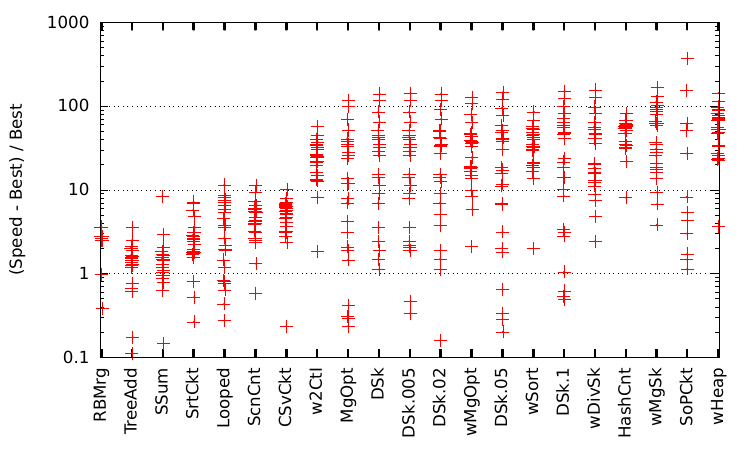}
\caption{\label{narrowestFigEnd}Suboptimality.  \CensusIncome, $N \leq 32$, Similarity queries.}
\end{figure}

See Figs.~\ref{narrowestFigStart}--\ref{narrowestFigEnd}.

\begin{figure}
\includegraphics[width=\textwidth]{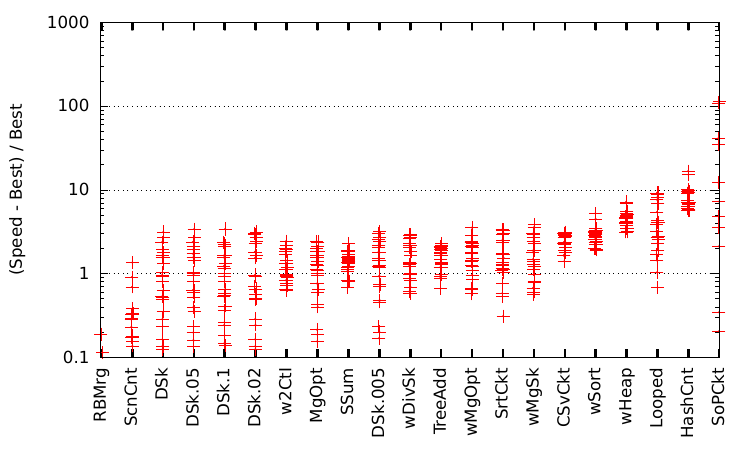}
\caption{Suboptimality.  \IMDBthree, $N \leq 32$, Similarity queries.}
\end{figure}

\begin{figure}
\includegraphics[width=\textwidth]{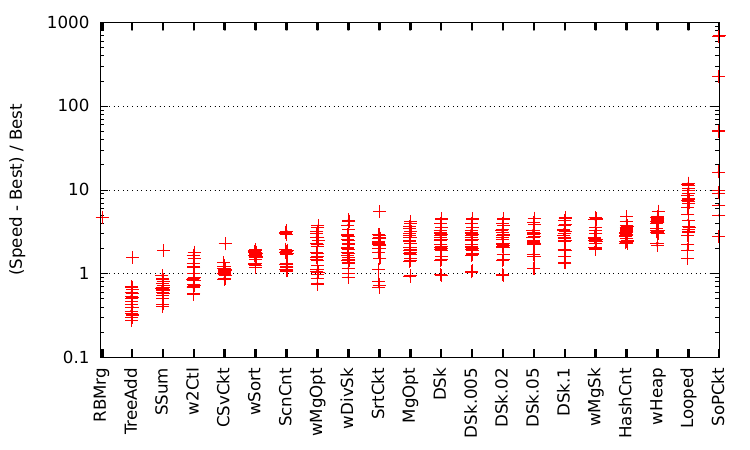}
\caption{Suboptimality.  \PGDVD, $N \leq 32$, Similarity queries.}
\end{figure}

\section{\scncnt\  on BitSet}
\label{app:scancnt-impl}

The code for \scncnt\ on BitSet inputs is particularly
simple.

\begin{lstlisting}
  Arrays.fill(counts,0);
  for (BitSet bs : inputsAsBitSets)
      for (int k = bs.nextSetBit(0); k >= 0;
                   k = bs.nextSetBit(k+1))
          counts[k]++;
  scanCountAns.clear();
  for (int k=0; k < counts.length; ++k)
      if (counts[k] >= K) scanCountAns.set(k);
\end{lstlisting}

\end{document}